\documentclass[prd,onecolumn]{revtex4}%
\usepackage{epsfig}
\usepackage{amsmath}
\usepackage{verbatim}
\usepackage{graphicx}%
\usepackage{amsfonts}%
\usepackage{amssymb}
\def\slash#1{\setbox0=\hbox{$#1$}                  \dimen0=\wd0                                    
\setbox1=\hbox{/} \dimen1=\wd1                  \ifdim\dimen0>\dimen1                              
\rlap{\hbox to \dimen0{\hfil/\hfil}}            #1                                           
\else                                              \rlap{\hbox to \dimen1{\hfil$#1$\hfil}}         /                                            \fi}

\begin{document}

\title{Direct instantons, topological charge screening and QCD glueball sum rules}
\author{Hilmar Forkel}
\affiliation{IFT - Universidade Estadual Paulista, Rua Pamplona, 145, 01405-900 Sao Paulo,
SP, Brazil }
\affiliation{Institut f\"{u}r Theoretische Physik, Universit\"{a}t Heidelberg, D-69120
Heidelberg, Germany }

\begin{abstract}
Nonperturbative Wilson coefficients of the operator product expansion (OPE)
for the spin-0 glueball correlators are derived and analyzed. A systematic
treatment of the direct instanton contributions is given, based on realistic
instanton size distributions and renormalization at the operator scale. In the
pseudoscalar channel, topological charge screening is identified as an
additional source of (semi-) hard nonperturbative physics. The screening
contributions are shown to be vital for consistency with the anomalous axial
Ward identity, and previously encountered pathologies (positivity violations
and the disappearance of the $0^{-+}$ glueball signal) are traced to their
neglect. On the basis of the extended OPE, a comprehensive quantitative
analysis of eight Borel-moment sum rules in both spin-0 glueball channels is
then performed. The nonperturbative OPE coefficients turn out to be
indispensable for consistent sum rules and for their reconciliation with the
underlying low-energy theorems. The topological short-distance physics
strongly affects the sum rule results and reveals a rather diverse pattern of
glueball properties. New predictions for the spin-0 glueball masses and decay
constants and an estimate of the scalar glueball width are given, and several
implications for glueball structure and experimental glueball searches\ are discussed.

\end{abstract}
\maketitle
\preprint{IFT-P.039/2003}

\section{Introduction}

Among exotic hadrons, i.e. those which evade classification according to the
constituent quark model, glueballs \cite{gel72,GbRew,pdg02} occupy an extreme
position. In fact, they contain no valence quarks at all and are the only
hadrons which would persist in a world without quarks. Even the exclusion of
sea quarks from QCD, as in quenched lattice simulations, is not expected to
alter their properties drastically \footnote{From the perspective of hadronic
gluon dynamics, mesons built out of heavy quarks are an opposite extreme in
which gluons mainly produce weak abelian-type Coulomb binding.}. Hence
glueball structure provides a unique source of information on nonperturbative
gluon dynamics and may even shed light on the often elusive gluon component of
the light classical hadrons (and potentially hybrids). A suggestive way to
access this information is to investigate the role of known coherent gluon
fields, with instantons as the most prominent example, in glueball structure.

The operator product expansion (OPE)\ \cite{wil69} of glueball correlation
functions provides an effective analytical framework for such investigations.
In fact, it exhibits several exceptional features of the gluonium channels
already at the qualitative level. The probably most instructive one is
directly associated with the defining characteristic of the OPE, i.e. its
factorization of the short-distance correlators into contributions from hard
and soft field modes. Indeed, since the glueball's gluon content should be
mostly nonperturbative, one might expect it to manifest itself primarily in
the soft contributions, i.e. in the gluon condensates. Surprisingly, this is
not the case: the condensate contributions are unusually weak
\cite{nov279,nov280,nar98} and cannot fully reflect the nonperturbative nature
of low-lying gluonia. This suggests that a major part of the nonperturbative
physics is relatively hard and thus resides in the Wilson coefficients. The
present paper contains a detailed analysis of such contributions and a
comprehensive study of their impact on spin-0 glueball properties.

Indications for the onset of nonperturbative physics in the scalar glueball
correlator at unusually short distances, and the ensuing departure from
asymptotic freedom, date back to the pioneering days of the QCD sum rule
approach \cite{nov81}. The prototypical candidates for such hard,
nonperturbative physics, direct instantons, describe tunneling processes which
rearrange the QCD vacuum topology \cite{sch98} in localized space-time regions
small enough to affect the $x$-dependence of the correlators over distances
$\left|  x\right|  \ll\Lambda_{QCD}^{-1}$. Although this physics enters the
Wilson coefficients, it was up to recently ignored in glueball (and other) sum
rules, with the exception of an early explorative estimate in Ref.
\cite{shu82}.

The development of instanton ``liquid'' models \cite{sch98} allowed several
bulk properties of the instanton size distribution in the vacuum to be
estimated, and the emerging scales were later supported by lattice simulations
\cite{lat-isize,rin99}. On the basis of these scales, the mentioned
exploration of instanton contributions to a $0^{++}$ glueball sum rule
\cite{shu82} found them to be of considerable size and indicated their
potential for improving the consistency with the underlying low-energy
theorem. However, the involved approximations and especially the neglect of
the crucial instanton-induced continuum contributions did not allow for
reliable estimates of glueball properties.

Only recently,\ the exact direct instanton contributions to the spin-0
glueball OPE (to leading order in $\hbar$) were obtained and the
instanton-induced continuum contributions derived \cite{for01}. On the basis
of the resulting ``instanton-improved OPE'' (IOPE) the first quantitative QCD
sum-rule analysis of direct instanton effects in scalar glueballs became
possible. This analysis resolved long-standing consistency problems of the
$0^{++}$ sum rules with purely perturbative Wilson coefficients and led to new
predictions for the scalar glueball mass and ``decay constant'' $f_{S}$
\cite{for01}.\ In particular, the direct instanton contributions were found to
increase the value of $f_{S}$ about threefold and supported earlier
indications for an exceptionally small scalar glueball size \cite{def92,sch95}%
. They also suggested a prominent role of instantons in the binding of the
$0^{++}$ glueball and generated scaling relations between its main properties
and bulk features of the instanton size distribution \cite{for01}. A
subsequent analysis of the related Gaussian sum rules \cite{har01}, based on
the same instanton contributions, confirmed some of these results and studied
more detailed parametrizations of the phenomenological correlator representation.

Previous implementations of direct instanton contributions to hadron
correlators, including those in the $0^{++}$ glueball channel, relied on
several standard approximations. However, there are reasons to suspect that
these approximations may cause artefacts in the sum-rule results. We therefore
provide a more thorough and systematic treatment of the direct-instanton
sector below. Our subsequent comprehensive sum-rule analysis will indeed
reveal a significant impact of these improvements on the predicted glueball
properties. In addition, we will extend the study of nonperturbative Wilson
coefficients to other glueball channels. While nonperturbative short-distance
contributions are small in the tensor channel (mainly due to the absence of
leading instantons corrections), they turn out to generate a rather complex
pattern of new physics in the pseudoscalar channel. The analysis of this
physics and the ensuing predictions for $0^{-+}$ glueball properties are
further central objectives of our investigation and will provide additional
insights into the role of (semi-) hard nonperturbative physics in glueball structure.

The paper is structured as follows. In Sec. \ref{gbcorrs} we discuss pertinent
general features of the glueball correlators and qualitative aspects of the
instanton contributions. We also set up the Borel sum-rule framework. In the
following Sec. \ref{pertcoeffs}, we start our discussion of the IOPE by
summarizing the known perturbative contributions to the Wilson coefficients.
The next section contains a detailed derivation and analysis of the direct
instanton contributions to both spin-0 glueball channels. Here we also
implement for the first time realistic instanton size distributions and we
explicitly renormalize the ensuing Wilson coefficients at the operator scale.
The exceptional strength of the direct instanton contributions to the spin-0
glueball correlators provides an ideal testing ground for these improvements,
which we expect to be useful in other hadron channels as well. (Previous work
on direct-instanton contributions in the classical meson and baryon channels,
and in particular the program starting with Ref. \cite{for93}, treated all
instantons as being of the same size.)

In Sec. \ref{topscr} we identify and implement new contributions to the IOPE
coefficients in the $0^{-+}$ glueball correlator. As the instantons, these
contributions are of nonperturbative origin and associated with the topology
of the gluon fields. They arise from the screening of topological charge in
the QCD vacuum and can be derived almost model-independently by requiring
consistency with the axial anomaly \cite{div80}. We present compelling
evidence for the screening contributions to be an indispensable complement to
the direct instanton contributions.

In Sec. \ref{qsra}, we embark on a comprehensive quantitative analysis of the
developed IOPE, including perturbative, direct-instanton and screening
contributions. We start by discussing several characteristic features of its
Borelmoments and their relevance for predicting glueball properties.
Subsequently, we study the origin and role of the various subtraction
constants (which appear in the lowest-moment sum rules) as well as their
impact on the sum-rule analysis. We then match the IOPE Borel moments to their
phenomenological counterparts, which include both perturbative and instanton
contributions to the duality continuum and either one or two isolated, narrow
resonances. The eight ensuing Borel sum rules are analyzed numerically, the
predictions for $0^{++}$ and $0^{-+}$ glueball properties are obtained and
their significance for glueball structure and experimental glueball searches
is discussed. Finally, in Sec. \ref{sumconcl}, we summarize our main results
and conclusions.

\section{Glueball correlation functions}

\label{gbcorrs}

The QCD correlations functions in the main glueball channels are
\begin{equation}
\Pi_{G}\left(  x\right)  =\left\langle 0|T\,O_{G}\left(  x\right)
O_{G}\left(  0\right)  |0\right\rangle ,
\end{equation}
where the local, composite operators $O_{G}$ with $G\in\left\{  S,P,T\right\}
$ are the standard gluonic interpolating fields, i.e. those with the lowest
mass dimension which carry the quantum numbers of the scalar ($0^{++}$),
pseudoscalar ($0^{-+}$) and tensor ($2^{++}$) glueballs:
\begin{align}
O_{S}\left(  x\right)   &  =\alpha_{s}G_{\mu\nu}^{a}\left(  x\right)
G^{a\mu\nu}\left(  x\right)  ,\label{sipf}\\
O_{P}\left(  x\right)   &  =\alpha_{s}G_{\mu\nu}^{a}\left(  x\right)
\tilde{G}^{a\mu\nu}\left(  x\right)  ,\label{pipf}\\
O_{T}\left(  x\right)   &  =\Theta_{\mu\nu}^{a}\left(  x\right)  .
\label{tipf}%
\end{align}
Here $\Theta_{\mu\nu}^{a}$ is the energy-momentum stress tensor of QCD and
$\tilde{G}$ is the dual of the (Minkowski) gluon field strength,
\begin{equation}
\tilde{G}_{\mu\nu}\equiv\frac{i}{2}\varepsilon_{\mu\nu\rho\sigma}G^{\rho
\sigma}.
\end{equation}
Note that $O_{S}$ ($O_{P}$) is proportional to the Yang-Mills action density
(topological charge density \footnote{In mathematical terms, this is the
Pontryagin density or second Chern class of the principle fibre bundle in
which the gauge fields live.}) of the gluon fields. The interpolators create
vacuum disturbances with the appropriate quantum numbers whose propagation
properties we will study below in the framework of a short-distance expansion.
The factors of $\alpha_{s}$ ensure renormalization group (RG) invariance (at
least to leading order in $\alpha_{s}=g_{s}^{2}/\left(  4\pi\right)  $
\footnote{RG-invariance of $O_{S}$ to all orders in $\alpha_{s}$ can be
achieved with the help of the QCD $\beta$ function by defining $O_{S}\left(
x\right)  =-4\pi^{2}\beta\left(  \alpha_{s}\right)  /\left(  \alpha_{s}%
b_{0}\right)  G_{\mu\nu}^{a}\left(  x\right)  G^{a\mu\nu}\left(  x\right)  $
(with $b_{0}=11N_{c}/3-2N_{f}/3$ and for massless quarks), from which our
interpolator is recovered to leading order in the perturbative expansion
$\beta\left(  \alpha_{s}\right)  =-b_{0}\alpha_{s}^{2}/\left(  2\pi\right)
^{2}+O\left(  \alpha_{s}^{3}\right)  $.}). The corresponding Fourier
transforms will be written as%
\begin{equation}
\Pi_{G}(Q^{2})=i\int d^{4}x\,e^{iqx}\left\langle 0|T\,O_{G}\left(  x\right)
O_{G}\left(  0\right)  |0\right\rangle \label{Qcorr}%
\end{equation}
with $Q^{2}\equiv-q^{2}$. The above expressions (and the following ones, if
not stated otherwise) refer to Minkowski space-time.

\subsection{Low-energy theorems}

\label{lets}

The zero-momentum limits of both spin-0 glueball correlators are governed by
low-energy theorems (LETs) which provide additional first-principle
information and useful consistency checks for IOPE and sum rule analysis. In
the scalar channel, the LET belongs to a class which was derived by Shifman,
Vainshtein and Zakharov on the basis of renormalization group and scaling
arguments \cite{let}. These ``dilatation'' theorems apply to the zero-momentum
limit of the correlators
\begin{equation}
\Pi_{\mathcal{O}}\left(  -q^{2}\right)  =i\int d^{4}xe^{iqx}\left\langle
TO_{S}\left(  x\right)  \mathcal{O}\left(  0\right)  \right\rangle _{npert},
\end{equation}
where $\mathcal{O}$ is an arbitrary local color-singlet operator. The
subscript refers to UV regularization by subtraction of the high-frequency
(perturbative) contributions, i.e.
\begin{equation}
\Pi_{\mathcal{O}}\left(  0\right)  =\frac{1}{\pi}\int_{0}^{\infty}\frac{ds}%
{s}\left[  \operatorname{Im}\Pi_{\mathcal{O}}\left(  s\right)
-\operatorname{Im}\Pi_{\mathcal{O}}^{high-freq}\left(  s\right)  \right]  .
\label{letsubtr}%
\end{equation}
The associated low-energy theorems read
\begin{equation}
\Pi_{\mathcal{O}}\left(  0\right)  =\frac{8\pi d_{\mathcal{O}}}{b_{0}%
}\left\langle \mathcal{O}\right\rangle
\end{equation}
where $d_{\mathcal{O}}$ is the canonical dimension of $\mathcal{O}$ and
$b_{0}=11N_{c}/3-2N_{f}/3$ is the lowest-order coefficient in the perturbative
expansion of the QCD $\beta$-function.

This general class includes as a special case with $\mathcal{O}=O_{S}$ the LET
which governs the low-energy behavior of the scalar glueball correlator:
\begin{equation}
\Pi_{S}\left(  Q^{2}=0\right)  =\frac{32\pi}{b_{0}}\left\langle \alpha
_{s}G^{2}\right\rangle . \label{sLET}%
\end{equation}
The appearance of the gluon condensate and the sizeable factor in front render
the numerical value of $\Pi_{S}\left(  Q=0\right)  $ both large and rather
uncertain. The present range of values for the gluon condensate is about
$\left\langle \alpha_{s}G^{2}\right\rangle \sim$ 0.25-0.75 GeV$^{4}$. With the
value $\langle\alpha_{s}G^{2}\rangle\equiv\left(  0.07\pm0.01\right)
\,\mathrm{GeV}^{4}$ used in \cite{nar98}, for example, one obtains an upper
limit $\Pi_{S}\left(  0\right)  \simeq11.17\left\langle \alpha_{s}%
G^{2}\right\rangle =0.78\,\mathrm{GeV}^{4}$ while $\Pi_{S}\left(  0\right)
\simeq0.4-0.6\,\mathrm{GeV}^{4}$ is probably more realistic. In the whole
range of acceptable values, however, the low-energy theorem provides by far
the largest soft contribution to the $k=-1$ scalar glueball sum rule (see below).

The zero-momentum limit of the pseudoscalar glueball correlator is likewise
governed by a low-energy theorem, although of a rather different nature
\footnote{In Ref. \cite{hal98} a somewhat heuristic attempt was made to
generalize the above type of low-energy theorem to the pseudoscalar sector in
pure Yang-Mills theory. Since we are working within full QCD, whose three
light quark flavors alter the behavior of the $0^{-+}$ correlator drastically,
we have no use for these results here.}. It is usually stated for the
zero-momentum limit of the correlator of the topological charge density%
\begin{equation}
Q\left(  x\right)  =\frac{\alpha}{8\pi}G_{\mu\nu}^{a}\tilde{G}^{a,\mu\nu
}=\frac{1}{8\pi}O_{P}\left(  x\right)  ,
\end{equation}
i.e. for the topological susceptibility
\begin{equation}
\chi_{t}=i\int d^{4}x\,\left\langle 0|T\,Q\left(  x\right)  Q\left(  0\right)
|0\right\rangle =\frac{1}{\left(  8\pi\right)  ^{2}}\Pi_{P}\left(
Q^{2}=0\right)  .
\end{equation}
In QCD with three light flavors and $m_{u,d}\ll m_{s}$, the low-energy theorem
then reads \cite{pslet}%
\begin{equation}
\chi_{t}=\frac{m_{u}m_{d}}{m_{u}+m_{d}}\left\langle \bar{q}q\right\rangle .
\end{equation}
This expression is based on the chiral Ward identities for the (anomalous)
flavor-singlet and -octet axial currents and generalizes the classic
large-$N_{c}$ result of Di Vecchia and Veneziano \cite{div80}. As a
consequence, we have%
\begin{equation}
\Pi_{P}\left(  Q^{2}=0\right)  =\left(  8\pi\right)  ^{2}\frac{m_{u}m_{d}%
}{m_{u}+m_{d}}\left\langle \bar{q}q\right\rangle \label{pLET}%
\end{equation}
which reduces to $\Pi_{P}\left(  0\right)  =0$ in the chiral limit$.$ In
Section \ref{qsra} (and \ref{subconst} in particular) we will analyze the
consistency of IOPE and glueball sum-rule results with the LETs (\ref{sLET})
and (\ref{pLET}).

\subsection{Qualitative aspects of instanton contributions}

\label{qualas}

Two properties of the spin-0 glueball correlators indicate on general grounds
that they will receive relatively large direct instanton contributions: (i)
the underlying spin-0 interpolators (\ref{sipf}) and (\ref{pipf}) couple
exceptionally strongly to instantons \cite{nov81} and (ii) the leading-order
(in $\hbar$) instanton contributions are enhanced by inverse powers of the
strong coupling (cf. Eq. (\ref{gin})).

In addition, the instanton contributions show a strong dependence on spin and
parity of the glueball correlator. This is reminiscent of the situation
encountered in light meson and baryon correlators, where the remarkable
topological, chiral, flavor and spin-color structure of the quark zero-modes
\cite{tho76} in the instanton background generates a characteristic channel
dependence pattern \cite{shu93,for203} (which is difficult to reproduce even
in more sophisticated quark models \cite{for403}). Although the zero modes and
quarks in general play a much smaller role in the glueball correlators, an
equally (if not more) distinctive channel dependence exists among them, too.
It is rooted in the (anti-) self-duality of the (anti-) instanton's field
strength,%
\begin{equation}
G_{\mu\nu}^{\left(  I,\bar{I}\right)  }=\pm\tilde{G}_{\mu\nu}^{\left(
I,\bar{I}\right)  }, \label{sdual}%
\end{equation}
(in Euclidean space-time) and therefore of purely gluonic nature. (Anti-)
self-dual gluon fields have color-electric and -magnetic fields of equal size,
$E_{i}^{a}=\pm B_{i}^{a}$. As a consequence of the Bianchi identity, they form
a subset of all solutions of the Euclidean Yang-Mills equation.

The impact of self-duality on the glueball correlators is twofold, as can be
read off directly from the interpolating fields (\ref{sipf}) - (\ref{tipf}).
First, during continuation to Minkowski space the Euclidean self-duality
equation (\ref{sdual}) picks up a factor of $i$, so that%
\begin{equation}
O_{P}^{\left(  I,\bar{I}\right)  }\left(  x\right)  =\pm iO_{S}^{\left(
I,\bar{I}\right)  }\left(  x\right)  .
\end{equation}
Therefore, direct instanton (or, more generally, self-dual) contributions to
the pseudoscalar glueball correlator are equal in size and opposite in sign to
those of the scalar correlator. This property provides a strong link between
the IOPEs of both channels, with far-reaching consequences for our analysis.
Moreover, self-duality implies that instanton-induced power corrections
(condensates) to the IOPE of both $0^{++}$ and $0^{-+}$ correlators are
restricted to a few terms and cancel in their sum \cite{nov279}.
Double-counting of soft instanton physics in both Wilson coefficients and
condensates of the IOPE is therefore excluded (with a few potential exceptions
to be discussed below). Furthermore, since direct-instanton induced
interactions turn out to be attractive in the scalar glueball channel, they
must be repulsive in the pseudoscalar channel.

Another obvious property of self-dual fields is that their energy,
proportional to $E^{2}-B^{2}$, is zero (as expected for vacuum fields). Hence
their energy-momentum tensor vanishes, i.e.
\begin{equation}
O_{T}^{\left(  I,\bar{I}\right)  }\left(  x\right)  =0.
\end{equation}
This implies the absence of leading direct instanton contributions to the
tensor glueball correlator. As a consequence, we have nothing to add to the
standard sum-rule analyses (with solely perturbative Wilson coefficients) in
the $2^{++}$ channel \cite{nov81,dom86,nar98}, except for a comment: the soft
(i.e. condensate) contributions to the tensor correlator are conspicuously
small, higher-dimensional power corrections vanish \cite{nov380} and
instanton-induced power corrections are absent \cite{nov280}. Nonperturbative
contributions to the IOPE of this correlator should therefore primarily reside
in the Wilson coefficients of higher-dimensional operators (starting with the
gluon condensate). However, direct instanton (and other hard, nonperturbative)
contributions, although not forbidden by self-duality, are suppressed by
minimally four factors of $\Lambda_{QCD}/Q\sim0.2$ where $Q\sim1$ GeV is the
typical momentum scale in sum rule analyses. Nonperturbative contributions to
the IOPE coefficients are therefore expected to be small, too, perhaps related
to the relatively weak binding in the tensor channel observed on the lattice
\cite{lee00}. In fact, the hierarchy of interactions induced by self-dual
gluon fields (i.e. attractive, absent and repulsive in the $0^{++}$, $2^{++}$
and $0^{-+}$\ channels, respectively) agrees with the level ordering among the
lowest-lying glueball states in the quenched lattice spectrum
\cite{sch95,lee00}.

\subsection{Spectral representation, Borel moments and sum rules}

In order to make contact with the glueball information in the IOPE, we match
it to the spectral representation of the correlators, i.e. we construct sum
rules. The spectral representation in the hadronic basis is conveniently
written in the form of a dispersion relation,%
\begin{equation}
\Pi_{G}\left(  Q^{2}\right)  =\frac{1}{\pi}\int_{0}^{\infty}%
ds\frac{\operatorname{Im}\Pi_{G}\left(  -s\right)  }{s+Q^{2}}, \label{disprel}%
\end{equation}
where $\operatorname{Im}\Pi_{G}\left(  -s\right)  $ is the imaginary part of
the correlator at time-like momenta and where the necessary number of
subtractions is implied but not written explicitly. Among the lowest-lying
on-shell intermediate states which contribute to the spectral function
$\rho_{G}\left(  s\right)  \equiv\operatorname{Im}\Pi_{G}\left(  -s\right)
/\pi$ are the lightest glueball in the $G$ channel (probably with admixtures
of light quarkonium) and possibly light mesons with the same quantum numbers
and sufficiently strong couplings to the gluonic interpolators. Beyond the
isolated resonances, the multi-hadron continuum will set in at a threshold
$s_{t}$. We therefore write the phenomenological representation of the
spectral function as%
\begin{equation}
\operatorname{Im}\Pi_{G}^{\left(  ph\right)  }\left(  -s\right)
=\operatorname{Im}\Pi_{G}^{\left(  res\right)  }\left(  -s\right)
+\operatorname{Im}\Pi_{G}^{\left(  cont\right)  }\left(  -s\right)  .
\label{phspecdens}%
\end{equation}

A simple and efficient description of most hadronic spectral functions, which
is particularly suited for QCD sum rule applications (and beyond, e.g. for the
parametrization and interpretation of lattice data \cite{chu93,han95}), can be
achieved with one or two narrow resonances and a duality continuum
\cite{shi79}. For the IOPE sum rules in the spin-0 glueball channels, we will
limit ourselves to maximally two isolated resonances which we describe in
zero-width approximation, i.e. as poles%
\begin{equation}
\operatorname{Im}\Pi_{G}^{\left(  res\right)  }\left(  -s\right)  =\pi
f_{G1}^{2}m_{G1}^{4}\delta\left(  s-m_{G1}^{2}\right)  +\pi f_{G2}^{2}%
m_{G2}^{4}\delta\left(  s-m_{G2}^{2}\right)
\end{equation}
with $\langle0|O_{G}\left(  0\right)  |Gi\left(  k\right)  \rangle
=f_{Gi}m_{Gi}^{2}$. Invoking local parton-hadron duality, the hadronic
continuum is replaced by its quark-gluon counterpart. (A different description
of the resonance region, in terms of Goldstone-boson pairs, can be found in
Ref. \cite{shi81}. The resonance position is then roughly estimated by
integrating the Goldstone-boson continuum up to the bound imposed by the
low-energy theorem (\ref{sLET}).) The duality continuum is obtained from the
discontinuities of the IOPE by analytical continuation to timelike momenta,
i.e.
\begin{equation}
\operatorname{Im}\Pi_{G}^{\left(  cont\right)  }\left(  -s\right)
=\theta\left(  s-s_{0}\right)  \operatorname{Im}\Pi_{G}^{\left(  IOPE\right)
}\left(  -s\right)  . \label{dualcontin}%
\end{equation}
The step function ensures that the continuum is restricted to the
invariant-mass region where it is dual to higher-lying resonance and
multi-hadron contributions. This duality region starts at the effective
threshold $s_{0}$.

For accurate QCD sum rules it is generally insufficient to match
phenomenological and IOPE representations of the correlators directly in
momentum space. However, substantial improvement can be achieved by a Borel
transform \cite{shi79}
\begin{equation}
\left[  \hat{B}\Pi\left(  Q^{2}\right)  \right]  \left(  \tau\right)
=\lim_{n,Q^{2}\rightarrow\infty}\frac{\left(  Q^{2}\right)  ^{n+1}}{n!}\left(
\frac{-d}{dQ^{2}}\right)  ^{n}\Pi\left(  Q^{2}\right)  ,\qquad\frac{Q^{2}}%
{n}\equiv\tau^{-1} \label{btrf}%
\end{equation}
(in the convention of Ref. \cite{bel82}) on both sides of the sum rules. The
Borel transform improves the ``convergence'' of the IOPE, eliminates
subtraction terms and, most importantly, implements an exponential continuum
suppression which puts more emphasis on the glueball states. For the
quantitative analysis of the spin-0 glueball channels one usually considers a
family of sum rules, based on the Borel moments
\begin{equation}
\mathcal{L}_{G,k}\left(  \tau\right)  =\hat{B}\left[  \left(  -Q^{2}\right)
^{k}\Pi_{G}(Q^{2})\right]  \left(  \tau\right)  ,\text{ \ \ \ \ \ }%
k\in\left\{  -1,0,1,2\right\}  . \label{bmoms}%
\end{equation}
The selection of these four moments, which differ in their weighting of the
intermediate-state spectrum, has practical reasons: higher powers $k>2$ reduce
the sum rule\ reliability (e.g. by shrinking the fiducial regions, see Section
\ref{srsetup}) without yielding much further information, while $k<-1$ would
introduce additional subtraction constants which are not determined by the
low-energy theorems. (The new subtraction term in the $k=-2$ sum rule of the
$0^{-+}$ glueball channel is proportional to the derivative of the topological
charge correlator at $Q^{2}=0$ and therefore relevant for understanding the
proton spin content \cite{sho92} (cf. Sec. \ref{srresults}).) The Borel
moments satisfy the recurrence relation%
\begin{equation}
\mathcal{L}_{G,k+1}\left(  \tau\right)  =-\frac{\partial}{\partial\tau
}\mathcal{L}_{G,k}\left(  \tau\right)  .
\end{equation}

By applying the operator in Eq. (\ref{bmoms}) to the dispersion relation
(\ref{disprel}) with the spectral function (\ref{phspecdens}), we obtain the
phenomenological representation of the Borel moments as
\begin{align}
\mathcal{L}_{G,k}^{\left(  ph\right)  }\left(  \tau;m_{Gi},f_{Gi}%
,s_{0}\right)   &  =f_{G1}^{2}m_{G1}^{4+2k}e^{-m_{G1}^{2}\tau}+f_{G2}%
^{2}m_{G2}^{4+2k}e^{-m_{G2}^{2}\tau}\nonumber\\
&  -\delta_{k,-1}\Pi_{G}^{\left(  ph\right)  }(0)+\frac{1}{\pi}\int_{s_{0}%
}^{\infty}dss^{k}\operatorname{Im}\Pi_{G}^{\left(  IOPE\right)  }\left(
-s\right)  e^{-s\tau}. \label{phenspec}%
\end{align}
In order to determine the hadronic parameters $m_{Gi},f_{Gi}$ and $s_{0}$, one
matches these moments to their IOPE counterparts $\mathcal{L}_{G,k}^{\left(
IOPE\right)  }$ (to be calculated below) in the fiducial $\tau$-region where
both sides are expected to be reliable. This leads to the sum rules
\begin{equation}
\mathcal{L}_{G,k}^{\left(  IOPE\right)  }\left(  \tau\right)  =\mathcal{L}%
_{G,k}^{\left(  ph\right)  }\left(  \tau;m_{Gi},f_{Gi},s_{0}\right)  .
\end{equation}
After the standard renormalization group improvement, these sum rules are
conveniently rewritten in terms of the continuum-subtracted Borel moments of
the IOPE,
\begin{equation}
\mathcal{R}_{G,k}\left(  \tau;s_{0}\right)  \equiv-\delta_{k,-1}\Pi
_{G}^{\left(  IOPE\right)  }(0)+\frac{1}{\pi}\int_{0}^{s_{0}}dss^{k}%
\operatorname{Im}\Pi_{G}^{\left(  IOPE\right)  }\left(  -s\right)  e^{-s\tau}.
\label{csubtrbmoms}%
\end{equation}
The subtraction constants $\Pi_{G}^{\left(  IOPE\right)  }(0)$ are generated
by the nonperturbative contributions to the Wilson coefficients (see below).
In terms of the $\mathcal{R}_{G,k}$, the glueball sum rules attain their final
form
\begin{equation}
\mathcal{R}_{G,k}\left(  \tau;s_{0}\right)  =-\delta_{k,-1}\Pi_{G}^{\left(
ph\right)  }(0)+f_{G1}^{2}m_{G1}^{4+2k}e^{-m_{G1}^{2}\tau}+f_{G2}^{2}%
m_{G2}^{4+2k}e^{-m_{G2}^{2}\tau} \label{gbsrs}%
\end{equation}
which isolates the glueball contributions, the subtraction term for $k=-1$ and
possibly an additional resonance on the right hand side.

The values of the subtraction constants $\Pi_{G}^{\left(  ph\right)  }(0)$ are
determined by the low-energy theorems (\ref{sLET}) and (\ref{pLET}). (The
perturbative UV contributions to the IOPE are, as in Eq. (\ref{letsubtr}),
removed by renormalization of the perturbative Wilson coefficients, cf.
Section \ref{pwcs}.) The large constant $\Pi_{S}^{\left(  ph\right)  }(0)$ is
known to dominate the scalar $k=-1$ sum rule with purely perturbative Wilson
coefficients \cite{nov280}. This leads to a much slower decay with $\tau$ than
in the $k\geq0$ sum rules and thus to a much smaller glueball mass prediction
(well below 1 GeV). The direct instanton contributions \cite{for01} overcome
this mutual inconsistency which had plagued the $0^{++}$ glueball sum rules
since their inception. We wil furtherl elaborate on the role of the
subtraction constants in Section \ref{subconst}.

\section{IOPE 1: perturbative Wilson coefficients}

\label{pertcoeffs}

Our main tool for the QCD-based calculation of the spin-0 glueball correlators
at short distances is the instanton-improved operator product expansion
(IOPE). The general expression for the IOPE at large, spacelike momenta
$Q^{2}\equiv-q^{2}\gg\Lambda_{QCD}$, \
\begin{equation}
\Pi_{G}(Q^{2})=\sum_{d=0,4,...}\tilde{C}_{d}^{\left(  G\right)  }\left(
Q^{2};\mu\right)  \left\langle \hat{O}_{d}\right\rangle _{\mu}\text{ },
\end{equation}
exhibits the characteristic factorization of contributions from hard and soft
field modes at the operator renormalization scale $\mu$: hard modes with
momenta $\left|  k\right|  >\mu$ contribute to the momentum-dependent Wilson
coefficients $\tilde{C}_{d}\left(  Q^{2};\mu\right)  $ while soft modes with
$\left|  k\right|  \leq\mu$ contribute to the vacuum expectation values
(``condensates'') of the operators $\hat{O}_{d}$ of dimension $d$ which are
renormalized at $\mu$.

The perturbative contributions to the Wilson coefficients constitute the
conventional OPE and will be discussed in the present section. In addition,
the Wilson coefficients receive crucial nonperturbative contributions from
direct instantons, i.e. those with sizes $\rho\lesssim\mu^{-1}$ \footnote{In
studies of\ the OPE at large orders $d$, instantons of a different type were
employed as a tool for assessing the impact of near-origin singularities on
the asymptotic OPE behavior \cite{chi97}. (In the context of an operational,
OPE-based definition of quark-hadron duality such effects are referred to as
``local duality violations''.)}. Their evaluation is the subject of Section
\ref{dirinst}. Additional hard and nonperturbative contributions to the IOPE
of the $0^{-+}$ correlator, due to topological charge screening, are
identified and evaluated in Section \ref{topscr}.

\subsection{Perturbative OPE coefficients}

\label{pwcs}

In this section we summarize what is known about the perturbative
contributions to the Wilson coefficients $\tilde{C}_{d}^{\left(  S,P\right)
}\left(  Q^{2}\right)  $ of the scalar and pseudoscalar glueball correlators.
At present, perturbative contributions are available up to maximally three
loops, and for operators up to dimension $d=8$. The accordingly truncated OPE
has the form
\begin{equation}
\Pi_{G}^{\left(  pc\right)  }(Q^{2})\simeq\tilde{C}_{0}^{\left(  G\right)
}+\tilde{C}_{4}^{\left(  G\right)  }\left\langle \hat{O}_{4}\right\rangle
+\tilde{C}_{6}^{\left(  G\right)  }\left\langle \hat{O}_{6}\right\rangle
+\tilde{C}_{8}^{\left(  G\right)  }\left\langle \hat{O}_{8}\right\rangle
\end{equation}
where $G=S,P$ denotes the $\left(  0^{++},0^{-+}\right)  $ glueball channel.

Up to $O\left(  \alpha_{s}^{2}\right)  $ (i.e. up to 3 loops - the two powers
of $\alpha_{s}$ from the interpolating fields are not counted here), the
coefficient of the unit operator has the generic $Q^{2}$ dependence%

\begin{equation}
\tilde{C}_{0}^{\left(  G\right)  }=Q^{4}\ln\left(  \frac{Q^{2}}{\mu^{2}%
}\right)  \left[  A_{0}^{\left(  G\right)  }+A_{1}^{\left(  G\right)  }%
\ln\left(  \frac{Q^{2}}{\mu^{2}}\right)  +A_{2}^{\left(  G\right)  }\ln
^{2}\left(  \frac{Q^{2}}{\mu^{2}}\right)  \right]
\end{equation}
were we have omitted irrelevant subtraction polynomials in $Q^{2}$. Since the
next-to-next-to-leading order (N$^{2}$LO, i.e. $O\left(  \alpha_{s}%
^{2}\right)  $) corrections are known only for the unit-operator coefficients
$\tilde{C}_{0}^{\left(  G\right)  }$, strict compliance with the perturbative
expansion would require their exclusion. Nevertheless, due to their
exceptional size, the power-suppression of their higher-dimensional
counterparts $\tilde{C}_{d\geq4}^{\left(  G\right)  }$ and their special role
in the duality continuum, they are usually taken into account.

The radiative corrections to the gluon-condensate coefficient $\tilde{C}_{4}$
are known up to $O\left(  \alpha_{s}\right)  $. The NLO corrections are of
particular importance in $\tilde{C}_{4}$ since the leading $O\left(
\alpha_{s}^{0}\right)  $ term is momentum-independent and therefore
contributes only to the lowest ($k=-1$) Borel-moment sum rule. The overall
$Q^{2}$ dependence of $\tilde{C}_{4}$ is
\begin{equation}
\tilde{C}_{4}^{\left(  G\right)  }\left\langle \hat{O}_{4}\right\rangle
=B_{0}^{\left(  G\right)  }+B_{1}^{\left(  G\right)  }\ln\left(  \frac{Q^{2}%
}{\mu^{2}}\right)  .
\end{equation}
The 3-gluon condensate term has the generic form%
\begin{equation}
\tilde{C}_{6}\left\langle \hat{O}_{6}\right\rangle =\frac{1}{Q^{2}}\left[
C_{0}+C_{1}\ln\left(  \frac{Q^{2}}{\mu^{2}}\right)  \right]
\end{equation}
while the 4-gluon condensates produce a term of the form%
\begin{equation}
\tilde{C}_{8}\left\langle \hat{O}_{8}\right\rangle =D_{0}\frac{1}{Q^{4}}.
\end{equation}

Collecting all four contributions, we arrive at the general expression for the
$Q^{2}$-dependence of the OPE with perturbative Wilson-coefficients in both
spin-0 glueball correlators,
\begin{align}
\Pi_{i}^{\left(  pc\right)  }(Q^{2})  &  \simeq\left[  A_{0}+A_{1}\ln\left(
\frac{Q^{2}}{\mu^{2}}\right)  +A_{2}\ln^{2}\left(  \frac{Q^{2}}{\mu^{2}%
}\right)  \right]  Q^{4}\ln\left(  \frac{Q^{2}}{\mu^{2}}\right)  +B_{0}%
+B_{1}\ln\left(  \frac{Q^{2}}{\mu^{2}}\right) \nonumber\\
&  +\left[  C_{0}+C_{1}\ln\left(  \frac{Q^{2}}{\mu^{2}}\right)  \right]
\frac{1}{Q^{2}}+D_{0}\frac{1}{Q^{4}}. \label{ope}%
\end{align}
During renormalization group improvement in Sec. \ref{pertBorelMom} the
coefficients $A_{i}-D_{i}$ will acquire a logarithmic $Q^{2}$-dependence due
to the presence of the running coupling $\alpha_{s}$ (and of the anomalous
dimension $\gamma_{3g}$ in the case of the $C_{i}$).

In view of potentially large quarkonium admixtures to physical glueball states
we emphasize that the perturbative Wilson coefficients receive explicit quark
loop contributions, both from radiative corrections and through the
perturbative $\beta$-function during RG-improvement. Especially in the scalar
channel \footnote{In Ref. \cite{asn92} it has been argued that the results of
the standard-OPE sum rules (without direct instanton contributions) for the
pseudoscalar glueball for $N_{f}=0$ is close to that of real QCD ($N_{f}=3$)
because the leading perturbative Wilson coefficient has a weak $N_{f}%
$-dependence (weaker than in the scalar channel) and since mixing with
quarkonium seems to be smaller in the $0^{-+}$ than in the $0^{++}$ sector
\ \cite{nar284}.} the perturbative $N_{f}$-dependence ($N_{f}$ is the number
of light quark flavors with $m_{q}\lesssim\Lambda_{QCD}$) can be significant
(see also Eq. (\ref{sLET})). The soft quark contributions to the condensates
and their $N_{f}$-dependence are very likely larger but more difficult to estimate.

\subsubsection{$0^{++}$ channel}

At present, the most accurate perturbative coefficients of the unit-operator,
renormalized in the $\overline{MS}$ scheme and with threshold effects
included, can be found in Refs. \cite{che97,har01}. For three light quark
flavors (i.e. $N_{c}=N_{f}=3$) they read
\begin{align}
A_{0}^{\left(  S\right)  }  &  =-2\left(  \frac{\alpha_{s}}{\pi}\right)
^{2}\left[  1+\frac{659}{36}\frac{\alpha_{s}}{\pi}+247.480\left(
\frac{\alpha_{s}}{\pi}\right)  ^{2}\right]  ,\qquad\\
A_{1}^{\left(  S\right)  }  &  =2\left(  \frac{\alpha_{s}}{\pi}\right)
^{3}\left[  \frac{b_{0}}{4}+65.781\frac{\alpha_{s}}{\pi}\right]  ,\qquad
A_{2}^{\left(  S\right)  }=-10.1250\left(  \frac{\alpha_{s}}{\pi}\right)
^{4}.
\end{align}
In addition to the rather recently calculated $O\left(  \alpha_{s}^{2}\right)
$ contributions, these expressions contain several corrections to the
$O\left(  \alpha_{s}\right)  $ contributions which were implemented in older
sum rules analyses. In Ref. \cite{bag90}, the quark loop contributions were
omitted (corresponding to $N_{f}=0$) and $A_{0}^{\left(  S\right)  }$
contained a small error \cite{kat82} which was later corrected \cite{kat96}
and implemented into $A_{0}^{\left(  S\right)  }$ \cite{nar98}. The $O\left(
\alpha_{s}\right)  $ contribution to $A_{0}^{\left(  S\right)  }$ was
corrected once more in Ref. \cite{che97}. As a consequence, the numerical
coefficient given above is about 25\% larger than the one used in Ref.
\cite{nar98}.

The coefficients $B_{i}$ of the lowest-dimensional nontrivial operator, i.e.
the gluon condensate, receive important radiative corrections which were
calculated in Ref. \cite{bag90} for $N_{f}=0$. Including the quark-loop
contributions for $N_{f}=3$ \cite{har01}, they become
\begin{equation}
B_{0}^{\left(  S\right)  }=4\alpha_{s}\left[  1+\frac{175}{36}\frac{\alpha
_{s}}{\pi}\right]  \left\langle \alpha_{s}G^{2}\right\rangle ,\qquad
B_{1}^{\left(  S\right)  }=-\frac{\alpha_{s}^{2}b_{0}}{\pi}\left\langle
\alpha_{s}G^{2}\right\rangle .
\end{equation}
($\left\langle \alpha_{s}G^{2}\right\rangle \equiv\left\langle \alpha
_{s}G_{\mu\nu}^{a}G^{a\mu\nu}\right\rangle $) As a consequence, the $O\left(
\alpha_{s}\right)  $ contribution to $B_{0}$ is about 20\% larger than the one
used in Ref. \cite{nar98}. Note that the $O\left(  1\right)  $ coefficient is
$Q^{2}$-independent. Hence, the leading power correction will enter all but
the lowest Borel-moment (i.e. $\mathcal{R}_{S,-1}$) sum rules solely via
radiative corrections.

The 3-gluon condensate term, again including $O\left(  \alpha_{s}\right)  $
radiative corrections, was first calculated for $N_{f}=0$ \cite{bag90} and
adapted to $N_{f}=3$ in Ref. \cite{har01}, with the result
\begin{equation}
C_{0}^{\left(  S\right)  }=8\alpha_{s}^{2}\left\langle gG^{3}\right\rangle
,\qquad C_{1}^{\left(  S\right)  }=0.
\end{equation}
($\left\langle gG^{3}\right\rangle \equiv\langle gf_{abc}G_{\mu\nu}^{a}%
G_{\rho}^{b\nu}G^{c\rho\mu}\rangle$) Both Refs. \cite{bag90} and \cite{nar98}
use the $N_{f}=0$ value $C_{1}^{\left(  S\right)  }=-58\alpha_{s}%
^{3}\left\langle gG^{3}\right\rangle $ \cite{bag90} instead. (Ref.
\cite{nar98} otherwise sets $N_{f}=3$.) Up to $O\left(  \alpha_{s}\right)  $,
there is no explicit $N_{f}$-dependence in $C_{0}^{\left(  S\right)  }$.

The contributions from the highest-dimensional ($d=8$) operators appear in the
combination
\begin{equation}
\left\langle \alpha_{s}^{2}G^{4}\right\rangle _{S}:=14\left\langle \left(
\alpha_{s}f_{abc}G_{\mu\rho}^{b}G_{\nu}^{\rho c}\right)  ^{2}\right\rangle
-\left\langle \left(  \alpha_{s}f_{abc}G_{\mu\nu}^{b}G_{\rho\lambda}%
^{c}\right)  ^{2}\right\rangle
\end{equation}
of 4-gluon condensates and were calculated to leading order in the pioneering
paper \cite{nov279}:%
\begin{equation}
D_{0}^{\left(  S\right)  }=8\pi\alpha_{s}\left\langle \alpha_{s}^{2}%
G^{4}\right\rangle . \label{d0s}%
\end{equation}
Estimates of the numerical value of $\left\langle \alpha_{s}^{2}%
G^{4}\right\rangle $ are customarily based on the vacuum factorization
approximation \cite{nov279},%
\begin{equation}
\left\langle \alpha_{s}^{2}G^{4}\right\rangle _{S}\simeq\frac{9}{16}%
\langle\alpha_{s}G^{2}\rangle^{2}.
\end{equation}
The uncertainty in this estimate does barely affect the quantitative sum rule
analysis since the impact of the coefficient (\ref{d0s}) is almost negligible.

\subsubsection{$0^{-+}$ channel}

The lowest-order perturbative contributions to the unit-operator coefficient
of the $0^{-+}$ correlator where first obtained in Ref. \cite{nov279}.
Subsequently, the calculation was extended to $O\left(  \alpha_{s}\right)  $
\cite{asn92} and augmented by the 3-loop corrections \cite{che97} in Ref.
\cite{zha03} (where also an error in the $O\left(  \alpha_{s}\right)  $
contribution was corrected). The result is
\begin{align}
A_{0}^{\left(  P\right)  }  &  =-2\left(  \frac{\alpha_{s}}{\pi}\right)
^{2}\left[  1+20.75\frac{\alpha_{s}}{\pi}+305.95\left(  \frac{\alpha_{s}}{\pi
}\right)  ^{2}\right]  ,\text{ \ \ \ \ \ }\\
A_{1}^{\left(  P\right)  }  &  =2\left(  \frac{\alpha_{s}}{\pi}\right)
^{3}\left[  \frac{9}{4}+72.531\frac{\alpha_{s}}{\pi}\right]  ,\text{
\ \ \ \ \ }A_{2}^{\left(  P\right)  }=-10.1250\left(  \frac{\alpha_{s}}{\pi
}\right)  ^{4}.
\end{align}
(Ref. \cite{asn92} contains a misprint of the sign of the $\ln^{2}$-term in
its Eq. (2).) The leading-order contributions to $A_{0,1}$ agree, as expected,
with those of the scalar glueball correlator. The NLO corrections of Ref.
\cite{che97} affect $A_{0}$ only weakly but increase $A_{1}$ by almost a
factor of four. The N$^{2}$LO contribution increases $A_{0}$ by about a factor
of two. Finally, $A_{2}\sim10^{-3}$ for $\alpha/\pi\sim0.1$ and thus is
practically negligible compared to $A_{0}$ and $A_{1}$.

The gluon-condensate contribution was calculated to lowest order in Ref.
\cite{nov279} and up to $O\left(  \alpha_{s}\right)  $ in Ref. \cite{asn92},
with the result
\begin{equation}
B_{0}^{\left(  P\right)  }=4\pi\frac{\alpha}{\pi}\left\langle \alpha_{s}%
G^{2}\right\rangle ,\text{ \ \ \ \ \ }B_{1}^{\left(  P\right)  }=9\pi\left(
\frac{\alpha_{s}}{\pi}\right)  ^{2}\left\langle \alpha_{s}G^{2}\right\rangle .
\end{equation}
(The sign of the LO contribution to $B_{0}$ is corrected and the NLO
contribution is not determined by the renormalization group \cite{asn92}).

The $\left\langle gG^{3}\right\rangle $ term was calculated to lowest order in
Ref. \cite{nov279}. The 2-loop contribution was first obtained for $N_{f}=0$
\cite{asn92} and later for $N_{f}=3$ \cite{zha03}, with the result
\begin{equation}
C_{0}^{\left(  P\right)  }=-8\pi^{2}\left(  \frac{\alpha_{s}}{\pi}\right)
^{2}\left\langle gG^{3}\right\rangle ,\text{ \ \ \ \ \ }C_{1}^{\left(
P\right)  }=0.
\end{equation}
(For $N_{f}=0$ one has $C_{1}^{\left(  P\right)  }=58\pi^{2}\left(  \alpha
_{s}/\pi\right)  ^{3}\left\langle gG^{3}\right\rangle $ instead \cite{asn92}.)

For the $d=8$ contributions, which is associated with the combination
\begin{equation}
\left\langle g^{4}G^{4}\right\rangle _{P}=2g_{s}^{4}f^{abc}f^{ade}\left\langle
G_{\mu\nu}^{b}G_{\alpha\beta}^{c}G^{d,\mu\nu}G^{e,\alpha\beta}+10G_{\mu\alpha
}^{b}G^{c,\alpha\nu}G^{d,\mu\beta}G_{\beta\nu}^{e}\right\rangle
\end{equation}
of 4-gluon condensates, one finds \cite{nov279}
\begin{equation}
D_{0}^{\left(  P\right)  }=\frac{1}{4}\frac{\alpha_{s}}{\pi}\left\langle
g_{s}^{4}G^{4}\right\rangle _{P},
\end{equation}
which is (explicitly and to the given $O\left(  \alpha_{s}\right)  $) $N_{f}%
$-independent. (Ref. \cite{zha03} contains a typographical error in the
definition of $\left\langle g_{s}^{4}G^{4}\right\rangle _{P}$.) As in the
scalar case, the condensate $\left\langle g_{s}^{4}G^{4}\right\rangle _{P}$ is
conventionally factorized \cite{nov279} according to%
\begin{align}
\left\langle \left(  f^{abc}G_{\mu\nu}^{b}G_{\alpha\beta}^{c}\right)
^{2}\right\rangle  &  \simeq\frac{5}{16}\left\langle G_{\mu\nu}^{a}G^{a,\mu
\nu}\right\rangle ^{2},\\
\left(  f^{abc}G_{\mu\alpha}^{b}G^{c,\alpha\nu}\right)  ^{2}  &
\simeq\frac{1}{16}\left\langle G_{\mu\nu}^{a}G^{a,\mu\nu}\right\rangle ^{2},
\end{align}
so that%
\begin{equation}
\left\langle g^{4}G^{4}\right\rangle _{P}=30\pi^{2}\left\langle \alpha
_{s}G^{2}\right\rangle ^{2}%
\end{equation}
and thus finally%
\begin{equation}
D_{0}^{\left(  P\right)  }=\frac{15\pi}{2}\alpha_{s}\left\langle \alpha
_{s}G^{2}\right\rangle ^{2}.
\end{equation}
As in the scalar case, the uncertainty in $D_{0}$ has practically no impact on
the quantitative sum-rule analysis.

We note in passing that the perturbative contributions to $\tilde{C}_{0}$
increase in both spin-0 glueball correlators with the order of $\alpha_{s}$.
The size ratios of the $O\left(  \alpha_{s}^{0}\right)  $ : $O\left(
\alpha_{s}^{1}\right)  $ : $O\left(  \alpha_{s}^{2}\right)  $ corrections to
$A_{0}^{\left(  P\right)  }$ are about $1:2:3$, for example (at $\alpha
_{s}/\pi\simeq0.1$). This type of behavior is often encountered in QCD
perturbation series and might indicate the approach to the asymptotic regime.

\subsection{Borel moments, continuum subtraction and RG improvement}

\label{pertBorelMom}

The perturbative Wilson coefficients enter the glueball sum rules
(\ref{gbsrs}) through their contributions to the continuum-subtracted Borel
moments (\ref{csubtrbmoms}). In the present section we outline the calculation
of these moments in the dispersive representation. The unit-operator
coefficients contain the strongest UV divergences and therefore require the
maximal number of subtractions, which turns out to be three (see below):
\begin{equation}
\Pi_{G}^{\left(  cont\right)  }\left(  Q^{2}\right)  =\Pi_{G}\left(  0\right)
-Q^{2}\Pi_{G}^{\prime}\left(  0\right)  +\frac{1}{2}Q^{4}\Pi_{G}^{\prime
\prime}\left(  0\right)  -\frac{Q^{6}}{\pi}\int_{s_{0}}^{\infty}%
ds\frac{\operatorname{Im}\Pi_{G}\left(  -s\right)  }{s^{3}\left(
s+Q^{2}\right)  }.
\end{equation}
The further evaluation of the continuum contributions requires explicit
expressions for the imaginary parts of the perturbative Wilson coefficients at
timelike momenta. Those are obtained from the coefficients of the last section
whose $q$-dependence resides in a combination of powers $\left(  q^{2}\right)
^{n}$ and logarithms $\ln\left(  -q^{2}/\mu^{2}\right)  $. The imaginary parts
of the pure power corrections (i.e. the $B_{0}$, $C_{0}$ and $D_{0}$
terms)\ are concentrated at $s=0$ and thus do not contribute to the duality
continuum, while the logarithms acquire an imaginary part $-\pi$ at time-like
momenta. For $s>0$ we therefore find%
\begin{equation}
\operatorname{Im}\Pi_{G}^{\left(  pc\right)  }\left(  -s\right)  =-\pi\left[
A_{0}s^{2}+2A_{1}s^{2}\ln\left(  \frac{s}{\mu^{2}}\right)  +A_{2}s^{2}\left(
3\ln^{2}\left(  \frac{s}{\mu^{2}}\right)  -\pi^{2}\right)  +B_{1}-\frac{C_{1}%
}{s}\right]  . \label{imsope}%
\end{equation}

The next step is to calculate the Borel moments of $\Pi_{G}^{\left(
pc\right)  }\left(  Q^{2}\right)  $ according to Eq. (\ref{bmoms}) and to
subtract the continuum contributions, which leads to \
\begin{equation}
\mathcal{L}_{G,k}^{\left(  pc\right)  }\left(  \tau;s_{0}\right)
=\mathcal{L}_{G,k}^{\left(  pc\right)  }\left(  \tau\right)  -\frac{1}{\pi
}\int_{s_{0}}^{\infty}dss^{k}\operatorname{Im}\Pi_{G}^{\left(  pc\right)
}\left(  -s\right)  e^{-\tau s}. \label{lcontinsubtr}%
\end{equation}
After inserting the imaginary part, Eq. (\ref{imsope}), the perturbative
continuum can be evaluated explicitly. The resulting, final expressions for
$\mathcal{L}_{G,k}\left(  \tau;s_{0}\right)  $ are conventionally written in
terms of the partial sums
\begin{equation}
\rho_{k}\left(  x\right)  =e^{-x}\sum_{n=0}^{k}\frac{x^{n}}{n!} \label{rhok}%
\end{equation}
and the exponential integral%
\begin{equation}
E_{1}\left(  x\right)  =\int_{1}^{\infty}dt\frac{e^{-xt}}{t}.
\end{equation}
They can be found in Appendix \ref{bpmoms}.

Before implementing the continuum-subtracted Borel moments into the sum rules
(\ref{gbsrs}), one still has to perform the standard renormalization group
(RG) improvement. Using the leading-order perturbative $\beta$-function, RG
improvement amounts to the replacements \cite{nov279}%
\begin{align}
\mu^{2}  &  \rightarrow\frac{1}{\tau},\\
\alpha &  \rightarrow\bar{\alpha}_{s}\left(  1/\tau\right)  =\frac{-4\pi
}{b_{0}\ln\left(  \Lambda^{2}\tau\right)  },\\
\left\langle gG^{3}\right\rangle  &  \rightarrow\left(  \frac{\bar{\alpha}%
_{s}\left(  1/\tau\right)  }{\bar{\alpha}_{s}\left(  \mu^{2}\right)  }\right)
^{7/11}\left\langle gG^{3}\right\rangle .
\end{align}
(Formally, the inclusion of NLO corrections to the $\beta$-function would be
more consistent for the evolution of the $O\left(  \alpha_{s}^{2}\right)  $
contributions to the unit-operator coefficients. However, their impact is
negligible in the kinematical region relevant for the sum rules.) Following
the notation of Eq. (\ref{csubtrbmoms}), the resulting Borel moments will be
renamed $\mathcal{R}_{G,k}^{\left(  pc\right)  }\left(  \tau,s_{0}\right)  $.
(It is common practice to perform the RG-improvement after the Borel
transform, i.e. after the subtraction terms are eliminated. Interchanging this
order leads to the same results, up to corrections of higher order in
$\alpha_{s}$.)

The final expressions for the continuum-subtracted Borel moments, as generated
by the perturbative IOPE coefficients, are ($\gamma=0.5772$ is Euler's
constant and the small contributions from the $A_{2}$ coefficients are omitted):

\begin{enumerate}
\item
\begin{align}
\mathcal{R}_{G,-1}^{\left(  pc\right)  }\left(  \tau,s_{0}\right)   &
=-\frac{A_{0}\left(  \tau\right)  }{\tau^{2}}\left[  1-\rho_{1}\left(
s_{0}\tau\right)  \right]  +\frac{2A_{1}\left(  \tau\right)  }{\tau^{2}%
}\left[  \gamma-1+E_{1}\left(  s_{0}\tau\right)  +e^{-\tau s_{0}}+\ln\left(
s_{0}\tau\right)  \rho_{1}\left(  s_{0}\tau\right)  \right] \nonumber\\
&  -B_{0}\left(  \tau\right)  +B_{1}\left(  \tau\right)  \left[  \gamma
+E_{1}\left(  s_{0}\tau\right)  \right]  -C_{0}\left(  \tau\right)  \tau
+C_{1}\left(  \tau\right)  \tau\left[  \gamma-1-\frac{e^{-\tau s_{0}}}{\tau
s_{0}}+E_{1}\left(  s_{0}\tau\right)  \right]  -\frac{D_{0}\left(
\tau\right)  }{2}\tau^{2} \label{rm1pc}%
\end{align}

\item
\begin{align}
\mathcal{R}_{G,0}^{\left(  pc\right)  }\left(  \tau,s_{0}\right)   &
=-\frac{2A_{0}}{\tau^{3}}\left[  1-\rho_{2}\left(  s_{0}\tau\right)  \right]
+\frac{4A_{1}}{\tau^{3}}\left[  \gamma-\frac{3}{2}+E_{1}\left(  s_{0}%
\tau\right)  +\rho_{0}\left(  s_{0}\tau\right)  +\frac{1}{2}\rho_{1}\left(
s_{0}\tau\right)  +\ln\left(  s_{0}\tau\right)  \rho_{2}\left(  s_{0}%
\tau\right)  \right] \nonumber\\
&  -\frac{B_{1}}{\tau}\left[  1-\rho_{0}\left(  s_{0}\tau\right)  \right]
+C_{0}-C_{1}\left[  \gamma+E_{1}\left(  s_{0}\tau\right)  \right]  +D_{0}%
\tau\label{r0pc}%
\end{align}

\item
\begin{align}
\mathcal{R}_{G,1}^{\left(  pc\right)  }\left(  \tau,s_{0}\right)   &
=-\frac{6A_{0}}{\tau^{4}}\left[  1-\rho_{3}\left(  s_{0}\tau\right)  \right]
+\frac{12A_{1}}{\tau^{4}}\left[  \gamma-\frac{11}{6}+E_{1}\left(  s_{0}%
\tau\right)  +\rho_{0}\left(  s_{0}\tau\right)  +\frac{1}{2}\rho_{1}\left(
s_{0}\tau\right)  \right. \nonumber\\
&  +\left.  \frac{1}{3}\rho_{2}\left(  s_{0}\tau\right)  +\ln\left(  s_{0}%
\tau\right)  \rho_{3}\left(  s_{0}\tau\right)  \right]  -\frac{B_{1}}{\tau
^{2}}\left[  1-\rho_{1}\left(  s_{0}\tau\right)  \right]  +\frac{C_{1}}{\tau
}\left[  1-\rho_{0}\left(  s_{0}\tau\right)  \right]  -D_{0} \label{r1pc}%
\end{align}

\item
\begin{align}
\mathcal{R}_{G,2}^{\left(  pc\right)  }\left(  \tau,s_{0}\right)   &
=-\frac{24A_{0}}{\tau^{5}}\left[  1-\rho_{4}\left(  s_{0}\tau\right)  \right]
+\frac{48A_{1}}{\tau^{5}}\left[  \gamma-\frac{25}{12}+E_{1}\left(  s_{0}%
\tau\right)  \right.  +\rho_{0}\left(  s_{0}\tau\right)  +\frac{1}{2}\rho
_{1}\left(  s_{0}\tau\right) \nonumber\\
&  +\left.  \frac{1}{3}\rho_{2}\left(  s_{0}\tau\right)  +\frac{1}{4}\rho
_{3}\left(  s_{0}\tau\right)  +\ln\left(  s_{0}\tau\right)  \rho_{4}\left(
s_{0}\tau\right)  \right]  -\frac{2B_{1}}{\tau^{3}}\left[  1-\rho_{2}\left(
s_{0}\tau\right)  \right]  +\frac{C_{1}}{\tau^{2}}\left[  1-\rho_{1}\left(
s_{0}\tau\right)  \right]  . \label{r2pc}%
\end{align}
\end{enumerate}

We conclude this section with a comment on the perturbative $N_{f}$-dependence
and the choice $N_{f}=3$ adopted above. At present, the identification of all
experimental glueball candidates remains controversial \cite{GbRew,pdg02} and
lattice simulations including\ light quarks are still limited to relatively
small lattices and poor statistics \cite{har02}. Hence, quenched lattice
calculations \cite{lee00} are currently the most reliable source for
information on the glueball spectrum. One might therefore be tempted to argue
that setting $N_{f}=0$ in the perturbative Wilson coefficients would allow for
a more accurate comparison with the quenched lattice results.

However, the values of all nonperturbative input parameters, i.e. the
condensates, the subtraction constants and the bulk scales of the instanton
size distribution, are deduced more or less directly from observables and
therefore refer to the physical world with three active flavors. The $N_{f}%
$-dependence of all these quantities, furthermore, is not known well enough to
allow for a reliable extrapolation to $N_{f}=0$. The realistic case $N_{f}=3$,
besides being physically more relevant and allowing for quarkonium mixing
effects, is therefore the only consistent choice for the perturbative Wilson coefficients.

\section{IOPE 2: direct instantons}

\label{dirinst}

We now turn to our main objective, the analysis of nonperturbative
contributions to the Wilson coefficients. In the present section we study
their prototype, direct instantons (with sizes $\rho<\mu^{-1}$) \cite{for203}
\footnote{By now there exists a rather large body of work on direct instanton
contributions to IOPE and Borel sum rules in classical hadron channels
\cite{for203} which uncovered important new instanton effects in hadron
structure (in the nucleon channel e.g. the stabilization of the chirally-odd
nucleon sum rule \cite{for93}, the emergence of a new, stable sum rule for the
nucleon magnetic moments, the reconciliaton of sum-rule and chiral
perturbation theory predictions for the neutron-proton mass difference, a
nonperturbative dynamical mechanism of isospin violation \cite{for97} etc.).},
which mediate fast tunneling rearrangements of the vacuum. Direct instanton
contributions to the scalar glueball correlators and Borel sum rules were
evaluated in Ref. \cite{for01} and found to be instrumental in resolving
mutual inconsistencies among the sum rules and with the underlying low-energy
theorem. Moreover, they generate more stable and reliable sum rules, scaling
relations between glueball and instanton properties and new predictions for
the fundamental glueball properties. (See also subsequent work on the related
Gaussian sum rules \cite{har01}.)

After a few introductory remarks on direct instantons and the semiclassical
approximation, we will discuss pertinent aspects of \ the instanton size
distribution $n\left(  \rho\right)  $ in the QCD vacuum. In particular, we
comment on the shortcomings of the ''spike'' approximation to $n\left(
\rho\right)  $ which underlied all previous studies of direct instanton
effects, and then improve upon it by constructing more realistic finite-width
parametrizations which embody the essential features of the instanton size
distribution as revealed by instanton liquid model and lattice simulations.

Armed with realistic size distributions, we then embark on the calculation of
direct instanton contributions to both spin-0 correlators, starting from
$x$-space, and discuss their properties in some detail. In the course of this
discussion we derive several new expressions for the instanton contributions.
We also outline the derivation of the instanton-induced continuum
contributions \cite{for01} and obtain the continuum-subtracted Borel moments
for general instanton size distributions. Next we discuss, on the basis of
some analytical results, generic effects of finite-width distributions and
their impact on IOPE and sum rules. Finally, we renormalize the
direct-instanton-induced Wilson coefficients at the operator scale $\mu$
(which was ignored in previous studies of direct instanton effects).

To start with, let us recall some basic features of the semiclassical
approximation which underlies the evaluation of the direct instanton
contributions. (For an introduction to direct instantons and more details see
Ref. \cite{for203}.) To leading order in $\hbar$, the (Euclidean) functional
integrals representing the correlation functions are evaluated at the relevant
action minima, i.e. the instanton and anti-instanton solutions closest to the
space-time points linked by the correlator, with field strength (in
nonsingular Lorentz gauge) \cite{bel75}%
\begin{equation}
G_{\mu\nu}^{\left(  I\right)  ,a}\left(  x\right)  =\frac{-4\rho^{2}}{g_{s}%
}\frac{\eta_{a\mu\nu}}{[\left(  x-x_{0}\right)  ^{2}+\rho^{2}]^{2}}
\label{gin}%
\end{equation}
($x_{0}$ and $\rho$ are position and size of the instanton, respectively, and
$\eta_{a\mu\nu}$ is the 't Hooft symbol \cite{tho76}). The functional
integrals then reduce to integrals over the collective coordinates of the
saddle points, i.e. over $x_{0,\mu}$, $\rho$ and the color orientation $U$.
Quantum effects break the conformal symmetry of classical Yang-Mills theory
and thereby generate a nontrivial measure $n\left(  \rho\right)  $, i.e. the
instanton size distribution, for the integral over $\rho$. The factor
$g_{s}^{-1}$ in Eq. (\ref{gin}) enhances the direct instanton contributions to
the correlators (\ref{Qcorr}) by a factor $\alpha_{s}^{-2}$ compared to the
leading perturbative contributions. Since $\alpha_{s}/\pi\sim0.1$ at typical
sum-rule momenta $Q\sim1$ GeV, this enhancement is substantial and partially
explains the important role of direct instantons in the spin-0 glueball channels.

The restriction of the semiclassical approximation to the saddle points
associated with the nearest (anti-) instanton is justified by the hierarchy of
IOPE and instanton scales. The distances $\left|  x\right|  \sim\left|
Q^{-1}\right|  \lesssim0.2-0.3$ fm $\ll\Lambda_{QCD}^{-1}$ accessible to the
IOPE, in particular, are much smaller than the average instanton separation
$\bar{R}\sim1$ fm (which is inferred from to the instanton density $\bar
{n}_{I+\bar{I}}\sim1$ fm$^{-4}$ in the vacuum, see below). The relative
diluteness of the instanton medium, characterized by the ratio $\bar{\rho
}/\bar{R}\sim0.3$, further reduces the impact of multi-instanton correlations.
Multi-instanton corrections were calculated explicitly in the pseudoscalar
meson channel (which strongly couples to instantons) in Ref. \cite{dor97} and
indeed found to be negligible. Nevertheless, we will discover a unique
exception to this rule, generated by short-range correlations among
topological charges (including those carried by instantons) in the QCD vacuum,
in Sec. \ref{topscr}.

\subsection{Instanton size distribution}

\label{isdistrib}

As alluded to above, the evaluation of direct instanton contributions to the
IOPE coefficients requires the instanton (and anti-instanton) size
distribution $n_{I,\bar{I}}\left(  \rho\right)  $ as an input. By definition,
$n_{I}\left(  \rho\right)  $ specifies the average number$\,N_{I}=n_{I}\left(
\rho\right)  d\rho d^{4}x_{0}$ of instantons with sizes between $\rho$ and
$\rho+d\rho$ and positions between $x_{0}$ and $x_{0}+dx_{0}$ in the vacuum,
i.e.%
\begin{equation}
n_{I}(\rho)=\frac{d^{5}N_{I}}{d\rho d^{4}x_{0}}.
\end{equation}
Integrating over $\rho$, one obtains the instanton density%
\begin{equation}
\bar{n}_{I}=\int_{0}^{\infty}d\rho n_{I}(\rho)=\frac{d^{4}N_{I}}{d^{4}x_{0}%
}\sim\frac{N_{I}}{V_{4}}%
\end{equation}
where $V_{4}$ is the space-time volume. Since the vacuum (in the thermodynamic
limit) contains on average equal numbers of instantons and anti-instantons,
the same relations hold for the anti-instanton size distribution
$n_{\bar{I}}\left(  \rho\right)  $ and density $\bar{n}_{\bar{I}}$. We
therefore omit the subscripts $I,\bar{I}$ and define%
\begin{equation}
n\left(  \rho\right)  \equiv n_{I}\left(  \rho\right)  =n_{\bar{I}}\left(
\rho\right)  ,
\end{equation}
which implies $\bar{n}=\bar{n}_{I}=\bar{n}_{\bar{I}}$. As a consequence, the
total number of pseudoparticles (instantons \textit{and} anti-instantons) in a
vacuum volume $V_{4}$ is $2\bar{n}V_{4}$.

In the IOPE context, the status of the size distribution $n\left(
\rho\right)  $ is similar to that of the condensates: both are generated
mostly by long-wavelength physics and characterize universal (i.e.
hadron-channel independent) bulk properties of the vacuum. Hence the
associated scales have to be imported from other sources (e.g. from
phenomenology or the lattice). Due to their definition in terms of QCD
amplitudes this is unambiguously possible, and due to their small number and
universal character the IOPE's predictive power is only moderately reduced.

The essential scales of $n\left(  \rho\right)  $ are set by its two leading
moments, i.e. the overall instanton density%
\begin{equation}
\bar{n}=\int_{0}^{\infty}d\rho n\left(  \rho\right)  \label{meandens}%
\end{equation}
and the average instanton size
\begin{equation}
\bar{\rho}=\frac{1}{\bar{n}}\int_{0}^{\infty}d\rho\rho n\left(  \rho\right)  .
\label{meansize}%
\end{equation}
A large body of successful phenomenology in the instanton liquid model
\cite{sch98} has settled on the benchmark values $\bar{\rho}\simeq0.33$ fm and
$\bar{n}\simeq0.5$ fm$^{4}$ which we will use throughout the paper. These
scales imply a mean separation $\bar{R}\sim1$ fm between instantons and are
inside errors consistent with the still exploratory lattice determinations
\cite{lat-isize,rin99} (see Ref. \cite{gar00} for a critical assessment).

The simplest parametrization which is able to embody these scales is the
``spike'' distribution $n_{sp}(\rho)=\bar{n}\delta\left(  \rho-\bar{\rho
}\right)  $ \cite{shu82}. Since the IOPE coefficients are not expected to be
particularly sensitive to details of $n\left(  \rho\right)  $, the spike
approximation has so far been exclusively relied upon in evaluating direct
instanton contributions. Nevertheless, it is clearly an oversimplification and
produces several artefacts, as we will see below. Moreover, a few features of
$n\left(  \rho\right)  $ beyond the moments $\bar{n}$ and $\bar{\rho}$ appear
by now well enough settled to be implemented in more realistic
parametrizations. Information on the behavior of $n\left(  \rho\right)  $ at
small $\rho$ is available from perturbation theory in the instanton background
\cite{tho76,cdg} and implies the power-law behavior
\begin{equation}
n\left(  \rho\right)  \overset{\rho\rightarrow0}{\longrightarrow}A\rho
^{b_{0}-5} \label{smallrhon}%
\end{equation}
(where $b_{0}=11N_{c}/3-2N_{f}/3$). Furthermore, lattice results and instanton
vacuum model arguments \cite{ilg81,dia84,sch98,rin99,kam00} indicate that the
large-$\rho$ tail of $n\left(  \rho\right)  $ has a Gaussian falloff,%
\begin{equation}
n\left(  \rho\right)  \overset{\rho\rightarrow\infty}{\longrightarrow
}Be^{-C\left(  \rho/\bar{\rho}\right)  ^{2}}. \label{largerhon}%
\end{equation}
Note, incidentally, that this strong suppression of large instantons (with
sizes $\rho\gg\bar{\rho}$) is in marked contrast to earlier dilute-gas
assumptions \cite{cdg} which proved inconsistent (i.e. IR-unstable).

An additional benefit of averaging the instanton contributions over a
realistic size distribution is that it affords a simple and gauge-invariant
way to implement the operator renormalization scale $\mu$, as we will discuss
in Sec. \ref{muimplem}. In the following subsections we are going to introduce
two specific finite-width parametrizations. Both of them are uniquely
determined by their behavior at small and large $\rho$ and by their two
leading moments $\bar{n}$ and $\bar{\rho}$. This is a useful restriction since
additional details of $n\left(  \rho\right)  $ are neither reliably known nor
should they have significant impact on the IOPE coefficients.

\subsubsection{Spike distribution}

Before discussing finite-width distributions, we mention a few obvious
properties of the simplest approximation to the instanton size distribution,
the zero-width or spike distribution%
\begin{equation}
n_{sp}\left(  \rho\right)  =\bar{n}\delta\left(  \rho-\bar{\rho}\right)  ,
\label{spike}%
\end{equation}
which sets the sizes of all instantons equal to their average value $\bar
{\rho}$ and has up to now been exclusively used in evaluating direct instanton
contributions. Its two lowest moments determine this distribution uniquely.
Moreover, it becomes exact in the large-$N_{c}$ limit of instanton liquid
vacuum models.

Most other features of the spike distribution, however, differ qualitatively
from more realistic finite-width parametrizations: the median is equal to its
mean, all higher (centralized) moments, including the variance
\begin{equation}
\left\langle \left(  \rho-\bar{\rho}\right)  ^{2}\right\rangle =0,
\end{equation}
vanish identically, and the support both at small and large $\rho$ is exactly
zero. As we will see in Sections \ref{analytres} and \ref{qsra}, these
properties induce several artefacts, including unphysical oscillations in the
instanton-induced spectral functions at timelike momenta.

\subsubsection{Exponential-tail distribution}

The minimal parametrization of a size distribution with power-law behavior at
small $\rho$ and an exponential decay at large $\rho$ is\textbf{ }%
\begin{equation}
n_{\exp}\left(  \rho\right)  =\frac{\left(  n+1\right)  ^{n+1}}{n!}%
\frac{\bar{n}}{\bar{\rho}}\left(  \frac{\rho}{\bar{\rho}}\right)  ^{n}%
\exp\left(  -\left(  n+1\right)  \frac{\rho}{\bar{\rho}}\right)  .
\end{equation}
All its additional properties and scales are determined by the two lowest
moments (\ref{meandens}) and (\ref{meansize}). Although the exponential
large-$\rho$ tail is less favored by lattice and instanton liquid results than
the Gaussian one to which we turn in the following section, it is nevertheless
useful because it allows the $\rho$-integral in the instanton-induced spectral
function to be done analytically. The resulting expressions provide a
convenient benchmark for analyzing generic effects of finite-width distributions.

For $N_{f}=3$, instanton-background perturbation theory fixes the power of the
small-$\rho$ behavior at $n=4$ (cf. Eq. (\ref{smallrhon})), i.e.
\begin{equation}
n_{\exp}\left(  \rho\right)  =\frac{5^{5}}{2^{3}3}\frac{\bar{n}}{\bar{\rho}%
}\left(  \frac{\rho}{\bar{\rho}}\right)  ^{4}\exp\left(  -\frac{5\rho}%
{\bar{\rho}}\right)  \text{ \ \ \ \ \ for }N_{f}=3. \label{nexp}%
\end{equation}
The variance of the exponential distribution,
\begin{equation}
\left\langle \left(  \rho-\bar{\rho}\right)  ^{2}\right\rangle =\frac{1}%
{\bar{n}}\int n_{\exp}(\rho)\rho^{2}\,d\rho-\bar{\rho}^{2}=\frac{1}{5}%
\bar{\rho}^{2},
\end{equation}
implies a half-width of about $0.7\bar{\rho}$, similar to lattice and
instanton liquid results \cite{sch98,rin99}. This is reassuring since the
width of $n_{\exp}$ is not an independent parameter but determined by
$\bar{\rho}$. The maximum of the distribution is reached at%
\begin{equation}
\rho_{peak}=\frac{4\bar{\rho}}{5}%
\end{equation}
(i.e. $\rho_{peak}\simeq4/3$ GeV$^{-1}$ for the standard value $\bar{\rho
}\simeq1/3$ fm). A parametrization for $n\left(  \rho\right)  $ with a similar
exponential tail,
\begin{equation}
n\left(  \rho\right)  =\frac{N}{\rho}\exp\left(  -\frac{\alpha}{\rho}%
-\beta\rho\right)  ,
\end{equation}
has been used in a study of the asymptotic OPE behavior \cite{chi97}. As our
distribution (\ref{nexp}), it allows relevant $\rho$-integrals to be done
analytically. However, it fails to reproduce the power-law behavior at small
$\rho$ and cannot be made to reproduce both $\bar{n}$ and $\bar{\rho}$, thus
introducing additional, insufficiently constrained parameters.

\subsubsection{Gaussian-tail distribution}

A size distribution with power behavior at small $\rho$ and a Gaussian tail at
large $\rho$ can be obtained from the minimal ansatz%
\begin{equation}
n\left(  \rho\right)  =N\rho^{n}\exp\left(  -A\frac{\rho^{2}}{\bar{\rho}^{2}%
}\right)
\end{equation}
by requiring the two lowest moments to satisfy Eqs. (\ref{meandens}) and
(\ref{meansize}). This determines $A$ and $N$ uniquely, with the result
\textbf{ }%
\begin{equation}
n_{g}\left(  \rho\right)  =2\frac{\bar{n}}{\bar{\rho}}\left(  \frac{\rho}%
{\bar{\rho}}\right)  ^{n}\frac{\left[  \Gamma\left(  \frac{n+2}{2}\right)
\right]  ^{n+1}}{\left[  \Gamma\left(  \frac{n+1}{2}\right)  \right]  ^{n+2}%
}\exp\left[  -\left(  \frac{\Gamma\left(  \frac{n+2}{2}\right)  }%
{\Gamma\left(  \frac{n+1}{2}\right)  }\right)  ^{2}\left(  \frac{\rho}%
{\bar{\rho}}\right)  ^{2}\right]  . \label{ngauss}%
\end{equation}
Gaussian-tail distributions were found in the instanton liquid model (ILM)
\cite{dia84} as well as in other approaches \cite{sch98,ilg81,kam00,sch99}.
They are also supported by quenched lattice data \cite{rin99}. In the ILM, in
particular, one finds \cite{dia84}
\begin{equation}
n_{ILM}\left(  \rho\right)  =C\rho^{b_{0}-5}\exp\left(  -\frac{b_{0}-4}%
{2}\frac{\rho^{2}}{\bar{\rho}^{2}}\right)
\end{equation}
($b_{0}=11N_{c}/3-2N_{f}/3$) which approaches the spike approximation at large
$N_{c}$,
\begin{equation}
\lim_{N_{c}\rightarrow\infty}n_{ILM}\left(  \rho\right)  \propto\delta\left(
\rho-\bar{\rho}\right)  .
\end{equation}
At $N_{c}=N_{f}=3$ it becomes
\begin{equation}
n_{ILM}\left(  \rho\right)  =C\rho^{4}\exp\left(  -\frac{5}{2}\frac{\rho^{2}%
}{\bar{\rho}^{2}}\right)  ,
\end{equation}
to be compared with our parametrization (\ref{ngauss}) for $n=4$,%
\begin{equation}
n_{g}\left(  \rho\right)  =\frac{2^{18}}{3^{6}\pi^{3}}\frac{\bar{n}}{\bar
{\rho}}\left(  \frac{\rho}{\bar{\rho}}\right)  ^{4}\exp\left(  -\frac{2^{6}%
}{3^{2}\pi}\frac{\rho^{2}}{\bar{\rho}^{2}}\right)  \text{ \ \ \ \ \ \ \ }%
\left(  N_{f}=3\right)  . \label{ngn3}%
\end{equation}
The width of the ILM distribution is about 10\% larger. Equation (\ref{ngn3})
is the most realistic parametrization among those which we will consider.

The distribution (\ref{ngn3}) has the variance
\begin{equation}
\left\langle \left(  \rho-\bar{\rho}\right)  ^{2}\right\rangle =\left(
\frac{3^{2}5\pi}{2^{7}}-1\right)  \bar{\rho}^{2}\simeq\frac{1}{10}\text{ }%
\bar{\rho}^{2}%
\end{equation}
which corresponds to a half-width of about $\bar{\rho}/2$, somewhat smaller
than the half-width of the exponential-tail distribution (\ref{nexp}). The
position of the peak lies at%
\begin{equation}
\rho_{peak}=\frac{3}{4}\sqrt{\frac{\pi}{2}}\bar{\rho}\simeq0.94\bar{\rho}.
\end{equation}
Since $n_{g}$ is more symmetric than $n_{\exp}$, its $\rho_{peak}$ is somewhat larger.

Due to the stronger decay at large $\rho$, $n_{g}\left(  \rho\right)  $
contains a larger number of intermediate-size instantons with $\rho\sim
\bar{\rho}$. Since only instantons with $\rho\lesssim\mu^{-1}\sim2\bar{\rho}$
contribute to the IOPE coefficients ($\mu$ is the operator renormalization
scale), the Gaussian-tail distribution - cut off at $\mu^{-1}$ - will
typically result in larger instanton contributions to the Borel moments than
the exponential one.

\subsection{Direct instanton contributions in $x$-space}

\label{dirinstx}

In this section we derive, to leading order in the semiclassical expansion,
the direct instanton contributions to the IOPE coefficients $\tilde{C}%
_{0}^{\left(  S,P\right)  }$ of the spin-0 glueball correlators. We also
comment on contributions of higher orders in $\hbar$ and on coefficients of
higher-dimensional operators. The calculations, as well as the discussion of
the singularity structure and of the pointlike-instanton limit, are best
performed in Euclidean space-time, the instanton's natural habitat.

As outlined above, the semiclassical contributions from instanton-induced
saddle points to the functional integral of the glueball correlators are
obtained by evaluating their Wick expansion in the instanton background, e.g.
by means of the gluon background-field propagator of Ref. \cite{bro78}. Adding
up the contributions from the nearest instanton and anti-instanton one finds,
to leading order in $\hbar$,%
\begin{align}
\Pi_{S}^{\left(  I+\bar{I}\right)  }\left(  x^{2}\right)   &  =\sum
_{I,\bar{I}}\int d\rho n\left(  \rho\right)  \int d^{4}x_{0}\left\langle
TO_{S}\left(  x\right)  O_{S}\left(  0\right)  \right\rangle _{I+\bar{I}}\\
&  =\frac{2^{9}3^{2}}{\pi^{2}}\int d\rho n\left(  \rho\right)  \int d^{4}%
x_{0}\frac{\rho^{8}}{[\left(  x-x_{0}\right)  ^{2}+\rho^{2}]^{4}[x_{0}%
^{2}+\rho^{2}]^{4}}.
\end{align}
We recall that it suffices to perform all calculations in the scalar channel
since the pseudoscalar correlator receives, due to self-duality, identical
instanton contributions (in Euclidean space-time). A convenient analytical
expression for the above $x_{0}$-integral in terms of a hypergeometric
function \cite{abr},
\begin{equation}
\int d^{4}x_{0}\frac{\rho^{12}}{[\left(  x-x_{0}\right)  ^{2}+\rho^{2}%
]^{4}[x_{0}^{2}+\rho^{2}]^{4}}=\frac{\pi^{2}}{42}\,_{2}F_{1}\left(
4,6,\frac{9}{2},-\frac{x^{2}}{4\rho^{2}}\right)  ,
\end{equation}
is derived in Appendix \ref{iint}.

The singularity structure of the instanton contributions at fixed instanton
size $\rho$ can then be read off from
\begin{equation}
\Pi_{S}^{\left(  I+\bar{I}\right)  }\left(  x^{2};\rho\right)  =\frac{2^{8}%
3}{7}\frac{1}{\rho^{4}}\,_{2}F_{1}\left(  4,6,\frac{9}{2},-\frac{x^{2}}%
{4\rho^{2}}\right)  \label{pixrho}%
\end{equation}
by recalling the well-known analyticity properties of the hypergeometric
functions \cite{abr}: Eq. (\ref{pixrho}) has no singularities inside a circle
of radius $2\rho$ around $x=0$, but it has cuts emerging from the two branch
points $x=\pm2i\rho$. (These singularities have their origin in the (partially
gauge-dependent) denominators of the instanton's field strength (\ref{gin}).)
Contributions from these branch points dominate the Fourier transform of Eq.
(\ref{pixrho}) at large momentum transfer (see next section). The final result
for the instanton contributions is obtained by integrating Eq. (\ref{pixrho})
over all instanton sizes with the weight $n\left(  \rho\right)  $:
\begin{equation}
\Pi_{S}^{\left(  I+\bar{I}\right)  }\left(  x^{2}\right)  =\int d\rho n\left(
\rho\right)  \Pi^{\left(  I+\bar{I}\right)  }\left(  x^{2};\rho\right)
=\frac{2^{8}3}{7}\int d\rho\frac{n\left(  \rho\right)  }{\rho^{4}}\,_{2}%
F_{1}\left(  4,6,\frac{9}{2},-\frac{x^{2}}{4\rho^{2}}\right)  . \label{instx}%
\end{equation}

Note that the $\rho$-integration over physically reasonable\ instanton size
distributions, with the power-law behavior (\ref{smallrhon}) at small $\rho$
and $N_{f}\leq3$, does not produce singularities of Eq. (\ref{instx}) at
$x^{2}=0$. This implies, in particular, that the $x^{2}\rightarrow0$ limit is
finite (see below). From $_{2}F_{1}\left(  a,b,c,0\right)  =1$ one obtains%
\begin{equation}
\Pi_{S}^{\left(  I+\bar{I}\right)  }\left(  x^{2}=0\right)  =\frac{2^{8}3}%
{7}\int d\rho\frac{n\left(  \rho\right)  }{\rho^{4}}.
\end{equation}
The $x$-dependence of Eq. (\ref{pixrho}) shows that the instanton-induced
correlations are maximal at $x^{2}=0$ and strongly decay ($\propto\left|
x\right|  ^{-8}$ at fixed $\rho$) for $\left|  x\right|  \gg2\rho$, i.e. when
the arguments of the correlator cease to lie both within the bulk of the
instanton's localized action density. Note, furthermore, that the
$x$-dependence of the fixed-size instanton contribution reduces to a delta
function in the small-$\rho$ limit,
\begin{equation}
\lim_{\rho\rightarrow0}\Pi_{S}^{\left(  I+\bar{I}\right)  }\left(  x^{2}%
;\rho\right)  =\frac{2^{8}3}{7}\lim_{\rho\rightarrow0}\frac{1}{\rho^{4}}%
\,_{2}F_{1}\left(  4,6,\frac{9}{2},-\frac{x^{2}}{4\rho^{2}}\right)  =2^{7}%
\pi^{2}\delta^{4}\left(  x\right)  . \label{pointlim}%
\end{equation}
This pointlike-instanton approximation will be useful below.

Technically, the result (\ref{instx}) provides the leading nonperturbative
contributions to the IOPE\ coefficient $\tilde{C}_{0}^{\left(  S\right)
}\left(  x\right)  $ of the unit-operator in the scalar glueball channel.
(Strictly speaking, soft-mode contributions, in particular those due to large
instantons with $\rho>\mu^{-1}$, should still be excluded, cf. Sec.
\ref{muimplem}.) Of course, the coefficients of higher-dimensional operators
receive direct instanton contributions as well. The coefficient $\tilde{C}%
_{4}^{\left(  S\right)  }$ of the lowest-dimensional condensate $\left\langle
\alpha G^{2}\right\rangle $, for example, gets a correction from the process
in which one of the gluon fields emanating from the interpolator (\ref{sipf})
is soft and turns into a gluon condensate while the other one is hard and
propagates in the instanton background. The general form of this contribution
is
\begin{equation}
\Pi_{S,\left\langle \alpha G^{2}\right\rangle }^{\left(  I+\bar{I}\right)
}\left(  x^{2}\right)  \propto\left\langle \alpha_{s}G^{2}\right\rangle \int
d\rho n\left(  \rho\right)  \int d^{4}x_{0}\frac{\rho^{4}}{\left[  \left(
x-x_{0}\right)  ^{2}+\rho^{2}\right]  ^{4}}.
\end{equation}
Relative to the unit-operator term, this correction is suppressed by four
powers of $\varepsilon\equiv\left|  x\right|  \Lambda_{QCD}\lesssim0.2$. More
generally, power corrections associated with $d$-dimensional operators are
parametrically suppressed by a factor $\varepsilon^{d}$ and also by the
relatively large glueball mass scale (cf. Sec. \ref{qsra}). Therefore, we
neglect such contributions in the following \footnote{In the IOPE of the
pseudoscalar meson correlator, analogous instanton-enhanced power corrections
are associated with chiral-symmetry breaking operators and therefore turn out
to be important \cite{for201}.}.

The IOPE coefficients also receive instanton corrections from higher orders in
the semiclassical expansion. To $O\left(  \hbar\right)  $, these corrections
arise from Gaussian quantum fluctuations around the instanton fields
\cite{sch98} and have the form
\begin{equation}
\Pi_{G,O\left(  \hbar\right)  }^{\left(  I+\bar{I}\right)  }\left(
x^{2}\right)  \propto\alpha_{s}^{2}\left.  \left\langle G_{\mu\nu}^{\left(
I\right)  ,a}\left(  x\right)  \left[  D_{\mu}^{x}D_{\rho}^{y}D_{\nu\sigma
}\left(  x,y\right)  \right]  ^{ab}G_{\rho\sigma}^{\left(  I\right)
,b}\left(  y\right)  \right\rangle \right|  _{y=0}+...
\end{equation}
where $D_{\mu}^{x}$ is the covariant derivative in the adjoint representation
and $D_{\nu\beta}$ is the gluon propagator in the instanton background.
Radiative corrections of this sort are suppressed by at least two powers of
$\alpha_{s}$. Moreover, the average instanton is small on the QCD scale,
$\bar{\rho}\Lambda_{QCD}\ll1$, and the instanton action $S_{I}\left(
\bar{\rho}\right)  \sim10\hbar$ consequently large compared to the action of
the quantum corrections. We therefore do not expect such corrections to have
significant impact on our analysis and do not consider them further.

Finally, it is instructive to compare our result (\ref{instx}) with direct
instanton contributions to pseudoscalar meson correlators
\cite{dor97,for201,shu83}. The latter arise mostly from quarks propagating in
the zero-modes states of the covariant Dirac operator in the instanton
background \cite{tho76}. (The non-zero modes can normally be approximated by
plane waves.) In the meson sector, the distinct topological, chiral, flavor
and spin-color properties of these zero modes are responsible for most
features of the direct instanton contributions, including their strong channel
dependence. Clearly, the instanton-induced physics in the glueball correlators
is quite different in this respect. (It does not require, incidentally, to
account for ambient soft vacuum fields in the gluon propagators which, in
contrast, rather strongly affect the quark zero-mode propagation.)

\subsection{Borel moments}

\label{IBorelMom}

The next step in our program is to calculate the Borel moments (\ref{bmoms})
associated with the direct instanton contributions (\ref{instx}). This is best
done by first transforming to (Euclidean) momentum space,
\begin{equation}
\Pi_{S}^{\left(  I+\bar{I}\right)  }\left(  Q^{2}\right)  =\int d^{4}%
xe^{iQx}\Pi_{S}^{\left(  I+\bar{I}\right)  }\left(  x^{2}\right)
\end{equation}
or, more explicitly,
\begin{align}
\Pi_{S}^{\left(  I+\bar{I}\right)  }\left(  Q^{2}\right)   &  =\frac{2^{10}%
3\pi^{2}}{7}\int d\rho\frac{n\left(  \rho\right)  }{\rho^{4}}\int_{0}^{\infty
}drr^{3}\frac{J_{1}\left(  Qr\right)  }{Qr}\text{ }_{2}F_{1}\left(
4,6,\frac{9}{2},-\frac{r^{2}}{4\rho^{2}}\right) \\
&  =2^{5}\pi^{2}\int d\rho n\left(  \rho\right)  \left(  Q\rho\right)
^{4}K_{2}^{2}\left(  Q\rho\right)  , \label{piqinst}%
\end{align}
where $J_{1}\left(  z\right)  $ and $K_{2}\left(  z\right)  $ are Bessel and
McDonald functions \cite{abr}. In the last equation we have made use of the
integral \cite{gra}%
\begin{equation}
\int_{0}^{\infty}drr^{2}J_{1}\left(  Qr\right)  _{\text{ }2}F_{1}\left(
4,6,\frac{9}{2},-\frac{r^{2}}{4\rho^{2}}\right)  =\frac{7}{2^{5}3}Q^{5}%
\rho^{8}K_{2}^{2}\left(  Q\rho\right)  .
\end{equation}
(An alternative calculation of Eq. (\ref{piqinst}), which agrees with the
expression first obtained in \cite{nov280}, can be found in Appendix
\ref{iint}.)

It is instructive to examine the limits of $\Pi_{S}^{\left(  I+\bar{I}\right)
}\left(  Q^{2}\right)  $. Using for simplicity the spike distribution
(\ref{spike}) and the asymptotic behavior of the McDonald functions
\cite{abr}, one finds
\begin{equation}
\Pi_{S}^{\left(  I+\bar{I}\right)  }\left(  Q^{2}\right)  \overset
{Q^{2}\rightarrow\infty}{\longrightarrow}2^{4}\pi^{3}\bar{n}\left(  Q\bar
{\rho}\right)  ^{3}e^{-2Q\bar{\rho}}. \label{largeQ}%
\end{equation}
The exponential decay of the integrand at large $Q$ is expected since $\Pi
_{S}^{\left(  I+\bar{I}\right)  }\left(  x^{2};\rho\right)  $ is analytic in
the circle $\left|  x\right|  <2\rho$ around $x^{2}=0$. Its scale is set by
the singularities nearest to the real axis, i.e. the branch points at
$x=\pm2i\rho$. Physically, this can be understood as the phase-space
suppression encountered when distributing the hard momentum $Q$ over multiple
soft modes of the coherent instanton field with size $\rho$. While the
fixed-size instanton contribution (\ref{pixrho}) contains no asymptotic power
corrections (due to the absence of singularities at $x^{2}=0$), integration
over the instanton size does produce inverse powers of $Q$. However, they
start with $Q^{-5}$ (the power is determined by the large-$\rho$ behavior
(\ref{smallrhon}) of $n\left(  \rho\right)  $) and therefore do not interfere
with the power corrections of the truncated IOPE (cf. Eq. (\ref{ope})).

In the opposite limit, i.e. for $Q^{2}\rightarrow0$, the instanton
contributions turn into
\begin{equation}
\Pi_{S}^{\left(  I+\bar{I}\right)  }\left(  Q^{2}=0\right)  =2^{7}\pi^{2}\int
d\rho n\left(  \rho\right)  . \label{q0lim}%
\end{equation}
This term plays a key role, as an instanton-induced subtraction constant, in
the lowest (i.e. $k=-1$) Borel moments and in the associated spin-0 sum rules
(cf. Sec. \ref{subconst}). The obligatory removal of soft contributions from
instanton-induced Wilson coefficients (cf. Secs. \ref{muimplem} and
\ref{subconst}) de-emphasizes contributions from multi-instantons and other
soft vacuum fields (not taken into account in Eq. (\ref{instx})) and should
thereby improve the nearest-instanton approximation. Although the latter is
strictly valid only for $\left|  x\right|  \ll\bar{R}$, it often seems to work
over larger distances (as long as cluster decomposition does not become an
issue \cite{sch98})), due to the strong localization and small packing
fraction of the instantons \footnote{This conclusion is supported by quark and
gluon condensate estimates on the basis of the one-instanton approximation
which are not far from the phenomenological values, and by the fact that their
multi-instanton corrections (evaluated in the ILM) are typically of the order
of 10-20\%. (I thank Sasha Dorokhov for interesting discussions on this
point.)}. Nevertheless, one should not expect Eq. (\ref{q0lim}) to be a
complete representation of the correlator at $Q=0$, as can be seen, for
example, from the one-instanton approximation to the gluon condensate,
\begin{equation}
\left\langle \alpha G^{2}\right\rangle _{I+\bar{I}}=16\pi\int d\rho n\left(
\rho\right)  ,
\end{equation}
which, when multiplied by $32\pi/b_{0}$, yields only about half of Eq.
(\ref{q0lim}) and therefore does not saturate the low-energy theorem
(\ref{sLET}). The inconsistency with the low-energy theorem (\ref{pLET}) in
the pseudoscalar channel, incidentally, is far more dramatic and can be
overcome only by additional nonperturbative physics, cf. Secs. \ref{topscr}
and \ref{subconst}.

According to the general definition (\ref{bmoms}), the Borel moments of the
direct instanton contributions are%
\begin{equation}
\mathcal{L}_{S,k}^{\left(  I+\bar{I}\right)  }\left(  \tau\right)  =\hat
{B}\left[  \left(  -Q^{2}\right)  ^{k}\Pi_{S}^{\left(  I+\bar{I}\right)
}\left(  Q^{2}\right)  \right]  \label{ibrtrf}%
\end{equation}
(for $k\in\left\{  -1,0,1,2\right\}  $). In order to calculate these moments
explicitly, it is convenient to rewrite Eq. (\ref{piqinst}) in terms of an
integral representation for the McDonald function before applying the Borel
operator (\ref{btrf}). In Appendix \ref{iint} we outline this calculation for
$k=-1$, which results in \cite{for01}
\begin{equation}
\mathcal{L}_{S,-1}^{\left(  I+\bar{I}\right)  }\left(  \tau\right)  =-2^{6}%
\pi^{2}\int d\rho n\left(  \rho\right)  \xi^{2}e^{-\xi}\left[  \left(
1+\xi\right)  K_{0}\left(  \xi\right)  +\left(  2+\xi+\frac{2}{\xi}\right)
K_{1}\left(  \xi\right)  \right]  . \label{inb-1}%
\end{equation}
Above we have introduced the dimensionless variable $\xi=\rho^{2}/2\tau$. From
the lowest, ``generating'' Borel moment all higher moments are obtained by
differentiation with respect to $-\tau$,%
\begin{equation}
\mathcal{L}_{S,k+1}^{\left(  I+\bar{I}\right)  }\left(  \tau\right)
=-\frac{\partial}{\partial\tau}\mathcal{L}_{S,k}^{\left(  I+\bar{I}\right)
}\left(  \tau\right)  \text{ \ \ \ \ \ \ \ \ \ (}k\geq-1\text{),}
\label{recurs}%
\end{equation}
and explicitly read%

\begin{align}
\mathcal{L}_{S,0}^{\left(  I+\bar{I}\right)  }\left(  \tau\right)   &
=2^{8}\pi^{2}\int d\rho\frac{n\left(  \rho\right)  }{\rho^{2}}\xi^{5}e^{-\xi
}\left[  K_{0}\left(  \xi\right)  +\left(  1+\frac{1}{2\xi}\right)
K_{1}\left(  \xi\right)  \right]  ,\label{inb0}\\
\mathcal{L}_{S,1}^{\left(  I+\bar{I}\right)  }\left(  \tau\right)   &
=-2^{9}\pi^{2}\int d\rho\frac{n\left(  \rho\right)  }{\rho^{4}}\xi^{7}e^{-\xi
}\left[  \left(  2-\frac{9}{2\xi}\right)  K_{0}\left(  \xi\right)  +\left(
2-\frac{7}{2\xi}-\frac{3}{2\xi^{2}}\right)  K_{1}\left(  \xi\right)  \right]
,\label{inb1}\\
\mathcal{L}_{S,2}^{\left(  I+\bar{I}\right)  }\left(  \tau\right)   &
=2^{10}\pi^{2}\int d\rho\frac{n\left(  \rho\right)  }{\rho^{6}}\xi^{9}e^{-\xi
}\left[  \left(  4-\frac{44}{2\xi}+\frac{51}{2\xi^{2}}\right)  K_{0}\left(
\xi\right)  +\left(  4-\frac{40}{2\xi}+\frac{32}{2\xi^{2}}+\frac{12}{2\xi^{3}%
}\right)  K_{1}\left(  \xi\right)  \right]  . \label{inb2}%
\end{align}

For $\tau\rightarrow0$, at fixed $\rho$, all instanton-induced Borel moments
vanish exponentially. This is a direct reflection of the exponential large-$Q$
suppression in Eq. (\ref{largeQ}) and renders these moments practically
negligible for $\tau\lesssim0.2$ GeV$^{-2}$ (at $\rho=\bar{\rho}$). More
specifically, if $\bar{\xi}=\bar{\rho}^{2}/2\tau\gg1$ and if $n\left(
\rho\right)  $ is replaced by the spike approximation (\ref{spike}), Eqs.
(\ref{inb-1}) - (\ref{inb2}) reduce to
\begin{equation}
\mathcal{L}_{S,k}^{\left(  I+\bar{I}\right)  }\left(  \tau\right)
\overset{\tau\rightarrow0}{\longrightarrow}\left(  -1\right)  ^{k}2^{4}%
\pi^{5/2}\bar{n}\bar{\rho}^{7+2k}\frac{e^{-\bar{\rho}^{2}/\tau}}{\tau^{\left(
9+4k\right)  /2}}\left[  1+O\left(  \frac{\tau}{\bar{\rho}^{2}}\right)
\right]  . \label{smalltau}%
\end{equation}
In the case of $k=0$, for instance, Eq. (\ref{smalltau}) is reliable when
$\tau\lesssim0.4$ GeV$^{-2}$ but it underestimates the maximum at $\tau
\sim0.6$ GeV$^{-2}$ and has the wrong decay power at larger $\tau$ (about
twice the correct one, see below). The estimate of the instanton contributions
to the $k=-1$ sum rule in Ref. \cite{shu82} were based on the approximation
(\ref{smalltau}). Due to the wrong decay behavior and the missing continuum
contributions, however, it is unsuitable for reliable glueball mass and
coupling estimates. In fact, after subtracting the associated continuum
contributions (as in Eq. (\ref{lcontinsubtr}), see next section), the
exponential suppression (\ref{smalltau}) will turn into an enhancement and
generate increasing instanton contributions down to $\tau=0$.

At intermediate $\tau$ ($\tau\gtrsim0.2$ GeV$^{-2}$), $\mathcal{L}%
_{S,-1}^{\left(  I+\bar{I}\right)  }\left(  \tau\right)  $ rises quite steeply
towards its maximum. The associated, fast variations of its slope generate
large oscillations in the higher Borel moments (due to the $\tau$-derivative
in Eq. (\ref{ibrtrf})) which would increasingly impede reliable sum rule fits.
Again, this problem is overcome by subtracting the crucial instanton-induced
continuum contributions, which transforms the oscillations into a monotonic
decay. In the opposite limit, i.e. for $\tau\gg\bar{\rho}^{2}/2$, the $k\geq0$
Borel moments approach zero from above while $\mathcal{L}_{S,-1}^{\left(
I+\bar{I}\right)  }$ reaches the negative subtraction constant $-$ $\Pi
_{S}^{\left(  I+\bar{I}\right)  }\left(  0\right)  $. The general decay
behavior of the instanton contributions for large $\tau$ is an inverse power
law. This continues to hold for fairly general $n\left(  \rho\right)  $
(including (\ref{nexp}) and (\ref{ngn3})) as long as $\tau\gg\rho_{c}^{2}/2$
where $\rho_{c}$ is roughly the maximal size at which $n\left(  \rho\right)  $
has appreciable support. More explicitly, one has
\begin{equation}
\mathcal{L}_{S,k}^{\left(  I+\bar{I}\right)  }\left(  \tau\right)
\overset{\tau\rightarrow\infty}{\longrightarrow}-\delta_{k,-1}\Pi_{S}^{\left(
I+\bar{I}\right)  }\left(  0\right)  +\frac{a_{k}}{\tau^{k+3}}\int
d\rho\frac{n\left(  \rho\right)  }{\rho^{2\left(  k+1\right)  }}+O\left(
\tau^{-k-4}\right)  \label{largetlim}%
\end{equation}
with positive coefficients $a_{k}$. As a consequence, the instanton-induced
Borel moments decay as fast as their leading perturbative counterparts for
$\tau\rightarrow\infty$, and faster than all power corrections (cf. Section
\ref{pertBorelMom}). (Of course, this limit is not physical since the
short-distance expansion breaks down for $\tau\rightarrow\infty$.)

\subsection{Imaginary part and instanton-induced continuum}

The discontinuities of the IOPE at time-like momenta generate the duality
continuum (\ref{dualcontin}). In the spin-0 glueball channels, the
instanton-induced continuum contributions were first obtained in Ref.
\cite{for01} and shown to have crucial impact on consistency and predictions
of the $0^{++}$-glueball sum rules. In the following, we outline their
calculation and analyze some of their pertinent properties. Besides providing
new insight into the structure of the instanton contributions, this analysis
will prove useful in assessing the effects of realistic instanton size
distributions in subsequent sections. Finally, we assemble the
instanton-induced, continuum-subtracted Borel moments which enter the spin-0
glueball sum rules.

The imaginary part of the instanton contributions is obtained by analytically
continuing Eq. (\ref{piqinst}) in the complex $s=q^{2}=-Q^{2}+i\varepsilon$
plane. The behavior of the McDonald functions under analytical continuation
can be expressed as \cite{abr}%
\begin{equation}
K_{\nu}\left(  z\right)  =\left\{
\begin{tabular}
[c]{c}%
$\frac{i\pi}{2}e^{i\pi\nu/2}H_{\nu}^{\left(  1\right)  }\left(  ze^{i\pi
/2}\right)  $ \ \ \ \ \ \ \ \ \ \ \ \ \ \ for $-\pi<\arg z\leq\frac{\pi}{2}$\\
$\frac{-i\pi}{2}e^{-i\pi\nu/2}H_{\nu}^{\left(  1\right)  }\left(  ze^{-i\pi
/2}\right)  $ \ \ \ \ \ \ \ \ \ \ \ \ for $\frac{\pi}{2}<\arg z\leq\pi$%
\end{tabular}
\ \ \ \ \ \ \ \ \ \ \ \right.
\end{equation}
where $H_{\nu}^{\left(  1\right)  }$ is the Hankel function of the first kind,%
\begin{equation}
H_{\nu}^{\left(  1\right)  }\left(  z\right)  =J_{\nu}\left(  z\right)
+iY_{\nu}\left(  z\right)  .
\end{equation}
From the cut structure of the Hankel functions \cite{abr} one then finds%
\begin{equation}
\operatorname{Im}K_{2}^{2}\left(  -i\sqrt{s}\rho\right)  =-\frac{\pi^{2}}%
{2}J_{2}\left(  \sqrt{s}\rho\right)  Y_{2}\left(  \sqrt{s}\rho\right)
-\frac{2\pi}{\rho^{2}}\delta\left(  s\right)
\end{equation}
which leads to \cite{for01}%
\begin{equation}
\operatorname{Im}\Pi_{S}^{\left(  I+\bar{I}\right)  }\left(  -s\right)
=-2^{4}\pi^{4}s^{2}\int d\rho n\left(  \rho\right)  \rho^{4}J_{2}\left(
\sqrt{s}\rho\right)  Y_{2}\left(  \sqrt{s}\rho\right)  . \label{impinsts}%
\end{equation}

Several qualitative and quantitative features can be readily deduced from this
expression. First of all, for $s\rightarrow0$ (i.e. $s\ll\bar{\rho}^{-2}$) it
reduces to
\begin{equation}
\operatorname{Im}\Pi_{S}^{\left(  I+\bar{I}\right)  }\left(  -s\right)
\overset{s\ll\bar{\rho}^{-2}}{\longrightarrow}2^{3}\pi^{3}s^{2}\int d\rho
n\left(  \rho\right)  \rho^{4}\left[  1+\frac{1}{6}\rho^{2}s+O\left(
s^{2}\right)  \right]  ,
\end{equation}
which has the leading $s^{2}$-dependence of the free gluon loop. This implies,
for example, that the large-$\tau$ behavior of the corresponding,
continuum-subtracted Borel moments (see below) equals that of the leading
perturbative Wilson coefficient. Since the continuum subtraction does not
affect the falloff at large $\tau$, it also confirms the decay behavior
(\ref{largetlim}) established in the last section.

A more important property to notice is that the analytical continuation has
turned the exponentially small instanton contributions (\ref{piqinst}) at
large $Q^{2}$ and fixed $\rho$ into strong oscillations with increasing
amplitude at timelike $s$. This becomes more explicit at large $s$,
\begin{equation}
\operatorname{Im}\Pi^{\left(  I+\bar{I}\right)  }\left(  -s\right)
\overset{s\rightarrow\infty}{\longrightarrow}2^{4}\pi^{3}s^{3/2}\bar{n}%
\bar{\rho}^{3}\cos\left(  2\sqrt{s}\bar{\rho}\right)  , \label{imsplarges}%
\end{equation}
where we have used the spike approximation (\ref{spike}). Such a behavior is
familiar from the analytical continuation of semiclassical tunneling
amplitudes in quantum mechanics and can lead to a selective enhancement of
parts of the amplitude (when crossing Stokes lines) \cite{mig??}. Potential
problems of this sort (including the strong oscillations) will be tamed by
finite-width instanton size distributions (cf. Sec. \ref{analytres}). In any
case, the overall growth of the instanton-induced imaginary part $\propto
s^{3/2}$ at large $s$ (and fixed $\rho$) is weaker than that of the free gluon
loop (cf. Eq. (\ref{imsope})) and the dispersive integral exists without
subtractions, i.e.
\begin{align}
\Pi_{S}^{\left(  I+\bar{I}\right)  }\left(  Q^{2}\right)   &  =\frac{1}{\pi
}\int_{0}^{\infty}ds\frac{\operatorname{Im}\Pi_{S}^{\left(  I+\bar{I}\right)
}\left(  -s\right)  }{s+Q^{2}}\label{idisprel}\\
&  =-2^{4}\pi^{3}\int d\rho n\left(  \rho\right)  \rho^{4}\int_{0}^{\infty
}ds\frac{s^{2}J_{2}\left(  \sqrt{s}\rho\right)  Y_{2}\left(  \sqrt{s}%
\rho\right)  }{s+Q^{2}}.
\end{align}
The analytical evaluation of the $s$-integral indeed reproduces Eq.
(\ref{piqinst}).

As discussed above, the small-$Q^{2}$ limit of the correlator (\ref{piqinst})
lies outside the range of validity of both the nearest-instanton approximation
and the IOPE. Usually, this does not cause concern because the sum rules are
analyzed at momenta $Q^{2}\gtrsim1$ GeV $^{2}$ where both approximations
should work. However, the lowest Borel moment (with $k=-1$) by design contains
the $Q^{2}=0$ limit of the correlator as a subtraction constant. This term is
not eliminated by the Borel transform since the latter is applied to
$\Pi\left(  Q^{2}\right)  /Q^{2}$:
\begin{align}
\mathcal{L}_{S,-1}^{\left(  I+\bar{I}\right)  }\left(  \tau\right)   &
=\hat{B}\left[  -\frac{\Pi_{S}^{\left(  I+\bar{I}\right)  }\left(  0\right)
}{Q^{2}}+\frac{\Pi_{S}^{\left(  I+\bar{I}\right)  }\left(  Q^{2}\right)
-\Pi_{S}^{\left(  I+\bar{I}\right)  }\left(  0\right)  }{-Q^{2}}\right]
\label{Lm1ia}\\
&  =-\Pi_{S}^{\left(  I+\bar{I}\right)  }\left(  0\right)  +\frac{1}{\pi}%
\int_{0}^{\infty}ds\frac{\operatorname{Im}\Pi_{S}^{\left(  I+\bar{I}\right)
}\left(  -s\right)  }{s}e^{-s\tau}, \label{Lm1ib}%
\end{align}
which of course also follows directly from Eq. (\ref{idisprel}) with $\hat
{B}\left[  Q^{-2}\left(  s+Q^{2}\right)  ^{-1}\right]  =s^{-1}\left[
1-\exp\left(  -s\tau\right)  \right]  $ and implies both
\begin{equation}
\lim_{\tau\rightarrow0}\mathcal{L}_{S,-1}^{\left(  I+\bar{I}\right)  }\left(
\tau\right)  =0
\end{equation}
(cf. Eq. (\ref{idisprel}) at $Q^{2}=0$) and in the opposite limit
\begin{equation}
\lim_{\tau\rightarrow\infty}\mathcal{L}_{S,-1}^{\left(  I+\bar{I}\right)
}\left(  \tau\right)  =-\Pi_{S}^{\left(  I+\bar{I}\right)  }\left(
Q^{2}=0\right)
\end{equation}
(in agreement with Eq. (\ref{largetlim})).

Note that the zero-momentum limit (\ref{q0lim})\ of the correlator can be
recovered from the dispersive representation (\ref{idisprel}) as%
\begin{equation}
\Pi_{S}^{\left(  I+\bar{I}\right)  }\left(  Q^{2}=0\right)  =-2^{4}\pi^{3}\int
d\rho n\left(  \rho\right)  \rho^{4}\int_{0}^{\infty}dssJ_{2}\left(  \sqrt
{s}\rho\right)  Y_{2}\left(  \sqrt{s}\rho\right)  =2^{7}\pi^{2}\int d\rho
n\left(  \rho\right)  ,
\end{equation}
where we have made use of the integral \cite{gra}%
\begin{equation}
\int_{0}^{\infty}ds\,sJ_{2}\left(  \sqrt{s}\rho\right)  Y_{2}\left(  \sqrt
{s}\rho\right)  =-\frac{8}{\pi\rho^{4}}.
\end{equation}
In order to avoid subtraction constants of this type, moments with negative
$k$ are usually not considered in sum-rule applications. In the spin-0
glueball channels, however, the $k=-1$ moments are particularly useful because
the subtraction terms on the phenomenological side (cf. Eq. (\ref{gbsrs})) are
determined by the low-energy theorems of Sec. \ref{lets}. As shown above, the
implementation of this additional first-principle information comes at the
price of dealing with the IOPE-induced subtraction constants, too. Their
interpretation, treatment and role in the $k=-1$ sum rules will be the subject
of Sec. \ref{subconst}.

The dispersive representation of the higher Borel moments follows from Eq.
(\ref{Lm1ib}) by taking the appropriate number of $\tau$-derivatives,
according to the recursion (\ref{recurs}). This results in
\begin{align}
\mathcal{L}_{S,k}^{\left(  I+\bar{I}\right)  }\left(  \tau\right)   &
=-\delta_{k,-1}\Pi_{S}^{\left(  I+\bar{I}\right)  }\left(  Q^{2}=0\right)
+\frac{1}{\pi}\int_{0}^{\infty}dss^{k}\operatorname{Im}\Pi_{S}^{\left(
I+\bar{I}\right)  }\left(  -s\right)  e^{-s\tau}\\
&  =-2^{7}\pi^{2}\delta_{k,-1}\int d\rho n\left(  \rho\right)  -2^{4}\pi
^{3}\int d\rho n\left(  \rho\right)  \rho^{4}\int_{0}^{\infty}dss^{k+2}%
J_{2}\left(  \sqrt{s}\rho\right)  Y_{2}\left(  \sqrt{s}\rho\right)  e^{-s\tau
}. \label{lkinst}%
\end{align}
Note that the UV-convergence of the dispersion integrals is ensured by the
Laplace kernel and that Eq. (\ref{lkinst}) is often easier to handle than the
individual integrated expressions (\ref{inb-1}) - (\ref{inb2}). Moreover, the
subtraction of the instanton-induced continuum (cf. Eq. (\ref{csubtrbmoms})),%
\begin{equation}
\mathcal{R}_{S,k}^{\left(  I+\bar{I}\right)  }\left(  \tau;s_{0}\right)
=\mathcal{L}_{S,k}^{\left(  I+\bar{I}\right)  }\left(  \tau\right)
-\frac{1}{\pi}\int_{s_{0}}^{\infty}dss^{k}\operatorname{Im}\Pi_{S}^{\left(
I+\bar{I}\right)  }\left(  -s\right)  e^{-s\tau},
\end{equation}
provides now immediately a compact expression for the instanton contributions
to all $0^{++}$ glueball sum rules:
\begin{equation}
\mathcal{R}_{S,k}^{\left(  I+\bar{I}\right)  }\left(  \tau;s_{0}\right)
=-2^{7}\pi^{2}\delta_{k,-1}\int d\rho n\left(  \rho\right)  -2^{4}\pi^{3}\int
d\rho n\left(  \rho\right)  \rho^{4}\int_{0}^{s_{0}}dss^{k+2}J_{2}\left(
\sqrt{s}\rho\right)  Y_{2}\left(  \sqrt{s}\rho\right)  e^{-s\tau}.
\label{rkinst}%
\end{equation}
(These contributions are not affected by perturbative (one-loop)
RG-improvement since the anomalous dimension of the interpolating field
$O_{S}\left(  x\right)  $ vanishes to this order.)

Depending on the choice for $n\left(  \rho\right)  $, the size of the
instanton-induced Borel moments (\ref{rkinst}) is either significantly larger
or comparable to the size of their perturbative counterparts (\ref{rm1pc}) -
(\ref{r2pc}). The instanton continuum turns out to be indispensable for the
sum rules to match at intermediate and small $\tau$ \cite{for01} since the
perturbative continuum cannot smoothly extend the exponentially vanishing
instanton contributions (\ref{lkinst}) towards $\tau\rightarrow0$ (cf. Eq.
(\ref{smalltau})).

\subsection{Analytical results for specific instanton size distributions}

\label{analytres}

In order to learn more about the impact of finite-width distributions
$n\left(  \rho\right)  $ on the spectral function (\ref{impinsts}) and the
derived Borel moments, it will prove useful to carry out the integration over
the instanton size analytically. As mentioned in Sec. \ref{isdistrib}, this
becomes possible when specializing to the exponential-tail distribution
(\ref{nexp}). First, however, we are going to establish the point of reference
for later comparison by briefly reviewing the results for the spike
distribution (\ref{spike}). Inserting it into the instanton-induced spectral
function (\ref{impinsts}) results in%
\begin{equation}
\operatorname{Im}\Pi_{S,spk}^{\left(  I+\bar{I}\right)  }\left(  -s\right)
=-2^{4}\pi^{4}\bar{n}\bar{\rho}^{4}s^{2}J_{2}\left(  \sqrt{s}\bar{\rho
}\right)  Y_{2}\left(  \sqrt{s}\bar{\rho}\right)  , \label{imspike}%
\end{equation}
which was used in \cite{for01} and subsequent work. The plot of this
expression in Fig. 1 (a) shows, besides the expected rather violent
oscillations with increasing amplitude (note the scale), a strong violation of
positivity for $s\gtrsim4$ GeV$^{2}$. (We use the standard instanton scales
$\bar{n}=0.75\times10^{-3}$ GeV$^{-4}$ and $\bar{\rho}=1.69$ GeV$^{-1}$
throughout the paper.) At small $s$, the spike-induced imaginary part starts
out quadratically,
\begin{equation}
\operatorname{Im}\Pi_{S,spk}^{\left(  I+\bar{I}\right)  }\left(  -s\right)
\overset{s\rightarrow0}{\longrightarrow}2^{3}\pi^{3}\bar{n}\left(  \bar{\rho
}^{2}s\right)  ^{2}. \label{spikeimsmalls}%
\end{equation}
At large $s$ the rise becomes slightly weaker and gets modulated by a harmonic
oscillation, cf. Eq. (\ref{imsplarges}). The direct instanton contributions to
the continuum-subtracted Borel moments simplify to%
\begin{equation}
\mathcal{R}_{S,k}^{\left(  I+\bar{I}\right)  }\left(  \tau\right)  =-2^{7}%
\pi^{2}\bar{n}\delta_{k,-1}-2^{4}\pi^{3}\bar{n}\bar{\rho}^{4}\int_{0}^{s_{0}%
}dss^{k+2}J_{2}\left(  \sqrt{s}\bar{\rho}\right)  Y_{2}\left(  \sqrt{s}%
\bar{\rho}\right)  e^{-s\tau}.
\end{equation}

Now we derive the analogous results on the basis of the more realistic
finite-width distribution (\ref{nexp}) with exponential large-$\rho$ tail.
Although a Gaussian falloff is favored by instanton-liquid and lattice
simulations (cf. Sec. \ref{isdistrib}) and modifies the detailed behavior of
the instanton-induced imaginary part at $s\lesssim\bar{\rho}^{-2}$, the
resulting expressions will be a useful benchmark for assessing qualitative
effects of realistic $n\left(  \rho\right)  $ (and for evaluating them
numerically). The exponential-tail distribution (\ref{nexp}) specializes the
imaginary part (\ref{impinsts}) to
\begin{equation}
\operatorname{Im}\Pi_{S,\exp}^{\left(  I+\bar{I}\right)  }\left(  -s\right)
=-\frac{2\cdot5^{5}\pi^{4}}{3}\frac{\bar{n}}{\bar{\rho}^{5}}s^{2}\int
_{0}^{\infty}d\rho\rho^{8}J_{2}\left(  \sqrt{s}\rho\right)  Y_{2}\left(
\sqrt{s}\rho\right)  e^{-5\rho/\bar{\rho}}.
\end{equation}
A somewhat lengthy but essentially straightforward calculation shows that the
$\rho$-integral can be expressed as a combination of three hypergeometric
functions,
\begin{align}
\operatorname{Im}\Pi_{S,\exp}^{\left(  I+\bar{I}\right)  }\left(  -s\right)
&  =\frac{5^{5}\pi^{4}}{2^{22}3}\frac{\bar{n}}{\bar{\rho}^{9}}s^{-9/2}\left\{
3^{4}5^{2}7^{2}11\left[  2^{7}s^{2}\bar{\rho}^{4}\,_{2}F_{1}\left(
\frac{9}{2},\frac{13}{2},5,-\frac{25}{4s\bar{\rho}^{2}}\right)  \right.
\right. \nonumber\\
&  +3\cdot5\cdot13\left(  -2^{5}3s\bar{\rho}^{2}\,_{2}F_{1}\left(
\frac{11}{2},\frac{15}{2},6,-\frac{25}{4s\bar{\rho}^{2}}\right)  \right.
+\left.  \left.  \left.  5^{3}11\,_{2}F_{1}\left(  \frac{13}{2},\frac{17}%
{2},7,-\frac{25}{4s\bar{\rho}^{2}}\right)  \right)  \right]  \right\}  .
\label{impiinstext}%
\end{align}
This spectral function is plotted in Fig. 1 (b). Comparison with its
counterpart (\ref{imspike}) in Fig. 1 (a) reveals that the finite-width
distribution has turned the problematic, spike-induced oscillations into a
monotonic falloff at large $s$.

More specifically, owing to $_{2}F_{1}\left(  a,b,c,0\right)  =1$, the first
and leading term in (\ref{impiinstext}) decays as
\begin{equation}
\operatorname{Im}\Pi_{S,\exp}^{\left(  I+\bar{I}\right)  }\left(  -s\right)
\overset{s\rightarrow\infty}{\longrightarrow}\frac{3^{3}5^{7}7^{2}11\pi^{4}%
}{2^{15}}\frac{\bar{n}}{\bar{\rho}^{5}}s^{-5/2}.
\end{equation}
The damping of the large-$s$ oscillations is a generic effect of finite-width
distributions. Nevertheless, the imaginary part (\ref{impiinstext}) still
changes sign once, due to the interference among the individual terms. The
associated positivity violations are much milder than those in Eq.
(\ref{imspike}), however, and their impact on the Borel moments in the
relevant $\tau$-region is strongly reduced. Moreover, the imaginary part of
the full IOPE is now nonnegative for all $s$, as it should be. We therefore
expect that extended size distributions\ will improve the consistency of the
Borel sum rules.

At small $s$, Eq. (\ref{impiinstext}) becomes
\begin{equation}
\operatorname{Im}\Pi_{S,\exp}^{\left(  I+\bar{I}\right)  }\left(  -s\right)
\overset{s\rightarrow0}{\longrightarrow}\frac{2^{7}3\cdot7\pi^{3}}{5^{3}}%
\bar{n}\left(  \bar{\rho}^{2}s\right)  ^{2}\,. \label{imlim11}%
\end{equation}
It is instructive to compare the above behavior to the $s\rightarrow0$ limit
(\ref{spikeimsmalls}) of the spike distribution: the quadratic $s$-dependence
is maintained but Eq. (\ref{imlim11}) is about three times larger. This
enhancement can be traced to the large-$\rho$ tail of $n_{\exp}\left(
\rho\right)  $ by expressing the small-$s$ behavior of the instanton-induced
spectral function for general $n\left(  \rho\right)  $ as
\begin{equation}
\operatorname{Im}\Pi_{S}^{\left(  I+\bar{I}\right)  }\left(  -s\right)
\overset{s\ll\bar{\rho}^{-2}}{\longrightarrow}2^{3}\pi^{3}\bar{n}\left\langle
\rho^{4}\right\rangle s^{2}, \label{iminstsmalls}%
\end{equation}
where $\left\langle \rho^{4}\right\rangle $ is the fourth moment of the
instanton size distribution. Comparison with Eq. (\ref{spikeimsmalls}) or
direct evaluation via Eq. (\ref{nexp}) shows that
\begin{equation}
\left\langle \rho^{4}\right\rangle _{\exp}=\frac{2^{4}3\cdot7}{5^{3}}\bar
{\rho}^{4},
\end{equation}
which relates the enhancement factor to the increased weight of instantons
with $\rho>\bar{\rho}$. We therefore anticipate this factor to be reduced when
the large-$\rho$ contributions are excluded from the IOPE coefficients in Sec.
\ref{muimplem}.

In view of the rather dramatic enhancement and modulation of the
instanton-induced spectral function at large $s$ by the spike distribution,
one might wonder how the latter can produce useful approximations to the
continuum-subtracted Borel moments (\ref{rkinst}) at all. The reason is
twofold: first, these moments are essentially Laplace transforms of the
spectral function,
\begin{equation}
\mathcal{R}_{S,k}^{\left(  I+\bar{I}\right)  }\left(  \tau;s_{0}\right)
=-2^{7}\pi^{2}\delta_{k,-1}\int d\rho n\left(  \rho\right)  -\frac{1}{\pi}%
\int_{0}^{s_{0}}dss^{k}\operatorname{Im}\Pi_{S}^{\left(  I+\bar{I}\right)
}\left(  -s\right)  e^{-s\tau},
\end{equation}
which implies that contributions from the $s\gg\tau^{-1}$ region are
exponentially suppressed. In addition, the large-$s$ contributions of the
spike-induced spectral function partially cancel each other due to the
modulating oscillations. The impact of the large-$s$ behavior on the Borel
moments is therefore strongly reduced (except at very small $\tau$ which are
irrelevant for the sum rule analysis, see below).

In order to explore the effects of finite-width distributions on the $\tau$-
and $s_{0}$-dependence of the Borel moments more quantitatively, we again
resort to the exponential-tail distribution which yields
\begin{align}
\mathcal{R}_{S,k,\exp}^{\left(  I+\bar{I}\right)  }\left(  \tau,s_{0}\right)
&  =-2^{7}\pi^{2}\delta_{k,-1}\int d\rho n_{\exp}\left(  \rho\right)
+\frac{5^{5}\pi^{3}}{2^{22}3}\frac{\bar{n}}{\bar{\rho}^{9}}\int_{0}^{s_{0}%
}ds\,s^{k-4}e^{-s\tau}\left\{  3^{4}5^{2}7^{2}11\left[  2^{7}s^{2}\bar{\rho
}^{4}\,_{2}F_{1}\left(  \frac{9}{2},\frac{13}{2},5,-\frac{25}{4s\bar{\rho}%
^{2}}\right)  \right.  \right. \nonumber\\
&  +3\cdot5\cdot13\left(  -2^{5}3s\bar{\rho}^{2}\,_{2}F_{1}\left(
\frac{11}{2},\frac{15}{2},6,-\frac{25}{4s\bar{\rho}^{2}}\right)  \right.
+\left.  \left.  \left.  5^{3}11\,_{2}F_{1}\left(  \frac{13}{2},\frac{17}%
{2},7,-\frac{25}{4s\bar{\rho}^{2}}\right)  \right)  \right]  \right\}  .
\label{riinexp}%
\end{align}
Obviously, all Borel moments reach finite limits for $\tau\rightarrow0$.
Towards $\tau\rightarrow\infty$ we find from Eq. (\ref{iminstsmalls}) that the
$\tau$-dependence of the Borel moments vanishes as%
\begin{equation}
\mathcal{R}_{S,k}^{\left(  I+\bar{I}\right)  }\left(  \tau,s_{0}\right)
\overset{\tau\rightarrow\infty}{\longrightarrow}-2^{7}\pi^{2}\delta_{k,-1}\int
d\rho n\left(  \rho\right)  +2^{3}\pi^{2}\left(  k+2\right)  !\frac{\bar
{n}\left\langle \rho^{4}\right\rangle }{\tau^{k+3}}\left[  1-\rho_{k+2}\left(
s_{0}\tau\right)  \right]  , \label{rinlargetau}%
\end{equation}
where we have used the definition (\ref{rhok}) for the partial sums $\rho
_{k}\left(  x\right)  $. Of course this result holds for all reasonable
instanton size distributions: specific choices for $n\left(  \rho\right)  $ do
not affect the $\tau$-dependence but just the overall magnitude of the moments
at $\tau\gg\bar{\rho}^{2}$.

The impact of finite-width distributions on the $s_{0}$-dependence of the
Borel moments (which generates the $0^{\pm}$-channel dependence of the direct
instanton contributions) will play an important role in our subsequent
discussion. From Eq. (\ref{riinexp}) we find that in the sum-rule relevant
$\left(  \tau,s_{0}\right)  $-region (specifically for $s_{0}\gtrsim2-4$
GeV$^{2}$ and $\tau\gtrsim0.2$ GeV$^{-1}$) the magnitude of the instanton
contributions decreases more strongly with increasing $s_{0}$ if the
exponential-tail distribution is employed. The direct instanton contributions
to the pseudoscalar glueball sum rule, where $s_{0}$ is typically a factor of
two larger than in the scalar channel, therefore become smaller than those to
the scalar sum rule. This reduces the instanton-induced repulsion in the
$0^{-+}$ channel without significantly weakening the important attraction in
the $0^{++}$ channel.

Furthermore, we observe that for sufficiently large $s_{0}$ the
instanton-induced Borel moments (\ref{riinexp}) turn negative at small $\tau$.
(For the $k=0$ moment this happens if $\tau\lesssim\tau_{c}\sim0.3$ GeV$^{-2}%
$. For $k>0$, $\tau_{c}$ increases due to the stronger $s^{k}$-weight of the
spectral function at larger $s$.) This positivity violation (which grows with
$s_{0}$ and therefore has more impact on the pseudoscalar sum rule) would
reduce the fiducial $\tau$-domain of the associated Borel sum rules (cf. Sec.
(\ref{srsetup})) but does not a priori prohibit a sum rule analysis.
Nevertheless, it exposes a shortcoming of Eq. (\ref{riinexp}) which can be
traced to the contributions from large instantons. In fact, we will argue in
the next section on general grounds that such contributions have to be
excluded from the IOPE coefficients. (The discussion of the far more serious
instanton-induced positivity violations at \textit{large} $\tau$ in the
$0^{-+}$ channel will be the subject of Sec. \ref{topscr}.)

Finally, we note that the above results remain qualitatively unchanged when
$n_{\exp}\left(  \rho\right)  $ is replaced by the Gaussian-tail distribution
(\ref{ngn3}). Even the quantitative results are generally very similar, as
expected from the arguments of Sec. \ref{isdistrib} and confirmed by numerical
integration of the corresponding $\rho$-integrals.

\subsection{Implementation of the operator renormalization scale}

\label{muimplem}

In the exact version of the (I)OPE, contributions to the Wilson coefficients
originate exclusively from hard field modes with momenta larger than the
operator renormalization scale $\mu$ \cite{wil69,nov85}. Potential
double-counting of soft modes and IR renormalons \cite{ben99} are thereby
excluded from the outset and the $\mu$-independence of (RG-invariant)
short-distance amplitudes becomes manifest. However, the gauge-invariant
implementation of IR cutoffs in the perturbative Wilson coefficients is
technically complex and commonly neglected in QCD sum-rule applications (with
a few exceptions \footnote{if IR divergencies are encountered, see e.g. Refs.
\cite{ir}}). This ``pragmatic'' approximation works well since in QCD soft
perturbative amplitudes are generally much smaller than their nonperturbative
counterparts (which generate the condensates).

However, it is a priori unclear whether the pragmatic neglect of IR cutoffs
works for nonperturbative Wilson coefficients as well, as tacitly assumed in
all previous work on direct instantons. In the following we will argue that
realistic, finite-width instanton size distributions permit an explicit and
straightforward implementation of such IR cutoffs for the direct instanton
contributions. This allows us to improve upon the ``pragmatic'' treatment of
the nonperturbative sector and to assess its conceptual basis and range of
validity. (The exceptional strength of the direct instanton contributions
makes the spin-0 glueball correlators a particularly suitable testing ground
for this renomalization.)

Our implementation of the renormalization scale $\mu$ is based on the
observation that an instanton with fixed size $\rho$ mainly contributes field
modes with momenta $\left|  k\right|  <\rho^{-1}$ to the correlator. The
contributions of large instantons with $\rho>\mu^{-1}$ and the fluctuations
around them are therefore almost exclusively soft and should be excluded from
the Wilson coefficients. This (partial, see below) cutoff procedure is gauge
invariant simply because the instanton size is. Since the renormalization
scale associated with the conventional condensate values (cf. Sec.
\ref{inputscales}) is not very accurately determined, it would not make sense
to treat $\mu^{-1}$ as a sharp cutoff. Instead, we implement it smoothly by
replacing the full instanton size distribution $n\left(  \rho\right)  $ with
\begin{equation}
\tilde{n}_{\mu}\left(  \rho\right)  \equiv\theta_{\beta}\left(  \mu^{-1}%
-\rho\right)  n\left(  \rho\right)  \label{nren}%
\end{equation}
where the soft step function $\theta_{\beta}$ can be chosen, e.g., in the form
of a Fermi distribution
\begin{equation}
\theta_{\beta}\left(  \mu^{-1}-\rho\right)  =\frac{1}{2}\left[  1-\tanh\left(
\frac{\beta}{2}\left(  \rho-\mu^{-1}\right)  \right)  \right]  .
\label{thetaf}%
\end{equation}
The ``diffuseness'' parameter $\beta$ sets the scale for the width of the
transition region. With $\beta\ll\mu$ the cutoff practically ceases to exist,
i.e. all instanton sizes are about evenly affected, whereas for $\beta\gg\mu$
a sharp cutoff is reached. For practical calculations we find values
$\beta\sim5\bar{\rho}^{-1}\sim3$ GeV to be an effective compromise between
these extremes. (For $\beta\gtrsim5$ GeV strong oscillations of the imaginary
part set in (even at $s>20$ GeV$^{2}$) while for $\beta\lesssim2$ GeV the
cutoff is already rather ineffective.) Alternatively, we will use%
\begin{equation}
\theta_{\beta}\left(  \mu^{-1}-\rho\right)  =\frac{1}{2}-\frac{1}{\pi}%
\arctan\left[  \beta\left(  \rho-\mu^{-1}\right)  \right]  , \label{thetaat}%
\end{equation}
which has a softer transition region and a bigger large-$\rho$ tail for equal
values of $\beta$, to estimate the sensitivity\ of the sum-rule results to the
details of the cutoff procedure.

The physical interpretation of the above $\mu$-implementation is very
transparent:\ the modified distribution (\ref{nren}) just excludes non-direct
instanton contributions to the IOPE coefficients since those are already
included in the condensates \footnote{Note that the large-$\rho$ cutoff
affects higher-$O\left(  \hbar\right)  $ corrections as well since it
effectively restricts the internal momenta of the quantum fluctuations.}.
Nevertheless, this does not exclude \textit{all} $\left|  k\right|  <\mu$
modes since even arbitrarily small instantons contain some soft contributions
\footnote{Another way to see that the large-$\rho$ cutoff cannot entirely
exclude soft contributions derives from the singularity structure of the
direct-instanton contributions (\ref{instx}) at $x=0$, as discussed in Section
\ref{dirinstx}. Indeed, potential instanton-induced power corrections are not
removed by the large-$\rho$ cutoff because the small-$x$ singularity structure
is not affected by large instantons (for physically sensible $n\left(
\rho\right)  $). (The nucleon correlator IOPE \cite{for93} provides a simple
example: instanton-induced power corrections are present despite the
large-$\rho$ ``cutoff'' implicit in the spike distribution.) Note,
incidentally, that the self-duality (\ref{sdual}) of the instanton's field
strength severely restricts potential instanton-induced power corrections in
quark-based correlators (to at most a few terms) \cite{dub81}.}. A complete
renormalization would naively amount to replacing%
\begin{equation}
G_{\mu}^{\left(  I\right)  }\left(  x\right)  \rightarrow\int_{\left|
k\right|  <\mu}\frac{d^{4}k}{\left(  2\pi\right)  ^{4}}e^{-ikx}\tilde{G}_{\mu
}^{\left(  I\right)  }\left(  k\right)
\end{equation}
where $\tilde{G}_{\mu}^{\left(  I\right)  }$ is the Fourier transform of the
instanton field. However, such a procedure is unacceptable since it violates
gauge invariance. In fact, devising a gauge-invariant cutoff is a complex task
since gauge transformations generally change the momentum content of the
fields. This problem is well-known in the application of Wilson's
renormalization group to gauge fields and has not yet been solved
satisfactorily. Nevertheless, there is reason to believe that our above,
simplified renormalization of the direct instanton contributions is
sufficiently complete since the overall momentum transfer $Q$ acts as an
additional IR\ cutoff \cite{moc97}, by suppressing field modes with momenta
$k\ll Q$.

An interesting question in this context concerns the direct-instanton induced
$\mu$-dependence. The $\mu$-dependence of both the perturbative coefficients
and the condensates (which arise from the anomalous-dimensions) is logarithmic
(cf. Appendix \ref{bpmoms}) and therefore rather weak \footnote{Additional
$\mu$-dependence of the condensates due to nonperturbative physics (e.g. due
to instantons \cite{bou03,kov03}) can of course not be excluded. Our results
and general sum-rule experience suggest, however, that it should be similarly
weak for $\mu\sim2-3\Lambda_{QCD}$.}. Since mutual cancellations among all
contributions are required to render the correlators $\mu$-independent, one
expects the $\mu$-dependence of the nonperturbative coefficients to be
similarly weak. This turns out to be indeed the case, mainly because
large-$\rho$ contributions are already strongly suppressed by the external
momentum scale $Q\gtrsim1$ GeV in Eq.\ (\ref{piqinst}) and by the exponential
tail of realistic instanton size distributions (cf. Eq. (\ref{largerhon})).
Hence the $\mu^{-1}$ cutoff does only reduce intermediate-size instanton
contributions significantly, implying a relatively weak $\mu$-dependence for
$\mu<\bar{\rho}^{-1}$ (which holds for $\mu=0.5$ GeV and $\bar{\rho}%
^{-1}\simeq0.6$ GeV adopted below).

One could contemplate, incidentally, to identify $\mu=Q$ in Eq. (\ref{nren})
by analogy with the RG improvement of the perturbative IOPE\ coefficients.
However, avoiding or summing large logarithms is not an issue here, the impact
on the sum rules would be limited by the moderate $\mu$-dependence, and the
additional $s$-dependence in the spectral integrals of Eq. (\ref{rkinst})
would obscure the comparison with spike distribution results. We will
therefore keep $\mu$ fixed in the quantitative analysis below.

The implementation of $\mu$ has several significant effects. One of them is
the reduction of the instanton-induced imaginary part for $s\ll\mu^{2}$ (where
the excluded soft-mode contributions had mainly contributed), and a second one
is the shift of its leading peak towards larger $s$. Both features can be
understood by noting that
\begin{equation}
\operatorname{Im}\Pi_{S}^{\left(  I+\bar{I}\right)  }\left(  -s\right)
\overset{s\ll\mu^{2}<\bar{\rho}^{-2}}{\longrightarrow}2^{3}\pi^{3}\bar{n}%
s^{2}\left[  \left\langle \rho^{4}\right\rangle _{\mu}+\frac{1}{6}\left\langle
\rho^{6}\right\rangle _{\mu}s+O\left(  s^{2}\right)  \right]
\label{inimrenorm}%
\end{equation}
for $s\ll\mu^{2}$, where we have defined the moments of the renormalized size
distribution $\tilde{n}\left(  \rho\right)  $ as
\begin{equation}
\left\langle \rho^{k}\right\rangle _{\mu}=\frac{1}{\bar{n}}\int_{0}^{\infty
}d\rho\rho^{k}\tilde{n}_{\mu}\left(  \rho\right)  . \label{kmom}%
\end{equation}
The reduced magnitude of\ the imaginary part (\ref{inimrenorm}) is obviously
due to $\theta_{\beta}\left(  \mu^{-1}-\rho\right)  <1$ for all $\rho$, which
implies $\left\langle \rho^{k}\right\rangle _{\mu}<\left\langle \rho
^{k}\right\rangle $. The shift of the peak is caused by the stronger reduction
of the higher moments: due to the asymmetry of $n\left(  \rho\right)  $,
impressed by its limiting behavior (cf. Eqs. (\ref{smallrhon}) and
(\ref{largerhon})), $\left\langle \rho^{k}\right\rangle /\bar{\rho}^{k}$
increases with increasing $k$. Since this enhancement originates from the
large-$\rho$ tail of the distribution, it is strongly reduced by
renormalization and the ratio $\left\langle \rho^{k}\right\rangle _{\mu
}/\left\langle \rho^{k}\right\rangle $ therefore decreases with $k$.

Another significant renormalization effect is the reduction of the number of
``active'' instantons in the Wilson coefficients, i.e.
\begin{equation}
\bar{n}=\int_{0}^{\infty}d\rho n\left(  \rho\right)  \rightarrow\int
_{0}^{\infty}d\rho\tilde{n}_{\mu}\left(  \rho\right)  \equiv\bar{n}%
_{dir}\equiv\zeta\bar{n} \label{ndir}%
\end{equation}
with $\zeta<1$, where $\bar{n}_{dir}$ correctly accounts for the density of
\textit{direct} instantons. This effect is missed when using the spike
distribution since it is normalized to the total instanton density, i.e. large
instantons with $\rho>\bar{\rho}$ are not excluded but simply counted as
instantons of average size. (This suggest to use $\bar{n}_{dir}$ instead of
$\bar{n}$ to improve direct instanton calculations on the basis of the spike
distribution.) The above reasoning indicates, incidentally, that direct
instanton calculations involving the spike distribution strictly make sense
only for $\bar{\rho}<\mu^{-1}$. Although satisfied by the (not very accurately
determined) standard values $\mu^{-1}\sim0.4$ fm and $\bar{\rho}\sim0.33$ fm,
this condition would be violated by larger choices for $\mu$ and/or $\bar
{\rho}$.

Summarizing the main results of this section, we find the exclusion of
non-direct instantons by an explicit large-$\rho$ cutoff to be mandatory when
implementing finite-width distributions. However, the ``pragmatic'' neglect of
$\mu$ \textit{together} with the use of the spike distribution (with
$\bar{\rho}<\mu^{-1}$) can be useful for an approximate estimate of the Borel
moments since large instantons (with $\rho\gg\bar{\rho}$) are rare in the QCD
vacuum and since their contributions are bounded by $Q^{-1}$ acting as an
additional cutoff. This justifies the tacit assumption underlying earlier
implementations of direct instantons. Nevertheless, the renormalization
generates several new and important effects, including an effective reduction
of the instanton density, which will significantly improve both sum rule
consistency and results (see Sec. \ref{qsra}).

\section{IOPE 3:\ topological charge screening}

\label{topscr}

The direct instanton contributions to $0^{++}$ and $0^{-+}$ glueball
correlators differ only in their sign (cf. Secs. \ref{qualas} and
\ref{dirinst}). As a consequence, the instanton-induced attraction in the
scalar channel turns into a strong repulsion in the pseudoscalar channel.
Therefore it is hardly surprising that the signal for the pseudoscalar
glueball disappears and that even the general positivity requirement is
violated \cite{zha03} when these repulsive contributions are added to the
perturbative Wilson coefficients. We demonstrate this by plotting the $k=0$
Borel moment in the $0^{-+}$ channel,
\begin{equation}
\mathcal{R}_{P,0}^{\left(  pc+I\right)  }\left(  \tau;s_{0}\right)
=\frac{1}{\pi}\int_{0}^{s_{0}}ds\left[  \operatorname{Im}\Pi_{P}^{\left(
pc\right)  }\left(  -s\right)  +\operatorname{Im}\Pi_{P}^{\left(
I+\bar{I}\right)  }\left(  -s\right)  \right]  e^{-s\tau}, \label{r0inpert}%
\end{equation}
obtained from both perturbative and instanton-induced Wilson coefficients (on
the basis of the spike distribution (\ref{spike})), in Fig. 2 (a). The
sum-rule relevant $s_{0}$-range starts at $s_{0}\gtrsim6-8$ GeV$^{2}$, where
the Borel moment indeed becomes negative for $\tau\gtrsim0.8$ GeV$^{-2}$, i.e.
in the middle of the fiducial domain (cf. Sec. \ref{qsra}). (The negative
areas of the Borel moment at smaller $s_{0}$ and $\tau$ are an artefact of the
spike distribution which will be removed by realistic instanton size
distributions, cf. Sec. \ref{qsra}.) Moreover, the $\tau$-slope becomes
positive for $\tau\gtrsim1$ GeV$^{-2}$, and both features make a match to the
decaying exponential of a resonance (cf. Eq. (\ref{gbsrs})) impossible.
Contrary to lattice results \cite{lee00,har02}, the corresponding sum rule
would therefore predict the absence of low-lying pseudoscalar glueballs. In
addition, we will find below that the low-energy theorem (\ref{pLET}) in the
$0^{-+}$ channel would be badly violated.

In search for the origin of these problems one should keep in mind that the
(up to the sign) identical expressions for the direct-instanton induced Borel
moments (\ref{rkinst}), involving the same approximations, play a highly
beneficial role in the $0^{++}$ sum rules and enhance their consistency
\cite{for01}. In fact, this might not be as paradoxical as it first appears:
although the moments (\ref{rkinst}) have the same magnitude in both channels,
the direct instanton contributions to the sum rules do not. The difference
lies in the value of the duality threshold $s_{0}$ which is typically more
than twice as large in the $0^{-+}$ than in the $0^{++}$ channel. (This is
mainly a reflection of the larger $0^{-+}$ glueball mass.) Their rather strong
$s_{0}$-dependence thus equips the direct instanton contributions with a
certain channel dependence.

As demonstrated above, however, this channel dependence is too weak to resolve
the mentioned deficiencies. Yet it is conceivable that the underlying
approximations, i.e. the spike distribution and the neglected renormalization,
may have underestimated the $s_{0}$-dependence. Partly for this reason we have
improved upon these standard approximations, by implementing realistic
instanton size distributions and the renormalization scale. It turns out that
their combined impact indeed mitigates the problems in the $0^{-+}$ glueball
channel, but it is too small to resolve any of them.

Thus we face for the first time the situation that direct instanton
contributions seem to worsen the consistency of a set of QCD sum rules and to
create serious new problems. Since our treatment of the direct instanton
sector leaves not much room for further improvement \footnote{As discussed in
Section \ref{dirinst}, multi-instanton effects are generally negligible, due
to the small instanton packing fraction. In the pseudoscalar meson channel,
where direct instanton contributions have exceptional strength \cite{for201},
this was shown explicitly \cite{dor97}.}, one is bound to conclude that
essential additional physics is still missing in the IOPE coefficients of the
$0^{-+}$ glueball correlator. Our main guide in the search for this physics,
which will be subject of the next section, is the fact that it apparently did
not show up in other hadron channels: previously considered sum rules were
found to be consistent and stable with perturbative and direct-instanton
induced Wilson coefficients only.

\subsection{Topological charge correlations}

As argued above, the failure of direct instanton contributions to generate
acceptable pseudoscalar glueball sum rules strongly suggests that equally
important - and therefore very likely nonperturbative - contributions to the
Wilson coefficients are still amiss. Since these contributions should affect
almost exclusively the $0^{-+}$ glueball correlator, one is led to search for
a unique property which singles this channel out among other hadron
correlators. The result is quite obvious: the interpolating field $O_{P}$ is
proportional to the topological charge density $Q$,
\begin{equation}
O_{P}\left(  x\right)  =8\pi Q\left(  x\right)  ,
\end{equation}
which implies that the $0^{-+}$ glueball correlator is proportional to the
topological charge correlator \footnote{As a consequence of the chiral anomaly
(see below), the same holds for the correlator of the divergence of the
flavor-singlet axial current (with the quantum numbers of the $\eta_{0})$ in
the chiral limit. This correlator is thus equally sensitive to topological
charge correlations.}. Hence it distinguishes in the strongest possible way
between instantons and anti-instantons and is highly sensitive to
instanton-anti-instanton correlations. (In contrast, $O_{S}\left(  x\right)  $
is proportional to the gluonic action density and therefore treats instantons
and anti-instantons on the same footing.)

The correlations between topological charges are generated by light quark
exchange or, equivalently at low energy, by the exchange of flavor-singlet
pseudoscalar mesons. The form of these interactions is dictated by the axial
$U\left(  1\right)  $ anomaly \cite{ano}
\begin{equation}
\partial_{\mu}\sum_{i=u,d,s}\bar{q}_{i}\gamma_{\mu}\gamma_{5}q_{i}%
=2\sum_{i=u,d,s}m_{i}\bar{q}_{i}i\gamma_{5}q_{i}+2N_{f}Q. \label{anom}%
\end{equation}
Compliance with the ensuing behavior under $U_{A}\left(  1\right)  $
transformations requires the lightest pseudoscalar flavor-singlet meson
$\eta_{0}$ to couple to the topological charge density as \cite{div80}%
\begin{equation}
\Delta\mathcal{L}=-i\gamma_{\eta_{0}}Q\left(  x\right)  \eta_{0}\left(
x\right)  \label{etaQ}%
\end{equation}
where the coupling $\gamma_{\eta_{0}}$ is often written as $\gamma_{\eta_{0}%
}=\sqrt{2N_{f}}/f_{\eta_{0}}.$ The form of this interaction and most of its
qualitative consequences do not depend on the specific origin of $Q\left(
x\right)  $. Instanton models and lattice simulations indicate that the
topological charge density in the QCD vacuum is rather strongly localized and
mostly carried by instantons \cite{sch98,tep00}, however, and we will adopt
this picture in some of the developments below. Owing to the coupling
(\ref{etaQ}), $\eta_{0}$ exchange generates strong correlations between
topological charges which add to the direct-instanton induced ones (of range
$\bar{\rho}$) evaluated in Sec. \ref{dirinst}. The quark chiralities entering
the anomaly equation (\ref{anom}) require the $\eta_{0}$-mediated interactions
to be attractive (repulsive) between topological charges of opposite (equal) sign.

Between isolated topological charges these interactions would be of long range
since the $\eta_{0}$ would be a light (quasi-\nolinebreak \nolinebreak )
Goldstone boson. In the topological-charge ensemble of the QCD vacuum,
however, the $\eta_{0}$-exchange forces are screened by the formation of
Debye-H\"{u}ckel clouds \cite{deb23} in which positive (negative) topological
charges surround themselves with negative (positive) ones. This collective
mechanism, in turn, renders the $\eta_{0}$-induced interactions short-ranged
by generating a screening mass for the $\eta_{0}$ \footnote{Note the direct
analogy with long-range Coulomb interactions which get screened into
short-range Yukawa interactions between the electric charges of an electrolyte
or a QED plasma, thereby turning the photon effectively massive.} (which
solves the $U\left(  1\right)  $ problem \cite{u1}). Topological charge
screening has been discussed in Refs. \cite{dow92,kik92} and investigated in
ILM \cite{shu95} as well as unquenched lattice \cite{chu00,has00} simulations.
Indirect evidence for strongly localized screening clouds has also been found
in the difference-to-sum ratio of the isovector scalar and pseudoscalar ILM
correlators \cite{fac03} where ``unquenching'' strongly suppresses the
instanton-induced spin-flip probability for a propagating quark at short distances.

In order to understand how topological charge screening affects the IOPE, one
has to compare the factorization scale $\mu$ with the characteristic scale of
the screening mechanism. Since the physical $\eta^{\prime}$ mass (i.e. the
$\eta_{0}$ mass after $\eta_{0}-\eta_{8}$ mixing is taken into account, see
below) is about twice as large as $\mu$, topological charge screening takes
place over distances of the order of the screening length
\begin{equation}
\lambda_{scr}\simeq\frac{1}{m_{\eta^{\prime}}}\sim0.2\text{ fm}<\mu^{-1}
\label{lamscr}%
\end{equation}
which are small on the IOPE scale. Hence screening contributes mainly to the
Wilson coefficients. This also follows directly from the observation that
$Q$-screening strongly influences the $x$-dependence of the topological charge
correlator at $\left|  x\right|  \lesssim\lambda_{scr}\ll\Lambda_{QCD}^{-1}$,
which resides exclusively in the Wilson coefficients. And since $Q$-screening
is a nonperturbative and collective mechanism originating from the light-quark
sector, it obviously cannot be described either by perturbative or by direct
instanton contributions. Thus we have reached our objective to uncover
nonperturbative contributions to the Wilson coefficients which go beyond the
direct instanton approximation and affect almost exclusively the $0^{-+}$
glueball correlator.

Under the premise that the topological vacuum charge density is mostly
associated with instantons, the screening corrections can be identified with
multi-instanton effects \footnote{At first, there seems to be a certain
analogy with the effective quark mass which instanton-induced quark zero modes
acquire due to interactions with ambient, soft vacuum fields. This effective
mass also goes beyond the isolated-instanton approximation. However, it
originates from soft subgraph corrections (of mean field type) to hard
subgraphs and therefore does not affect the $x$-dependence of the Wilson
coefficients. The topological charge screening corrections, in contrast, have
a very pronounced $x$- and channel dependence.}. In fact, topological charge
screening has been observed in instanton vacuum model calculations and found
to affect mostly the $0^{-+}$ glueball and $\eta^{\prime}$ channels
\cite{sch98,sch95}. Even if light-quark induced interactions and therefore
screening effects are neglected, phenomenological and lattice results in the
majority of all calculated hadron correlators are well reproduced. However, in
the $\eta^{\prime}$ and the related $0^{-+}$ glueball channel (and to a lesser
degree in the scalar-isovector $a_{0}$ meson channel) this approximation
clearly fails and the $\eta^{\prime}$ correlator turns negative.
``Unquenching'' of the light-fermion sector resolves these problems and
generates the $\eta^{\prime}$ \cite{sch96} and $0^{-+}$ glueball \cite{sch95}
resonances (although neglected interference terms between classical and
one-loop effects did not allow to determine $0^{-+}$ glueball properties in
Ref. \cite{sch95}).

\subsection{Screening contributions}

The screening contributions to the topological charge correlator (and
therefore to the $0^{-+}$ glueball correlator) can be calculated from the
pertinent low-energy approximation to the (Euclidean) QCD generating
functional%
\begin{equation}
Z\left[  \theta\right]  =\int D\left\{  A_{\mu},q,\bar{q},\zeta\right\}
\exp\left(  -S_{QCD}+i\int d^{4}x\theta\left(  x\right)  Q\left(  x\right)
\right)  \label{genfunc1}%
\end{equation}
in the presence of a classical source $\theta\left(  x\right)  $. The
correlation function of the topological charge density (and the related one
for the $0^{-+}$ glueball) is then obtained by functional differentiation,
\begin{equation}
\left\langle Q\left(  x\right)  Q\left(  0\right)  \right\rangle =-Z\left[
0\right]  ^{-1}\left.  \frac{\delta^{2}Z\left[  \theta\right]  }{\delta
\theta\left(  x\right)  \delta\theta\left(  0\right)  }\right|  _{\theta=0}.
\label{topcorr1}%
\end{equation}

We now outline the derivation of the appropriate low-energy approximation to
$Z\left[  \theta\right]  $. (More details can be found in Refs.
\cite{dow92,kik92}.) To begin with, one separates the domain of the gluonic
integration variable into the underlying multi-instanton background and the
non-topological remainder, over which one integrates first. (Of course, in
practice this could be done at best approximately, on the lattice, but the
part of the result which is relevant in our context will be obtained
indirectly from the $U_{A}\left(  1\right)  $ anomaly (\ref{anom}) and
effective field theory arguments.) It remains to sum over all classical
multi-instanton configurations and to integrate over their collective
coordinates $\left(  U,\rho,x_{0}\right)  _{i}$. Integrating over the color
orientations $U_{i}$ and sizes $\rho_{i}$ turns the $i$-th (anti-) instanton
into a colorless lump of topological charge density centered at $x_{0,i}$.
Equation (\ref{genfunc1}) then becomes the partition function of a medium of
localized topological charges (which we approximate as pointlike), interacting
via a multi-local ``potential'' $V$:%
\begin{equation}
Z\left[  \theta\right]  =\sum_{N_{+},N_{-}}\frac{\bar{n}^{N_{+}+N_{-}}}%
{N_{+}!N_{-}!}\left(  \prod_{i=1}^{N_{+}+N_{-}}\int d^{4}x_{0,i}\right)
\exp\left(  -V\left(  x_{0,1}...x_{0,N_{+}+N_{-}}\right)  +i\sum_{i=1}%
^{N}Q_{i}\theta\left(  x_{0,i}\right)  \right)  , \label{genfunc2}%
\end{equation}
where $\bar{n}$ is the average density of either positive or negative
topological charges. In order to find an approximate expression for $V$, we
recall that the lowest-lying excitations of the QCD vacuum are the
pseudoscalar Goldstone bosons. Since (in the chiral limit) the flavor-singlet
$\eta_{0}$ gets its mass from the topological gluon fields over which we did
not integrate yet, we expect it at this stage to be part of a degenerate
$U\left(  3\right)  $ flavor nonet. At low energies, the interactions $V$
between the topological charges will then be dominantly mediated by $\eta_{0}$
exchange,
\begin{equation}
V\left(  x_{0,1}...x_{0,N_{+}+N_{-}}\right)  \simeq\sum_{i<j}^{N_{+}+N_{-}%
}v_{\eta_{0}}\left(  x_{0,i},x_{0,j}\right)  ,
\end{equation}
with the coupling to $Q\left(  x\right)  $ determined by the axial anomaly
(cf. Eq. (\ref{etaQ})). Since the $\eta_{0}$ contains a superposition of both
quark chiralities with opposite sign, the anomaly implies that equal
(opposite) topological charges repel (attract) each other. We therefore have
\begin{equation}
v_{\eta_{0}}\left(  x_{0,i},x_{0,j}\right)  =q_{i}q_{j}\int\frac{d^{4}%
q}{\left(  2\pi\right)  ^{4}}\frac{\gamma_{\eta_{0}}^{2}}{q^{2}+m_{NG}^{2}%
}\exp\left(  iq\left(  x_{0,i}-x_{0,j}\right)  \right)  \label{veta0}%
\end{equation}
where $q_{i}=\pm1$ is the topological charge located at $x_{0,i}$ \ and
$m_{NG}^{2}$ is the part of the $\eta_{0}$ mass due to finite current quark
masses. After absorbing an infinite factor into the renormalization of
$\bar{n}$ and performing the remaining sums and integrals in Eq.
(\ref{genfunc2}), one ends up with the low-energy approximation
\begin{equation}
Z\left[  \theta\right]  \simeq\mathcal{N}\int D\eta_{0}\exp\left(
-S_{eff}\left[  \eta_{0},\theta\right]  \right)
\end{equation}
and the effective action \cite{pol77}%
\begin{equation}
S_{eff}\left[  \eta_{0},\theta\right]  =\int d^{4}x\left[  \frac{1}{2}\left(
\partial\eta_{0}\right)  ^{2}+\frac{1}{2}m_{NG}^{2}\eta_{0}^{2}+2\bar{n}%
\cos\left(  \gamma_{\eta_{0}}\eta_{0}+\theta\right)  \right]  . \label{seff}%
\end{equation}
Expanding the action (\ref{seff}) up to $O\left(  \eta_{0}^{2}\right)  $ one
finds that the interaction between the topological charges in the vacuum has
generated the expected screening contribution%
\begin{equation}
m_{scr}^{2}=2\bar{n}\gamma_{\eta_{0}}^{2} \label{mscr}%
\end{equation}
to the $\eta_{0}$ mass, $m_{\eta_{0}}^{2}=m_{NG}^{2}+m_{scr}^{2}$, which does
not vanish in the chiral limit and solves the $U\left(  1\right)  $ problem.
(The tree approximation in Eq. (\ref{veta0}) is reliable as long as $\bar
{n}\gamma_{\eta_{0}}^{4}\ll1$ and $\bar{n}/m_{NG}^{4}\gg1$.) With the help of
Eq. (\ref{topcorr1}) we finally obtain the low-energy approximation
\begin{equation}
\left\langle Q\left(  x\right)  Q\left(  0\right)  \right\rangle
=\frac{\Pi_{P}\left(  x\right)  }{\left(  8\pi\right)  ^{2}}\simeq2\bar
{n}\delta^{4}\left(  x\right)  -\left(  2\bar{n}\gamma_{\eta_{0}}\right)
^{2}\left\langle \eta_{0}\left(  x\right)  \eta_{0}\left(  0\right)
\right\rangle \label{leqcor}%
\end{equation}
to the topological charge correlator. Although we have referred to instantons
as the source of the topological charge density $Q\left(  x\right)  $ for
definiteness, the above arguments would hold with inessential modifications
(e.g. in the set of collective coordinates) for other localized,
topological-charge carrying gluon fields as well.

The first term in Eq. (\ref{leqcor}) is just the\ correlation due to the
topological charge ``cloud'' of a single instanton in the pointlike limit
(which we have adopted above for simplicity). This can be seen explicitly from
the direct instanton contribution (\ref{instx}) by implementing the
pointlike-instanton limit (\ref{pointlim}) with the help of the spike
distribution for $\bar{\rho}=0$, i.e. $n_{pl}\left(  \rho\right)  =\bar
{n}\delta\left(  \rho\right)  $. The result
\begin{equation}
\frac{\Pi_{P,\rho\rightarrow0}^{\left(  I+\bar{I}\right)  }\left(  x\right)
}{\left(  8\pi\right)  ^{2}}=\frac{1}{\left(  8\pi\right)  ^{2}}\int d\rho
n_{pl}\left(  \rho\right)  \Pi_{P}^{\left(  I+\bar{I}\right)  }\left(
x^{2};\rho\right)  =2\bar{n}\delta^{4}\left(  x\right)  \label{scr1}%
\end{equation}
is indeed identical to the first term in Eq. (\ref{leqcor}). The second term
in Eq. (\ref{leqcor}) is the expected Debye screening correction.

\subsection{Singlet-octet mixing, Borel moments and subtraction term}

\label{somix}

The screening contributions to the topological charge and $0^{-+}$ glueball
correlators in Eq. (\ref{leqcor}) do not yet contain the $\eta_{0}-\eta_{8}$
mixing corrections which arise from finite light-quark masses. These
corrections are rather substantial, however, and should be implemented into
the IOPE. This will also allow us to use the experimental $\eta^{\prime}$ and
$\eta$ meson masses and mixing angle, instead of extrapolations to the chiral
limit. In addition, it is more consistent with our use of the standard
condensate values below which were obtained from phenomenology and thus
correspond to realistic light quark masses \footnote{The calculation of the
quark-mass dependence of gluon condensates on the lattice is impeded by small
signal-to-noise ratios and problems with reaching physical light-quark masses.
The quark-mass dependence is expected to be non-negligible.}. The quark-mass
dependence of the perturbative Wilson coefficients, on the other hand,
originates from (dominantly strange) quark-loop corrections and should be negligible.

The screening contributions to Eq. (\ref{scr1}) can be adapted to physical
quark masses by expressing them in terms of the $\eta$ and $\eta^{\prime}$
mass eigenstates. These are\ related to the singlet $\left(  \eta_{0}\right)
$ and octet $\left(  \eta_{8}\right)  $ flavor eigenstates by
\begin{equation}
\left(
\begin{tabular}
[c]{l}%
$\left|  \eta_{0}\right\rangle $\\
$\left|  \eta_{8}\right\rangle $%
\end{tabular}
\right)  =\left(
\begin{tabular}
[c]{cc}%
$\cos\varphi$ & $-\sin\varphi$\\
$\sin\varphi$ & $\cos\varphi$%
\end{tabular}
\right)  \left(
\begin{tabular}
[c]{l}%
$\left|  \eta^{\prime}\right\rangle $\\
$\left|  \eta\right\rangle $%
\end{tabular}
\right)
\end{equation}
where $\varphi$ is the $\eta_{0}-\eta_{8}$ mixing angle. The $\eta_{0}$
correlator therefore becomes
\begin{equation}
\left\langle \eta_{0}\left(  x\right)  \eta_{0}\left(  0\right)  \right\rangle
=\cos^{2}\varphi\left\langle \eta^{\prime}\left(  x\right)  \eta^{\prime
}\left(  0\right)  \right\rangle +\sin^{2}\varphi\left\langle \eta\left(
x\right)  \eta\left(  0\right)  \right\rangle ,
\end{equation}
which allows us to rewrite the screening contributions to the topological
charge correlator (\ref{scr1}) in the form%
\begin{equation}
\left\langle Q\left(  x\right)  Q\left(  0\right)  \right\rangle
_{scr}=-\left(  2\bar{n}\gamma_{\eta_{0}}\right)  ^{2}\left[  \cos^{2}\varphi
D\left(  m_{\eta^{\prime}},x\right)  +\sin^{2}\varphi D\left(  m_{\eta
},x\right)  \right]  ,
\end{equation}
where we have used the Euclidean tree-level propagator
\begin{equation}
\left\langle \eta\left(  x\right)  \eta\left(  0\right)  \right\rangle
=D\left(  m_{\eta},x\right)  =\frac{m_{\eta}}{4\pi^{2}x}K_{1}\left(  m_{\eta
}x\right)
\end{equation}
for $\eta$ and the analogous one for $\eta^{\prime}$. Diagonalization of the
physical Goldstone boson mass matrix \cite{dow92,shu95} relates the screening
mass (\ref{mscr}) to the physical masses of the pseudoscalar mesons as%
\begin{equation}
m_{scr}^{2}=2\bar{n}\gamma_{\eta_{0}}^{2}=m_{\eta^{\prime}}^{2}+m_{\eta}%
^{2}-2m_{K}^{2}. \label{topmass}%
\end{equation}
(Note that Eq. (\ref{topmass}) reduces to $m_{scr}^{2}=m_{\eta_{0}}%
^{2}=m_{\eta^{\prime}}^{2}$ in the chiral limit.) After continuation to
Minkowski space-time, our final expression for the screening contribution to
the IOPE of the pseudoscalar glueball correlator becomes%
\begin{equation}
\Pi_{P}^{\left(  scr\right)  }\left(  x\right)  =F_{\eta^{\prime}}^{2}D\left(
m_{\eta^{\prime}},x\right)  +F_{\eta}^{2}D\left(  m_{\eta},x\right)
\label{scrcorr2}%
\end{equation}
where ($F_{\eta^{\prime}},F_{\eta})=16\pi\bar{n}\gamma_{\eta_{0}}\left(
\cos\varphi,\sin\varphi\right)  $

All mass and coupling parameters in Eq. (\ref{scrcorr2}) will be fixed at
their experimental values. For the $\eta$ and $\eta^{\prime}$ masses and the
mixing angle we use $m_{\eta}=547.30\pm0.12$ MeV, $m_{\eta^{\prime}}%
=957.78\pm0.14$ MeV \cite{pdg02} and $\varphi=22.0%
%TCIMACRO{\U{b0}}%
%BeginExpansion
{{}^\circ}%
%EndExpansion
\pm1.2%
%TCIMACRO{\U{b0}}%
%BeginExpansion
{{}^\circ}%
%EndExpansion
$ \cite{fel00}. The axial anomaly renders even the $Q$-$\eta_{0}$ coupling and
the overall coupling $2\bar{n}\gamma_{\eta_{0}}$ experimentally accessible,
with the result $\left(  2\bar{n}\gamma_{\eta_{0}}\right)  ^{2}%
=9.\,\allowbreak732\times10^{-4}$ GeV$^{6}$ \cite{fel00}. Alternatively, this
coupling could be estimated from the standard value $\bar{n}=0.5$
fm$^{-4}=\allowbreak7.\,\allowbreak53\times10^{-4}$ GeV$^{4}$ of the instanton
density and the experimental pseudoscalar meson masses according to
\begin{equation}
\left(  2\bar{n}\gamma_{\eta_{0}}\right)  ^{2}=2\bar{n}\left(  m_{\eta
^{\prime}}^{2}+m_{\eta}^{2}-2m_{K}^{2}\right)  ,
\end{equation}
which yields (with $m_{K^{0}}=497.67\pm0.31$ MeV \cite{pdg02}) the value
$\left(  2\bar{n}\gamma_{\eta_{0}}\right)  ^{2}=1.\,\allowbreak086\times
10^{-3}$ GeV$^{6}$. It is reassuring that both estimates are perfectly
consistent. We will use the rounded value $\left(  2\bar{n}\gamma_{\eta_{0}%
}\right)  ^{2}=10^{-3}$ GeV$^{6}$ in our quantitative analysis below, which
implies $F_{\eta^{\prime}}^{2}=0.543$ GeV$^{6}\,$ and $F_{\eta}^{2}%
=\allowbreak0.0886$ GeV$^{6}$.

From the Fourier transform of the screening contribution,
\begin{equation}
\Pi_{P}^{\left(  scr\right)  }\left(  Q\right)  =\frac{F_{\eta^{\prime}}^{2}%
}{Q^{2}+m_{\eta^{\prime}}^{2}}+\frac{F_{\eta}^{2}}{Q^{2}+m_{\eta}^{2}},
\end{equation}
we finally obtain the Borel moments
\begin{equation}
\mathcal{R}_{P,k}^{\left(  scr\right)  }\left(  \tau\right)  =-\delta
_{k,-1}\left(  \frac{F_{\eta^{\prime}}^{2}}{m_{\eta^{\prime}}^{2}%
}+\frac{F_{\eta}^{2}}{m_{\eta}^{2}}\right)  +F_{\eta^{\prime}}^{2}%
m_{\eta^{\prime}}^{2k}e^{-m_{\eta^{\prime}}^{2}\tau}+F_{\eta}^{2}m_{\eta}%
^{2k}e^{-m_{\eta}^{2}\tau}. \label{rkscr}%
\end{equation}

The $\tau$-independent term in Eq. (\ref{rkscr}) is the screening-induced
subtraction constant $-\Pi_{P}^{\left(  scr\right)  }\left(  0\right)  $ which
will play a pivotal rule in the $\mathcal{R}_{P,-1}$ sum rule analysis below.
Already at this stage, however, its necessity can be seen from yet another
inconsistency which one would encounter by restricting the nonperturbative
Wilson coefficients exclusively to direct instanton contributions. Indeed, the
latter generate a large subtraction constant (cf. Eq. (\ref{q0lim}))
\begin{equation}
\Pi_{P}^{\left(  I+\bar{I}\right)  }\left(  0\right)  =-2^{7}\pi^{2}\bar{n}
\label{psubtr}%
\end{equation}
(where we have adopted the spike distribution) which cannot be matched on the
phenomenological side of the sum rule since the low-energy theorem
(\ref{pLET}) dictates the zero-momentum limit of the physical correlator to be
of the order of the light current quark masses, i.e. about an order of
magnitude smaller. The additional screening contribution from Eq.
(\ref{rkscr}) cancels most of the direct instanton contributions, however, and
thereby restores consistency. The underlying mechanism becomes most
transparent in the chiral limit where the low-energy theorem requires the
subtraction term to vanish and the cancellation to become complete (cf. Eq.
(\ref{pLET})). Indeed, the zero-quark-mass limit of the screening
contribution,
\begin{equation}
\Pi_{P}^{\left(  scr\right)  }\left(  0\right)  =\frac{F_{\eta^{\prime}}^{2}%
}{m_{\eta^{\prime}}^{2}}=\frac{\left(  16\pi\bar{n}\gamma_{\eta_{0}}\right)
^{2}}{2\bar{n}\gamma_{\eta_{0}}^{2}}=2^{7}\pi^{2}\bar{n} \label{psubtrscr}%
\end{equation}
(obtained from the definitions of $F_{\eta}$ and $F_{\eta^{\prime}}$ and the
mass relation (\ref{topmass}) with $m_{\eta}=m_{K}=\varphi=0$), exactly
cancels the direct instanton contribution (\ref{psubtr}) and thereby restores
consistency with the low-energy theorem. The topological charge is totally
screened in the chiral limit since the massless Goldstone boson exchange
generates infinite-range interactions. The presence of the screening
contributions thus explains how the direct-instanton induced subtraction terms
can be of equal size in both spin-0 channels whereas the low-energy theorems
(\ref{sLET}) and (\ref{pLET}) require the size of their phenomenological
counterparts to differ by an order of magnitude. This provides compelling
evidence for the vital role of topological charge screening in the IOPE of the
$0^{++}$ glueball correlator.

Furthermore, the cancellation between the direct-instanton- and
screening-induced subtraction terms suggests a strategy for implementing the
IOPE scale $\mu$ into the screening contributions. As shown above, restricting
the size of the direct instantons to $\rho<\mu^{-1}$ implies the replacement
of $\bar{n}$ in Eq. (\ref{psubtr}) by $\bar{n}_{dir}=\zeta\bar{n}$ with
$\zeta<1$ (cf. Eq. (\ref{ndir})). Compliance with the low-energy theorem then
requires the same replacement in the screening contributions (\ref{psubtrscr})
and therefore in $\Pi_{P}^{\left(  scr\right)  }\left(  Q^{2}\right)  $. This
is physically reasonable since only instantons with $\rho<\mu^{-1}\sim
O\left(  \lambda_{scr}\right)  $ can take part in the short-range screening
mechanism. A conceptually similar restriction has already been implied by
using the pointlike (i.e. $\rho=0$) approximation in deriving Eq.
(\ref{scrcorr2}).

We conclude this section with a first look at the quantitative impact of the
screening contributions on the $0^{-+}$ Borel moments. To this end, we
contrast Fig. 2 (a) of the $k=0$ moment, based on the perturbative and
instanton-induced Wilson coefficients as given in Eq. (\ref{r0inpert}), with
the same plot but including the screening contributions, in Fig. 2 (b). The
previous deficiencies, i.e. both the negative sign and the positive slope of
the Borel moment in the $s_{0}\gtrsim7-8$ GeV$^{2}$ region, are clearly
resolved. In particular, $\mathcal{R}_{P,0}$ now decays monotonically\ and
approximately exponentially with $\tau$ and thus contains a clear signal for a
pseudoscalar glueball resonance (and the $\eta^{\prime}$, see below) which
will be analyzed quantitatively in Sec. \ref{qsra}. (We recall that the
negative regions of the Borel moment at smaller $s_{0}$ and $\tau$ are an
artefact of the spike distribution and will disappear when realistic instanton
size distributions are implemented, cf. Fig. 4 below.)

In summary, we have found strong evidence for the direct instanton
contributions alone to be a seriously incomplete description of the hard,
nonperturbative physics in the $0^{-+}$ glueball correlator. Sum rules based
on this selective choice of contributions \cite{zha03}, i.e. without the
screening contributions, are in several ways inconsistent. Reliable sum rules
can be obtained only if the complementary physics generated by topological
charge screening is properly included.

\section{Quantitative analysis of Borel moments and sum rules}

\label{qsra}

We now assemble the various IOPE contributions (\ref{rm1pc}) - (\ref{r2pc}),
(\ref{rkinst}) and (\ref{rkscr}) obtained above into our theoretical
prediction for the continuum-subtracted Borel moments (\ref{csubtrbmoms}),%
\begin{equation}
\mathcal{R}_{G,k}\left(  \tau;s_{0}\right)  =\mathcal{R}_{G,k}^{\left(
pc\right)  }\left(  \tau,s_{0}\right)  +\mathcal{R}_{G,k}^{\left(
I+\bar{I}\right)  }\left(  \tau;s_{0}\right)  +\delta_{G,P}\mathcal{R}%
_{P,k}^{\left(  scr\right)  }\left(  \tau\right)  , \label{fullbmoms}%
\end{equation}
which form the left-hand side of the Borel sum rules (\ref{gbsrs}). These
moments summarize the microscopic (i.e. quark-gluon level) information
contained in the IOPE and the local duality continuum. Some additional input
of a more implicit nature is needed, however, before quantitative glueball
properties can be extracted from them by means of a sum rule analysis: a
specific parametrization of the resonance sector (i.e. the number of isolated
resonances and their shapes and widths), the determination of the duality
threshold, the establishment of the ``fiducial'' $\tau$-domain in which all
underlying approximations are expected to be reliable, and finally the choice
of sum-rule optimization criteria.

Before tackling these issues in Sec. \ref{srsetup}, we will make an effort to
obtain as much qualitative insight as possible from the IOPE moments alone.
The main focus will be on pertinent features of the $\tau$- and $s_{0}%
$-dependence as well as the scales which govern it, and an analysis of the
various subtraction constants and their impact. The gained insights will help
both to select the optimal strategy for the sum-rule analysis and to better
understand the physics which underlies its predictions.

\subsection{Input parameters and scales}

\label{inputscales}

Before embarking on the quantitative analysis, we have to fix the values of
the various constants and scales which appear in the IOPE and its moments
(\ref{fullbmoms}). All of them will be standard (or in their standard range).
The dominant soft scale is set by the gluon condensate,%

\begin{equation}
\langle\alpha_{s}G^{2}\rangle\equiv\langle\alpha_{s}G_{\mu\nu}^{a}G^{a\mu\nu
}\rangle\simeq0.055\,\mathrm{GeV}^{4}%
\end{equation}
(this value lies in the middle of the phenomenologically acceptable range
$\langle\alpha_{s}G^{2}\rangle\sim0.035-0.075$ $\mathrm{GeV}^{4}$
\cite{shi79,nar98}), which determines the higher-dimensional gluon condensates
as%
\begin{align}
\langle gG^{3}\rangle &  \equiv\langle gf_{abc}G_{\mu\nu}^{a}G_{\rho}^{b\nu
}G^{c\rho\mu}\rangle\simeq-1.5\,\langle\alpha_{s}G^{2}\rangle^{3/2}\,,\\
\left\langle \alpha_{s}^{2}G^{4}\right\rangle _{S}  &  \simeq\frac{9}%
{16}\langle\alpha_{s}G^{2}\rangle^{2},\text{ \ \ \ \ \ \ }\left\langle
\alpha_{s}^{2}G^{4}\right\rangle _{P}\simeq\frac{15}{8}\left\langle \alpha
_{s}G^{2}\right\rangle ^{2}.
\end{align}
All condensate values refer to the renormalization scale $\mu\simeq
0.5\,\mathrm{GeV}$. The lattice estimate for the three-gluon condensate
\cite{g3lat} is sometimes replaced by the single-instanton estimate $\langle
gG^{3}\rangle\sim0.27$ $\mathrm{GeV}^{2}\langle\alpha_{s}G^{2}\rangle$.
(Adopting different values of the lowest-dimensional condensates (inside their
standard range) would affect mainly the predictions of the pseudoscalar
$k=-1,0$ sum rules where the relative impact of the power corrections is
largest.) Furthermore, the phenomenological subtraction constant in the
pseudoscalar channel, determined by the low-energy theorem (\ref{pLET}),
contains the quark condensate
\begin{equation}
\left\langle \bar{q}q\right\rangle \simeq-\left(  0.24\text{ }\mathrm{GeV}%
\right)  ^{3}.
\end{equation}
For the remaining IOPE parameters we adopt the values%
\begin{align}
m_{u}  &  \simeq m_{d}\simeq0.005\,\mathrm{GeV}\\
\Lambda_{QCD}  &  =0.2\,\mathrm{GeV}\\
\mu &  =0.5\,\mathrm{GeV}%
\end{align}
For the leading moments of the instanton size distribution we use the
canonical values \cite{sch98}%
\begin{align}
\bar{n}  &  \simeq\frac{1}{2}\text{ fm}^{-4}=\allowbreak7.\,\allowbreak
53\times10^{-4}\text{ GeV}^{4},\\
\bar{\rho}  &  \simeq\frac{1}{3}\text{ fm}=1.69\text{ GeV}^{-1}%
\end{align}
which also completely determine the finite-width size distributions discussed
in Section \ref{isdistrib}. Our quantitative predictions will be based on the
probably most realistic of these parametrizations, the Gaussian-tail
distribution (\ref{ngn3}) with the soft cutoff (\ref{thetaf}) and $\beta=3$
GeV. The corresponding direct-instanton fraction $\zeta$, defined in Eq.
(\ref{ndir}), is
\begin{equation}
\zeta_{g}\equiv\frac{\bar{n}_{dir,g}}{\bar{n}}=\frac{1}{\bar{n}}\int
_{0}^{\infty}d\rho n_{g}\left(  \rho\right)  \theta_{\beta}\left(  \mu
^{-1}-\rho\right)  \simeq0.66, \label{zetag}%
\end{equation}
which implies that about two thirds of all instantons are direct, i.e. small
enough to affect the Wilson coefficients. (The $\zeta$-values from the
exponential-tail distribution (\ref{nexp}) and/or the alternative cutoff
function (\ref{thetaat}) differ by about 1\%.) Note that both the screening-
and direct-instanton-induced subtraction constants are proportional to the
instanton density $\bar{n}$ and therefore\ equally attenuated by
renormalization (recall that $F_{\eta,\eta^{\prime}},m_{\eta,\eta^{\prime}%
}^{2}\propto\bar{n}$),
\begin{align}
\Pi_{P}^{\left(  scr\right)  }\left(  0\right)   &  \rightarrow\zeta\Pi
_{P}^{\left(  scr\right)  }\left(  0\right)  ,\\
\Pi_{S/P}^{\left(  I+\bar{I}\right)  }\left(  0\right)   &  \rightarrow
\zeta\Pi_{S/P}^{\left(  I+\bar{I}\right)  }\left(  0\right)  .
\end{align}
Finally, we recall the masses and couplings
\begin{align}
F_{\eta^{\prime}}^{2}  &  =0.543\text{ GeV}^{6},\text{ \ \ \ \ \ }F_{\eta}%
^{2}=\allowbreak0.0886\text{ GeV}^{6},\\
m_{\eta}  &  =0.55\text{ GeV},\text{ \ \ \ \ \ \ \ \ }m_{\eta^{\prime}%
}=0.96\text{ GeV}%
\end{align}
which determine the topological charge screening contributions (\ref{rkscr}).

\subsection{Qualitative behavior of the Borel moments}

\label{qualbehav}

In Figs. 3 and 4 we plot all four IOPE Borel moments (\ref{fullbmoms}) in both
spin-0 glueball channels. The direct instanton contributions were calculated
on the basis of the Gaussian-tail size distribution (\ref{ngn3}) and
renormalized according to Eq. (\ref{nren}) with $\beta=3$. These Borel moments
will be the theoretical input for our sum rule analysis in Sec.
\ref{srresults}. For instructive purposes, i.e. to exhibit several qualitative
features more prominently, the plots cover a region of the $\tau-s_{0}$ plane
which extends beyond the limits in which the sum-rule analysis is reliable.

An obvious feature of all Borel moments is that they are monotonically
increasing with $s_{0}$ and monotonically decreasing with $\tau$ in their
physically sensible domains. This behavior stabilizes multi-parameter sum rule
fits since the minimization routine is unlikely to get trapped in a local
minimum of the deviation measure (cf. Sec. \ref{srsetup}). Furthermore, the
monotonic increase of the $\tau$-slope with $s_{0}$, especially at small
$\tau$, ensures that the largest resonance mass squared remains below the
duality continuum threshold, as it should be.

Another general property of all Borel moments is their increasing $s_{0}%
$-independence for $\tau\gg s_{0}^{-1}$. This is a consequence of the Laplace
suppression factor $\exp\left(  -\tau s\right)  $ in the dispersion integrals
(\ref{csubtrbmoms}) (which therefore have practically no support at $s\sim
s_{0}$) and the damping of the instanton contributions at large $s$ by
realistic instanton size distributions (cf. Sec. \ref{analytres}).

All Borel moments show an approximately exponential decay with increasing
$\tau$ (for $s_{0}\gtrsim6$ GeV$^{2}$ in the lowest $0^{-+}$ moments, which
includes the physical region). Thus, all of them provide resonance signals in
the sum rules (\ref{gbsrs}). After normalizing the moments to a common scale,
the $\tau$-slopes of the $0^{-+}$ Borel moments are consistently larger than
those of their $0^{++}$ counterparts in the region $\tau\lesssim0.8$
GeV$^{-2}$ where the heaviest isolated resonance dominates. Thus the glueball
mass scale is considerably larger in the $0^{-+}$ than in the $0^{++}$
channel, in agreement with lattice results \cite{lee00,har02}. The same
conclusion can be drawn independently from the ratios $\mathcal{R}%
_{G,k+1}/\mathcal{R}_{G,k}$ of adjacent moments in each channel. For $k\geq0$
and if one pole dominates, they are about equal to the square of the resonance
mass and indeed consistently larger in the $0^{-+}$ channel.

The pseudoscalar moments (especially those with $k\geq0$) flatten out rather
suddenly in the $\tau$-direction for $\tau\gtrsim1$ GeV$^{-2}$. The visibly
slower decay in the large-$\tau$ region is a qualitative indication for the
appearance of a second resonance with a clear mass scale separation. The
quantitative sum rule analysis below will confirm that the screening
contributions have produced an additional signal for the $\eta^{\prime}$, as
expected from ``unquenching'' the instanton contributions.

Several qualitative features of the Borel moments can be traced to the impact
of finite-width distributions and renormalization on the direct instanton
contributions. To exhibit them, we compare the $k=0$ moments of Figs. 3 and 4
with those obtained from the traditional spike distribution, plotted in Fig.
5. Clearly, the neglect of interference between contributions from instantons
of similar size in the spike distribution distorts the moments at small $\tau
$: positivity and monotonic rise with $s_{0}$ disappear. Moreover, the
negative $\tau$-slope of $\mathcal{R}_{S,0}$ at small $\tau$ turns positive at
larger $s_{0}$ and creates a ``mountain ridge''. In addition, $\mathcal{R}%
_{S,0}$ itself turns negative in the latter region while $\mathcal{R}_{P,0}$
becomes negative mostly in the small-$s_{0}$ (and small-$\tau$) region. In
both channels, the maximal $\tau$-slope and consequently an upper bound on the
glueball mass predictions is reached at intermediate $s_{0}$.

The positivity violations generated by the spike distribution can be traced to
the large-$s$ oscillations in the instanton-induced imaginary part. As
discussed in Sec. \ref{analytres}, these oscillations affect mostly the
small-$\tau$ behavior of the moments and are a conceptually worrisome
artefact. Their practical impact is moderate, however, since the positivity
violations occur mostly outside of the fiducial $\tau$- and $s_{0}$-domains in
which the sum rule analysis takes place. This is in marked contrast to the
impact of the positivity violations which we have discussed in Sec.
\ref{topscr}, i.e. those which arise in the pseudoscalar Borel moments when
the screening contributions are ignored. Indeed, a glance at Fig. 2 (a) shows
that these occur mostly at intermediate and large $\tau$, i.e. exactly in the
region where the sum rules are matched. They would make a meaningful sum rule
analysis impossible and once more underline the necessity of the topological
charge screening contributions.

The renormalization of the direct instanton contributions generally increases
the size of the moments at small $\tau$ (due to the enhanced imaginary part at
large $s$, cf. Fig. 1 (b)) and reduces it at the upper end of the fiducial
$\tau$-domain (due to the effectively reduced instanton density $\bar
{n}\rightarrow\zeta\bar{n}$). Both effects tend to increase the resonance mass
predictions of the IOPE sum rules (cf. Eq. (\ref{gbsrs})). Another effect of
the improved treatment of the nonperturbative Wilson coefficients is a
substantial reduction in the overall size of the direct instanton
contributions. In addition, the perturbative contributions are enhanced by the
3-loop corrections (and additionally by the now favored, larger values for
$\Lambda_{QCD}$, cf. Sec. \ref{pwcs}). Hence the previous dominance of the
instanton-induced coefficients, found in Ref. \cite{for01} on the basis of the
two-loop radiative corrections and the spike distribution, is considerably
weakened. As a side effect, the derivation of the scaling relations between
glueball and instanton properties \cite{for01} is obscured since the latter
relied on the dominance of the instanton contributions at intermediate and
large $\tau$. Nevertheless, these relations remain suggestive, especially
because they are consistent with large-$N_{c}$ counting.

A semi-quantitative upper bound on the sum-rule predictions for the glueball
masses can be obtained from the moment ratios considered above. Its derivation
starts from the expressions
\begin{equation}
m_{G}^{\left(  k\right)  }\left(  \tau;s_{0}\right)  \equiv\sqrt
{\frac{\mathcal{R}_{G,k}\left(  \tau;s_{0}\right)  }{\mathcal{R}%
_{G,k-1}\left(  \tau;s_{0}\right)  +\delta_{k,0}\Pi_{G}^{\left(  ph\right)
}(0)}} \label{mtau}%
\end{equation}
($k=0-2$) which are obtained by setting $f_{G2}=0$ in the sum rules
(\ref{gbsrs}) and solving for $m_{G1}$. (An alternative approach would be to
take logarithmic $\tau$-derivatives of the Borel moments.) Upper bounds for
the glueball masses are then found by searching for the position in the
$\left(  \tau,s_{0}\right)  $ plane where $m_{G}^{\left(  k\right)  }$ is
least sensitive to variations in both $\tau$ and $s_{0}$. These are either
extrema or inflection points of Eq. (\ref{mtau}). Estimates of this type
\cite{nar98} are often used instead of a full sum-rule matching analysis
although they can accommodate only one resonance pole and are of limited
reliability. To give a specific example of this approach, we will use
$m_{S}^{\left(  2\right)  }$ to establish a bound on the scalar glueball mass.
From our Borel moments we obtain $m_{S}^{\left(  2\right)  }$ as plotted in
Fig. 6. One reads off an extremum (maximum) in $\tau$ at $\tau^{\ast}\sim0.2$
GeV$^{-2}$ for $s_{0}\lesssim6.5$ GeV$^{2}$ and an inflection point in $s_{0}$
at $s_{0}^{\ast}\sim4.5$ GeV. (In the analysis based on purely perturbative
Wilson coefficients, one finds instead an inflection point of $m_{S}^{\left(
2\right)  }$ in $\tau$ and an extremum (minimum) in $s_{0}$ \cite{nar98}.)
Together, they yield the upper bound
\begin{equation}
m_{S}\leq m_{S}^{\left(  2\right)  }\left(  \tau^{\ast};s_{0}^{\ast}\right)
\simeq1.7\text{ GeV}%
\end{equation}
for the scalar glueball mass. The analogous estimates from $m_{S}^{\left(
0,1\right)  }$ result in somewhat smaller bounds. Nevertheless, these bounds
should not be considered as a substitute for the sum rule results. Indeed, our
quantitative sum-rule analysis in Sec. \ref{srresults} will show that they
overestimate the $0^{++}$ glueball mass prediction by about 35\%.

Figure 6 furthermore reveals that $m_{S}^{\left(  2\right)  }$ has a stronger
$\tau$-dependence than the analogous expression for the spike size
distribution (cf. Ref. \cite{for01}, Fig. 3). This might indicate that a
one-resonance sum-rule analysis is somewhat less favored if realistic
instanton size distributions and renormalization of the instanton-induced
coefficients are taken into account.

Finally, the Figs. 3 and 4 show that the $k=-1$ Borel moments have a $\tau
$-independent offset which becomes visible at large $\tau$. It is rather large
and negative in $\mathcal{R}_{S,-1}$ while smaller and positive in
$\mathcal{R}_{P,-1}$. These offsets are due to the subtraction terms which the
nonperturbative IOPE coefficients generate. (For this reason, $\mathcal{R}%
_{G,-1}<0$ does not imply positivity violations.) Their match to the
subtraction constants on the phenomenological side of the sum rules is an
important consistency criterion which we are going to discuss in the next section.

\subsection{Subtraction constants}

\label{subconst}

By design, the $k=-1$ Borel moment included first-principle information
provided by the low-energy theorem (\ref{sLET}) into the $0^{++}$ glueball
sum-rule analysis \cite{nov280}. As pointed out in Sec. \ref{lets}, an equally
useful low-energy theorem (\ref{pLET}) exists in the pseudoscalar channel and
suggests to analyze the analogous $0^{-+}$ sum rule, based on the moment
$\mathcal{R}_{P,-1}$, as well. In order to prepare for this analysis, the
present section investigates the conceptual and quantitative impact of the
involved subtraction constants.

Since the perturbative UV contributions to the subtraction constants are
removed by definition in the low-energy theorems (cf. Sec. \ref{lets}) and by
renormalization in the perturbative IOPE coefficients, it remains to clarify
the role and treatment of the nonperturbative contributions. The
direct-instanton induced subtraction terms
\begin{equation}
\Pi_{S/P}^{\left(  I+\bar{I}\right)  }\left(  0\right)  =\pm2^{7}\pi^{2}%
\bar{n}_{dir}\simeq\pm\zeta\times0.95\text{ GeV}^{4} \label{subcdiri}%
\end{equation}
are part of the $k=-1$ Borel moments (\ref{inb-1}) and (\ref{rkinst}). Their
size is significant but smaller than the size of the $\tau$-dependent
contributions, in particular those due to direct instantons (at small and
intermediate $\tau$). Moreover, for $\zeta=1$ (i.e. for the spike
distribution) the instanton-induced subtraction constant in the scalar channel
is more the 50\% larger than (and of the same sign as) the phenomenological
LET\ value (cf. Eq. (\ref{sLET}))
\begin{equation}
\Pi_{S}^{\left(  ph\right)  }\left(  0\right)  =\frac{32\pi}{b_{0}%
}\left\langle \alpha_{s}G^{2}\right\rangle \simeq0.61\,\mathrm{GeV}^{4}.
\label{sletquanti}%
\end{equation}
It would remain larger even when the largest available value for the gluon
condensate, $\left\langle \alpha_{s}G^{2}\right\rangle \sim0.07$ GeV$^{4}$, is
used. In the pseudoscalar channel the discrepancy is yet more pronounced: the
value (\ref{subcdiri}) is more than an order of magnitude larger than the
LET\ value (\ref{pLET})
\begin{equation}
\Pi_{P}^{\left(  ph\right)  }\left(  0\right)  =\left(  8\pi\right)
^{2}\frac{m_{u}m_{d}}{m_{u}+m_{d}}\left\langle \bar{q}q\right\rangle
\simeq-0.022\,\,\mathrm{GeV}^{4} \label{pletquanti}%
\end{equation}
(and again of the same sign).

The impact on the $k=-1$ sum rules results from the fact that $\mathcal{R}%
_{G,-1}$ has to be fitted to the decaying resonance exponentials and the
``phenomenological'' subtraction constants (\ref{sletquanti}) and
(\ref{pletquanti}). As a general rule, the smaller the difference between the
IOPE subtraction constants and the LET values, which makes up the remaining
offset, the better will be the fit quality to the resonances (which is the
only intrinsic reliability measure for QCD sum rules). The impact of the
remaining imbalance can be rather subtle since it is largest towards the upper
boundary of the fiducial $\tau$-domain where the match to the exponential
resonance contributions becomes delicate.

A glance at the above scales confirms that such a match is ruled out in the
$0^{-+}$ channel, with its extreme discrepancy between LET and
direct-instanton induced subtraction constants, if the nonperturbative IOPE
coefficients arise exclusively from direct instantons. As we have shown in
Sec. \ref{somix}, however, this inconsistency is overcome by the crucial
topological charge screening correction%
\begin{equation}
\Pi_{P}^{\left(  scr\right)  }\left(  0\right)  =\zeta\left(  \frac{F_{\eta
^{\prime}}^{2}}{m_{\eta^{\prime}}^{2}}+\frac{F_{\eta}^{2}}{m_{\eta}^{2}%
}\right)  \simeq\zeta\times0.89\text{ GeV}^{4},
\end{equation}
which cancels most of the direct-instanton contribution (\ref{subcdiri}) and
brings the total IOPE subtraction constant in line with the small LET value
(\ref{pletquanti}). Previous analyses of $0^{-+}$ sum rules have discarded the
$k=-1$ moment and therefore missed valuable first-principle information from
the low-energy theorem as well as a useful consistency check.

One might wonder what happens to the rather delicate balance between the
$0^{-+}$ subtraction constants in the $N_{f}=0$ limit, i.e. in pure gauge
theory or in the quenched approximation. In this case the above cancellation
does not work since topological charge screening disappears. As is well-known,
however, the LET (\ref{pLET}) and thus the phenomenological value for the
zero-momentum correlator is strongly affected by the absence of light quarks,
too. Indeed, it becomes
\begin{equation}
\Pi_{P}^{\left(  ph,qn\right)  }\left(  0\right)  =-\left(  8\pi\right)
^{2}\chi_{t}^{\left(  qn\right)  }\simeq-0.66\text{ GeV}^{4},
\end{equation}
where we have used the standard (quenched) lattice value $\chi_{t}^{\left(
qn\right)  }\simeq(0.18$ GeV$)^{4}$ for the topological susceptibility
\cite{tep00}, which is in agreement with the Witten-Veneziano formula
\cite{wit79}. Thus we find again perfect consistency: the LET subtraction
constant\ becomes much larger and cancels the direct-instanton contributions
by itself (assuming that the latter are not strongly affected by quenching).

A special situation arises if the difference between the IOPE and LET values
of the subtraction constant becomes so strongly negative that even the $\tau
$-dependent contributions cannot prevent
\begin{equation}
\mathcal{R}_{G,-1}\left(  \tau;s_{0}\right)  +\Pi_{G}^{\left(  ph\right)
}\left(  0\right)  \label{rsum}%
\end{equation}
from turning negative inside the fiducial $\tau$-domain \footnote{When
$\tau\rightarrow\infty$ this will eventually happen for any negative
difference (and does, of course, not imply that the spectral function violates
the positivity bound).}. In this case the sum-rule fit to the (always
positive) resonance exponentials in Eq. (\ref{gbsrs}) is substantially
worsened even if the negative offset in (\ref{rsum}) remains relatively small.
Exactly this situation is encountered in the scalar $k=-1$ sum rule since%
\begin{equation}
\Pi_{S}^{\left(  ph\right)  }\left(  0\right)  -\Pi_{S}^{\left(
I+\bar{I}\right)  }\left(  0\right)  \simeq\left(  0.61\,-0.95\,\text{\ }%
\zeta\right)  \text{ GeV}^{4}<0.
\end{equation}
For $\zeta=1$ (spike distribution) the imbalance between the $\tau
$-independent terms is of about the same size with and without the direct
instanton contribution. The main effect of the instanton contribution is to
turn the sign negative, which causes the mentioned decline in sum-rule
consistency. One option to deal with this problem is to assume that unreliably
calculated soft contributions dominate the instanton-induced subtraction
constant (in particular when using the spike distribution), as discussed in
Sec. \ref{IBorelMom}, and therefore to discard it completely. This strategy
was adopted in Ref. \cite{for01} where the constant (\ref{subcdiri}) was
removed from the Borel moments (\ref{inb-1}) and (\ref{rkinst}). It strongly
improves the sum-rule quality and maintains the crucial mutual consistency
with the predictions of the $k\geq0$ sum rules and with the LET (\ref{sLET})
(which is badly violated if nonperturbative contributions to the Wilson
coefficients are ignored) \cite{for01}. This shows that the main reason for
achieving a large glueball mass scale and LET consistency in the
$\mathcal{R}_{S,-1}$ sum rule is the large overall size and slope of the
$\tau$-dependent direct instanton contributions (and not the relatively small,
instanton-induced subtraction constant, as suggested in Ref. \cite{har201}).

Of course, discarding the instanton-induced subtraction constant completely is
a very crude way of ``renormalization``. Our implementation of the IOPE
renormalization scale on the basis of realistic size distributions accounts
more accurately for the soft physics to be removed. Comparison of the above
scales shows that this significantly improves the consistency with the LET:
the difference between instanton- and LET-induced subtraction terms is reduced
by $\zeta<1$, and for realistic values $\zeta\simeq2/3$ (cf. Eq.
(\ref{zetag})) the sum (\ref{rsum}) remains positive over the whole fiducial
$\tau$-domain \footnote{Although their numerical values are quite close, one
should keep in mind that the physics entering $\Pi_{S}^{\left(
I+\bar{I}\right)  }\left(  0\right)  $ and $\Pi_{S}^{\left(  ph\right)
}\left(  0\right)  $ is mostly complementary:\ as part of a Wilson
coefficient, $\Pi_{S}^{\left(  I+\bar{I}\right)  }\left(  0\right)  $ receives
dominantly (semi-) hard contributions while $\Pi_{S}^{\left(  ph\right)
}\left(  0\right)  $ is renormalized by subtracting the hard (perturbative)
fluctuations and therefore dominated by soft modes (note the appearance of the
gluon condensate in the LET (\ref{sLET})).}. Note that this renormalization
maintains consistency with the LET\ in the pseudoscalar channel as well, while
simply discarding the instanton-induced subtraction constant would leave the
compensating screening corrections out of balance (cf. Sec. \ref{somix}).

In the context of this section it might be useful to recall that the large
``phenomenological'' subtraction constant had caused serious problems in the
$0^{++}$glueball sum rule as long as the nonperturbative Wilson coefficients
were ignored: the ensuing, weaker decay of the sum (\ref{rsum}) generated a
much smaller $0^{++}$glueball mass ``prediction'' (well below 1 GeV) which was
inconsistent with the predictions well beyond 1 GeV from the $k\geq0$ sum
rules. It was pointed out in Ref. \cite{for01} that the missing, strongly
decaying $\tau$-dependence can only reside in the Wilson coefficients and that
it should be of nonperturbative origin, suggesting (together with other
indications) direct instantons as its main source. The inclusion of the direct
instanton contributions indeed overcame the mutual inconsistencies among the
$0^{++}$ sum rule predictions and restored the consistency with the LET.

In summary, the direct-instanton induced subtraction terms and their
renormalization play a rather complex role in the $k=-1$ glueball sum rules.
In both channels they are essential for achieving consistency with the
underlying low-energy theorems. In the pseudoscalar sum rules, this
additionally requires strong cancellations with the indispensable topological
charge screening contributions. These cancellations explain, in particular,
how the equal size of the instanton-induced subtraction constants in both
spin-0 glueball channels can be reconciled with the conspicuous difference
between the LET values.

\subsection{Sum rule analysis setup}

\label{srsetup}

In order to extract glueball properties from the IOPE Borel moments by means
of a QCD sum-rule analysis, the parametrization of the phenomenological
spectral functions and the matching criteria have to be specified. Of course,
the complexity of the phenomenological side (measured by the number of
parameters to be predicted) is limited by the information content and
resolution power of the truncated short-distance expansion. Therefore,
judicious decisions are required about the amount of detail to include, e.g.
about the number of isolated resonances, their individual widths and shapes,
specific assumptions on quarkonium admixtures or even explicit multi-hadron
continua (beyond the local-duality approximation). Guided by the asymptotic
nature of the spacelike IOPE (i.e. the factorial and not Borel-summable growth
of the higher Wilson coefficients \cite{shi95}) and its truncation, we
restrict ourselves in the following analysis to at most two resonance poles in
zero-width approximation \footnote{Finite resonance widths require exponential
resolution and are therefore in principle inaccessible to the standard OPE
\cite{shi98}. The situation is more complex when including nonperturbative
Wilson coefficients which themselves introduce exponential contributions. In
any case, the zero-width approximation in QCD sum rules does not require the
corresponding, physical resonances to be particularly narrow.}\ and the
local-duality continuum, as anticipated in Eq. (\ref{gbsrs}).

Any quantitative sum-rule analysis requires a numerical measure $\delta$ for
the deviation between both sides over the discretized fiducial $\tau$-domain
(whose boundaries $\tau_{\min},\tau_{\max}$ will be determined below). The
iterative minimization of $\delta$ up to the desired accuracy is then
performed numerically. We will adopt the Belyaev-Ioffe measure \cite{bel82}
\begin{equation}
\delta=\frac{1}{N}\sum_{i=0}^{N}\ln\left(  \frac{\max\left(  \mathcal{R}%
_{G,k}^{\left(  pole\right)  }\left(  \tau_{i}\right)  -\delta_{k,-1}\Pi
_{G}^{\left(  ph\right)  }\left(  0\right)  ,\mathcal{R}_{G,k}\left(  \tau
_{i}\right)  \right)  +\xi}{\min\left(  \mathcal{R}_{G,k}^{\left(
pole\right)  }\left(  \tau_{i}\right)  -\delta_{k,-1}\Pi_{G}^{\left(
ph\right)  }\left(  0\right)  ,\mathcal{R}_{G,k}\left(  \tau_{i}\right)
\right)  +\xi}\right)
\end{equation}
with $N=100$ grid points and $\tau_{i}=\tau_{\min}+i\left(  \tau_{\max}%
-\tau_{\min}\right)  /N$ \footnote{Due to the generally monotonic decrease of
the glueball Borel moments with $\tau$, scale-invariant measures favor an
improved matching in the large-$\tau$ region. In two-resonance fits this
favors the smaller-mass resonance.}. The constant $\xi$ is an offset to be
added if otherwise the argument of the logarithm would become negative. An
important requirement on reliable sum rule fits is that they should be stable,
i.e. that the resulting hadron properties should be - inside the other typical
errors of the analysis - independent of the starting values for the iterative
minimization of $\delta$. As mentioned above, the monotonic behavior of the
IOPE Borel moments generally improves stability by reducing the likelihood for
less than optimal local minima of $\delta$. We have tested several alternative
expressions for $\delta$ (with, e.g., different weights for the deviations in
the large- and small-$\tau$ regions) and found them to change predictions of
stable fits maximally at the one-percent level.

The ``fiducial'' $\tau$-domain, in which the sum-rule analysis takes place, is
designed to optimally exploit the physical information in the IOPE without
leaving the region of validity of the involved approximations. Hence, one
seeks the maximal $\tau$-interval in which the sum rules can be expected to be
both reliable and predictive. Towards small $\tau$, the duality continuum in
the Borel moments (\ref{phenspec}) increasingly dominates the glueball signal.
In order to ensure that the sum rules remain sensitive to the glueball
properties, we therefore fix $\tau_{\min}\left(  k,s_{0}\right)  $ by the
standard requirement that the continuum contributions to the Borel moments
must not exceed the resonance contributions, i.e.
\begin{equation}
\frac{\mathcal{R}_{G,k}^{\left(  cont\right)  }\left(  \tau_{\min}%
;s_{0}\right)  }{\mathcal{R}_{G,k}\left(  \tau_{\min};s_{0}\right)  }=0.5.
\end{equation}
Generally, the higher Borel moments require larger values of $\tau_{\min}$.
The above criterion therefore also helps to assure consistency among sum rule
predictions from different moments. Furthermore, it typically reduces the
fiducial domain of the higher-$k$ sum rules and thereby renders their fits
somewhat less stable. (The results for $s_{0}$, in particular, become less
reliable with increasing $k$ since the $s_{0}$-dependence of the Borel moments
is largest at small $\tau$.)

The standard sum-rule criterion for determining the upper limit of the
fiducial domain, $\tau_{\max}$, is to restrict the contribution of the
highest-dimensional operator to maximally 10\% of the total OPE contribution,
i.e.%
\begin{equation}
\frac{\tilde{C}_{8}^{\left(  G\right)  }\left(  \tau_{\max};\mu\right)
\left\langle \hat{O}_{8}\right\rangle _{\mu}}{\sum_{d=0,4,6,8}\tilde{C}%
_{d}^{\left(  G\right)  }\left(  \tau_{\max};\mu\right)  \left\langle
\hat{O}_{d}\right\rangle _{\mu}}=0.1. \label{tmax1}%
\end{equation}
The neglected contributions from $d>8$ operators should therefore remain small
up to the onset of the asymptotic IOPE region. The above criterion is more
stringent at smaller $k$ where the condensates contributions have a relatively
larger impact. (The size of the topological charge screening contributions
also decreases with increasing $k$.)

In the presence of nonperturbative IOPE coefficients, the criterion for
$\tau_{\max}$ requires some additional thought. Even for exclusively
perturbative Wilson coefficients, the condition (\ref{tmax1}) it is not too
restrictive in the spin-0 glueball channels since the power corrections are
unusually small. When the large direct instanton contributions to the unit
operator are added, it can become almost ineffective. (In the pseudoscalar
channel their impact is reduced by cancellations with the screening
contributions and the larger $s_{0}$-values.) Another constraint on
$\tau_{\max}$, arising from the requirement that multi--direct-instanton
corrections to the Wilson coefficient should remain negligible, will then
become more stringent. In order to formulate this criterion quantitatively, we
adopt the rough estimate
\begin{equation}
\tau\leq\tau_{\max}\simeq\frac{1}{2}\left(  \bar{R}-2\bar{\rho}\right)
^{2}\sim1.5\text{ GeV}^{-2}. \label{tmaxbound}%
\end{equation}
For the higher Borel moments this criterion is typically more restrictive than
(\ref{tmax1}) since the relative impact of the power corrections at large
$\tau$ decreases with increasing $k$. In view of its approximate nature,
however, it is reassuring that the sum rule results are rather insensitive to
variations in $\tau_{\max}$. (The sensitivity to $\tau_{\min}$ is larger, as expected.)

Under the phenomenological parameters to be determined by the sum rule
analysis, the duality threshold $s_{0}$ plays a special role. It is not
associated with a glueball property but rather corresponds roughly to the
squared mass gap between ground state and first radially excited state.
Indications for a delayed onset of duality \cite{vai02} suggest, however, that
this rule of thumb is invalid in the scalar glueball channel. (These
indications could be tested on the lattice when reliable unquenched glueball
spectra become available.)\ Hence, our only roboust expectation for $s_{0}$ is
that it should be larger than the squared mass of the highest-lying isolated resonance.

Below, we will determine $s_{0}$ together with the glueball parameters from
the sum rule fits, which turns out to be possible even in the presence of two
isolated resonances. Nevertheless, it might be instructive to briefly comment
on alternative strategies for obtaining the duality threshold. One alternative
would be to divide the fitting procedure into two steps, by first constraining
$\sqrt{s_{0}}$ to exceed the largest resonance mass by a constant amount
$\Delta s$,%

\begin{equation}
\sqrt{s_{0}}=m_{G}+\Delta s \label{delm}%
\end{equation}
and by subsequently minimizing $\delta$ as a function of $\Delta s$. Such a
constant splitting is probably not an unreasonable assumption since the $\tau
$-slopes of the IOPE Borel moments at small $\tau$ increases rather
monotonically with $\sqrt{s_{0}}$ (cf. Figs 3 and 4) in the sum-rule relevant
$\tau$-region.\ (The couplings $f_{S}$ are much less $s_{0}$-dependent.)
Although this two-step procedure can accelerate the minimization procedure (in
particular in the two-resonance case), we did not find it necessary even in
the pseudoscalar channel. Another strategy, adopted e.g. in Ref. \cite{asn92},
determines $s_{0}$ by analyzing a related finite-energy sum rule (FESR) which
roughly expresses a duality constraint. In view of the likely delayed onset of
local duality this procedure might be misleading, however, in the scalar
glueball channel. In Refs. \cite{nar98,har01}, furthermore, $s_{0}$ is
required to render specific combinations of Borel moments minimally $\tau
$-sensitive. In Ref. \cite{nar98}, finally, an upper limit\ on the FESR value
for $s_{0}$ is obtained by locating the extrema or inflection points of moment
ratios like (\ref{mtau}).

\subsection{Results and discussion}

\label{srresults}

After having done the groundwork, we now proceed to the quantitative sum-rule
analysis. This amounts to matching the continuum-subtracted IOPE Borel moments
(\ref{fullbmoms}) to either one or two isolated resonances and, if $k=-1$, to
the phenomenological subtraction constant (cf. Eq. (\ref{gbsrs})). The
decision about how many resonances to include will be made in the scalar and
pseudoscalar channels individually, based on a comparative quantitative
analysis. Of course, the more flexible two-resonance parametrization is almost
bound to reduce the nominal fit errors at least slightly. Since this comes at
the price of two more parameters to be determined, however, we will resort to
the two-resonance parametrization only if it leads to a clear improvement of
the fits. This is a necessary requirement for stable and physically meaningful
predictions of additional resonance parameters.

\subsubsection{Scalar glueball}

The analysis of the $0^{++}$ glueball Borel sum rules on the basis of the
spike distribution (with perturbative coefficients up to $O\left(  \alpha
_{s}\right)  $) showed that a one-pole fit could almost perfectly match the
IOPE moments \cite{for01}. This result left little room for improvement and
therefore no indication for the presence of a second low-lying resonance with
strong coupling to the scalar gluonic interpolator (\ref{sipf}).

An extensive numerical survey of all four improved $0^{++}$ Borel sum rules,
based on the IOPE (\ref{fullbmoms}), reveals that they can be well fitted by
only one $0^{++}$ glueball pole, too. Again, there is no conclusive evidence
for the presence of another low-lying, strongly coupled $0^{++}$ resonance. A
sufficiently exhaustive analysis of the scalar sum rules can therefore be
based on the traditional one-pole plus duality continuum parametrization. The
main part of the following discussion will deal with the results of this
``benchmark''\ analysis. Nevertheless, there seems to be some indication for
additional low-lying strength, as foreshadowed in the analysis of the moment
ratio in Sec. \ref{qualbehav}. We will come back to this issue at the end of
this section.%

%TCIMACRO{\TeXButton{B}{\begin{table}[htbp] \centering}}%
%BeginExpansion
\begin{table}[htbp] \centering
%EndExpansion%
\begin{tabular}
[c]{|c|c|c|c|c|c|c|}\hline
$k$ & $\tau_{\min}$ (GeV$^{-2}$) & $\tau_{\max}$ (GeV$^{-2}$) & $m_{S}$
(GeV) & $f_{S}$ (GeV) & $\sqrt{s_{0}}$ (GeV) & $\delta\times10^{-3}%
$\\\hline\hline
$-1$ & $0.3$ & $1.3$ & $1.28$ & $1.02$ & $2.\,\allowbreak22$ & $1.06$\\\hline
$0$ & $0.6$ & $1.3$ & $1.16$ & $1.11$ & $1.\,\allowbreak75$ & $5.70$\\\hline
$1$ & $0.8$ & $1.5$ & $1.22$ & $1.04$ & $1.\,\allowbreak63$ & $1.63$\\\hline
$2$ & $1.0$ & $1.5$ & $1.39$ & $1.01$ & $1.\,\allowbreak82$ & $2.13$\\\hline
\end{tabular}
\caption{The fiducial domain, fitting error and predictions of the scalar glueball sum rule based on the k-th Borel moment.\label{ts}%
}
%TCIMACRO{\TeXButton{E}{\end{table}}}%
%BeginExpansion
\end{table}%
%EndExpansion

The numerical results of the one-resonance analysis of all four Borel-moment
sum rules are collected in Tab. \ref{ts}. They can be summarized in the
overall prediction%
\begin{equation}
m_{S}=1.25\pm0.2\text{ GeV},\text{ \ \ \ \ \ \ }f_{S}=1.05\pm0.1\text{ GeV}
\label{spredicts}%
\end{equation}
for the scalar glueball mass and coupling. The error assignment includes the
estimated uncertainties in the input parameter values. The corresponding sum
rule fits, in their individual fiducial $\tau$-domains, are shown in Fig. 7.

Clearly, the overall fit quality is more than satisfactory. As anticipated,
there seems to be no need for a second isolated (and rather narrow) resonance.
The 3-parameter fits are inside typical errors independent of the starting
values and also quite independent of the upper border of the $\tau$-domain.
(Nevertheless, far-off starting values for the threshold $s_{0}$ can
significantly slow down the minimization procedure since the $s_{0}%
$-dependence is rather shallow.) The instanton continuum contributions
decisively improve the individual and mutual consistency of the sum rules
(including the one associated with $\mathcal{R}_{S,-1}$) and of their results.
As expected, we find little sensitivity to details of the instanton size
distribution as long as its overall scales are kept fixed. The sum rule
consistency noticeably worsens, however, when finite-width distributions are
used without subtracting the large-$\rho$ contributions.

The $\mathcal{R}_{S,-1}$ sum rule, singled out by the presence of subtraction
terms (cf. Sec. \ref{subconst}), has played a notorious role in previous
glueball sum rule analyses and deserves some more specific comments. Since the
phenomenological subtraction constant is of significant size, the
contributions of the perturbative Wilson coefficients alone were unable to
generate a glueball mass of more than a few hundred MeV. As explained before,
the direct instanton contributions with their large slope (cf. Fig. 7) resolve
this problem \cite{for01}. The instanton-induced subtraction term, however, is
overestimated by the spike approximation to the instanton size distribution:
it so strongly overcompensates the phenomenological subtraction constant that
the sum rule becomes inconsistent (the matching error increases by about 2
orders of magnitude) \footnote{The same should probably hold for the
corresponding Gaussian sum rule which, perhaps for this reason, was excluded
in the analysis of Ref. \cite{har01}.}. As anticipated in Sec. \ref{subconst},
this problem is resolved by realistic $\rho$-distributions and renormalization
of the instanton contributions. Indeed, Fig. 7 shows that the fit quality of
the improved $\mathcal{R}_{S,-1}$ sum rule matches that of its higher-moment
counterparts, and all sum rules yield mutually consistent predictions for the
glueball properties.

The impact of the subtraction constants on the sum rule analysis increases
towards larger $\tau$ since the dominant $\tau$-dependent contributions
decrease. The sum-rule fit in Fig. 7 shows that the residual discrepancy
between LET- and IOPE-induced subtraction terms can be comfortably compensated
by the small condensate contributions which are enhanced in the $k=-1$ sum
rule. Even larger discrepancies, encountered e.g. when using smaller values of
the gluon condensate, could be accommodated. The\ $k=1$ sum rule would then
predict a larger value of $s_{0}$ and a marginally larger glueball mass.

In addition to the sum-rule fits for the total IOPE Borel moments, Fig. 7 also
shows the contributions from perturbative and nonperturbative Wilson
coefficients separately. Comparison with the analogous figures in Ref.
\cite{for01} (based on the spike distribution) confirms the discussion of Sec.
\ref{qualbehav}:\ the perturbative contributions are somewhat enhanced by the
$O\left(  \alpha_{s}^{2}\right)  $ corrections while the direct instanton
contributions are significantly reduced in the improved IOPE. (The duality
continuum threshold decreases somewhat, too.) As a result, perturbative and
nonperturbative coefficients are now of comparable size. Moreover, the slopes
of the instanton contributions are reduced. Both effects manifests themselves
in a 20\% smaller prediction for the scalar glueball mass. However, the about
threefold increase of the glueball decay constant $f_{S}$ predicted in Ref.
\cite{for01} (relative to previous sum rule analyses) remains intact. This
implies that the reduced size of the Borel moments on the left-hand side of
the sum rules (\ref{gbsrs}) is compensated by the increased perturbative
contributions and, in particular, by the smaller glueball-mass factors on the
right-hand side.

Our large result for $f_{S}$ predicts strongly increased partial widths for
the radiative decay of the heavy quarkonia $J/\psi$ and $\Upsilon$ (and
others) into scalar glueballs and therefore has relevance for experimental
glueball searches \cite{cak94}. In particular, it allows for an improved
analysis of the existing $\Upsilon\rightarrow\gamma f_{0}$ decay data of the
CLEO collaboration \cite{cleo} and the forthcoming larger samples from
CLEO-III. It will also be interesting to compare our prediction to the first
calculation of $f_{S}$ on the (quenched) lattice which is in progress
\cite{che03}. Since $f_{S}$ is related to the glueball wave function (or
Bethe-Salpeter amplitude) at the origin, our large value predicts a strongly
concentrated wave function and thus an unusually small size of the scalar
glueball. Similar results have been found in the instanton liquid model
\cite{sch95} and on the lattice \cite{def92}.

Our prediction for the central value of the scalar glueball mass is about
10-20\% smaller than the results of our previous analysis \cite{for01} based
on the spike distribution, and lies 10-40\% below the quenched lattice results
$m_{S}^{\left(  q\right)  }=1.4-1.8$ GeV \cite{lee00}. (All raw lattice data
agree within statistical errors (about 40 MeV). The much larger range quoted
above reflects the ambiguities in setting the mass scale in the absence of
direct experimental input (cf. the talk of Bali under Ref. \cite{har02}).) A
similar reduction (25\%) of the $0^{++}$ glueball mass was found in unquenched
lattice simulations (on still rather small lattices with very limited
statistics and quark masses of about 70 MeV) \cite{har02}, although this might
be dominantly a lattice artifact. On general grounds, however, one would
expect light-quark effects and quarkonium admixtures to lower the quenched
masses at least somewhat. Due to its smaller size, furthermore, the scalar
glueball should be particularly susceptible to the momentum dependence of the
sea-quark induced vacuum polarization which modifies the color-dielectric
properties of the vacuum at short distances \cite{wei82}. Since this part of
the vacuum polarization is neglected in the quenched approximation, one would
expect the quenched mass predictions to be less reliable in the scalar
channel. This expectation is supported, e.g., by the recent analysis of the
radial-excitation and Regge trajectories of the known isoscalar mesons
\cite{ani03}, including those newly established by the Crystal Barrel
Collaboration in proton-antiproton annihilation. The standard $0^{++}$
glueball candidates above 1 GeV, i.e. the $f_{0}\left(  1300\right)  $,
$f_{0}\left(  1500\right)  $ and $f_{0}\left(  1750\right)  $ resonances, are
found to lie solidly on $q\bar{q}$ trajectories and to fit well into a flavor
nonet classification. Instead, the $K$-matrix analysis predicts a light and
broad scalar glueball state in the 1200-1600 MeV region \cite{ani03},
compatible with our result. An even lighter (and similarly broad) glueball
state, centered around 1 GeV or a bit larger, is expected in mixing schemes
which assume only one $0^{++}$ multiplet below 1.8 GeV \cite{och03}.

The $k$-dependence of our predictions for the scalar glueball properties in
Tab. \ref{ts} shows a certain systematics. While the results for the coupling
are practically $k$-independent (within sum-rule accuracy), the predictions
for the mass increase\ with $k$ for $k\geq0$. Since $f_{S}$ is associated with
the integrated strength of the glueball, it enters all sum rules (\ref{gbsrs})
in the same (i.e. $k$-independent) power. The $k$-dependence of $f_{S}$ in
Tab. \ref{ts} thus gives an idea of the typical uncertainties of the matching
analysis. The variations in the mass predictions, on the other hand, are
larger and systematically increase with $k$. This suggest that the glueball
strength is distributed over a rather broad $s$-region: since the higher
moments weight the spectral function more strongly at larger $s$, they will
then predict a larger pole mass (cf. Eq. (\ref{gbsrs})). (The $k=-1$ sum rule
result is not conclusive in this regard since it receives additional
contributions from the subtraction terms.) Our predicted mass range can
therefore be regarded as a rough lower bound on the width of the scalar
glueball, i.e. $\Gamma_{S}\gtrsim0.3$ GeV, similar to the width found in the
$K$-matrix analysis \cite{ani03}.

The gap $\Delta s$ between glueball mass and continuum threshold, as defined
in Eq. (\ref{delm}), is relatively constant among the $k\geq0$ sum-rule
results,
\begin{equation}
\Delta s=0.5\pm0.1\text{ GeV,}%
\end{equation}
while it is about twice as large in the $k=-1$ sum rule ($\Delta s=0.94$ GeV),
due to the subtraction term. The assumption of a common shift $\Delta s$ for
all sum rules in a simplified analysis (cf. Sec. (\ref{srsetup})) would
therefore fail in the lowest moment. The relatively early onset of the
continuum may indicate, incidentally, that the first excited scalar glueball
state lies around 2 GeV. Although the quenched lattice spectrum predicts this
excitation well beyond 2 GeV and beyond the lowest $0^{-+}$ glueball state,
its mass might again be lowered by quarkonium admixtures.

We had argued above that the quality of the one-pole sum rule fits leaves
little room for physically significant improvement and provides no clear
evidence for an additional low-lying resonance with strong coupling to the
gluonic interpolating field. Nevertheless, a residual $\tau$-dependence of the
Borel moment ratio $m_{S}^{\left(  2\right)  }$ was found in Sec.
\ref{qualbehav}, and the fit quality of the $\mathcal{R}_{S,0}$ sum rule is
somewhat lower than that of the other ones (cf. Tab. \ref{ts} and Fig. 7).
Since this sum rule is probably most sensitive to low-lying strength, in
particular at large $\tau$ where it would be hidden in $\mathcal{R}_{S,-1}$ by
the subtraction constants, it is tempting to attribute its reduced fit quality
to some broad, low-lying structure missing on the phenomenological side.
Amusingly, this would be consistent with the $K$-matrix analysis of the
$0^{++}$ meson data \cite{ani03} which tentatively identifies the broad
``$\sigma$ resonance'' $f_{0}\left(  600\right)  $ as a probably gluon-rich
exotic state.

Unfortunately, a two-resonance analysis of the $\mathcal{R}_{S,0}$ sum rule
does not help to settle this issue: due to the high quality of the one-pole
fits, it can hardly reduce the deviation measure further and thus becomes
unstable. (If nevertheless performed, it seems to favor an only somewhat
smaller mass $m_{S1}\sim0.8-1.2$ GeV of the lowest-lying resonance and a
heavier glueball at $m_{S2}\sim1.8$ GeV.) We conclude that significant
low-lying $0^{++}$ strength with a sufficiently strong coupling to the gluonic
interpolator (\ref{sipf}), if it exists, is probably to broadly distributed to
be resolved by a sum rules analysis.

\subsubsection{Pseudoscalar glueball}

As in the scalar channel, we start by surveying the quantitative behavior of
all four sum rules in order to determine how many isolated resonances are
required. The result is opposite to that in the scalar channel:\ all sum
rules, especially those derived from the lower Borel moments, clearly favor
two resonances, with a large separation between their masses. The improvement
over the one-pole fits is substantial. In fact, the latter do not only produce
significantly larger errors but also tend to destabilize the sum rules since
either the stronger decay at small $\tau$ or the weaker one at large $\tau$,
but not both, can be matched to one resonance.

The emergence of a second, relatively low-lying isolated resonance is of
course expected. In fact, its traces were already visible in the qualitative
decay behavior of the Borel moments (cf. Sec. \ref{qualbehav}). Moreover, our
discussion in Sec. \ref{topscr} anticipated the emergence of large
$\eta^{\prime}$ intermediate-state contributions to the $0^{-+}$ glueball
correlator, as a consequence of the anomaly-induced $\eta^{\prime}$-coupling
to the topological charge. The results of the quantitative sum rule analysis,
listed in Table \ref{tp}, indeed confirm this expectation: the central mass
value of the lighter resonance reproduces the $\eta^{\prime}$ mass. Our
central value for the coupling $f_{\eta^{\prime}}$ is somewhat larger than its
phenomenological value $0.82$ GeV \cite{fel00}, likely because of $\eta
$-admixtures. (Note that we treat $f_{\eta^{\prime}}$ on the same footing as
the glueball decay constant, i.e. we do not extract the conventional factor
$1/\sqrt{2N_{f}}$.) In any case, the result for $f_{\eta^{\prime}}$ is
probably our least accurate sum rule prediction since it originates from
the\ small residue of the large-$\tau$ tail.

A two-resonance analysis, with five independent hadronic parameters to be
determined, generally stretches the sum-rule resolution to its limits. In
fact, there is no guarantee that such an analysis will be stable. The large
separation between the resonance masses in the $0^{-+}$ channel, however,
improves the situation decisively since it assigns mutually almost exclusive
roles to the two poles: the heavier $0^{-+}$ glueball has to fit the
small-$\tau$ region and consequently decays so fast that it cannot
significantly ``contaminate'' the large-$\tau$ tail, which is mostly generated
by the $\eta^{\prime}$. This scenario is corroborated by the fact that the
predicted $\eta^{\prime}$ properties are, in contrast to the glueball
properties, almost $s_{0}$-independent. A glance at Fig. 4 shows that the
small-$\tau$ behavior of the moments indeed varies much more strongly with
$s_{0}$ than the tails. It turns out that the clear mass separation renders
the five-parameter fits stable and makes a quantitative sum rule analysis possible.

As another consequence of the large gap between the resonance masses, the
relative strength of the $\eta^{\prime}$ and glueball signals, $\propto\left(
m_{\eta^{\prime}}/m_{P}\right)  ^{4+2k}$, decreases strongly with $k$ since
the Borel moments weight the large-$s$ region of the spectral function by a
factor $s^{k}$. For this reason, the glueball predictions of the higher-moment
sum rules become more stable and the quality of one-pole fits improves with
$k$, reflecting the diminishing impact of the $\eta^{\prime}$. Nevertheless,
the one-pole analysis remains inherently unstable since the resonance
exponential can either match the IOPE moments at large or small $\tau$, but
not over the whole fiducial region. Hence the two-pole fits continue to be
superior even for $k=2$. \ The impact of the condensate contributions,
incidentally, also decreases with $k$ since the higher $k$-moments are
obtained by derivatives with respect to $-\tau$ and thus decrease the size of
the power corrections at large $\tau$.%

%TCIMACRO{\TeXButton{B}{\begin{table}[htbp] \centering}}%
%BeginExpansion
\begin{table}[htbp] \centering
%EndExpansion%
\begin{tabular}
[c]{|c|c|c|c|c|c|c|c|c|}\hline
$k$ & $\tau_{\min}$ (GeV$^{-2}$) & $\tau_{\max}$ (GeV$^{-2}$) & $m_{\eta
^{\prime}}$ (GeV) & $f_{\eta^{\prime}}$ (GeV) & $m_{P}$ (GeV) & $f_{P}$
(GeV) & $s_{0}$ (GeV$^{2}$) & $\delta\times10^{-3}$\\\hline\hline
$-1$ & $0.2$ & $1.2$ & $0.97$ & $0.88$ & $2.12$ & $0.43$ & $6.79$ &
$3.22$\\\hline
$0$ & $0.25$ & $1.3$ & $0.81$ & $1.31$ & $2.32$ & $0.76$ & $9.14$ &
$3.89$\\\hline
$1$ & $0.55$ & $1.4$ & $1.05$ & $0.84$ & $2.08$ & $0.72$ & $6.63$ &
$1.59$\\\hline
$2$ & $0.55$ & $1.5$ & $1.08$ & $1.10$ & $2.20$ & $0.79$ & $7.31$ &
$4.86$\\\hline
\end{tabular}
\caption{The fiducial domain, fitting error and predictions of the pseudoscalar glueball sum rule based on the k-th Borel moment.\label{tp}%
}
%TCIMACRO{\TeXButton{E}{\end{table}}}%
%BeginExpansion
\end{table}%
%EndExpansion

The four $0^{-+}$ sum rule fits are displayed in Fig. 8 and produce the
results contained in Tab. \ref{tp}. Their central values yield the predictions%
\begin{align}
m_{P}  &  =2.2\pm0.2\text{ GeV},\text{ \ \ \ \ \ \ }f_{P}=0.6\pm0.25\text{
GeV,}\\
m_{\eta^{\prime}}  &  =0.95\pm0.15\text{ GeV},\text{ \ \ \ \ \ \ }%
f_{\eta^{\prime}}=1.05\pm0.25\text{ GeV,}%
\end{align}
again including an additional error due to estimated input parameter
uncertainties. (One should keep in mind that the $\eta^{\prime}$ pole probably
receives some strength from $\eta$-admixtures.) As in the scalar channel, the
figures also show the contributions from the perturbative and nonperturbative
IOPE coefficients separately.

In the graph for $\mathcal{R}_{P,-1}$, we have additionally plotted the
topological charge screening contribution alone (dotted line). Its negative
sign is a consequence of the screening-induced subtraction term. Nevertheless,
the total nonperturbative unit-operator coefficient (dashed line) is positive,
due to the (positive) instanton-induced subtraction term. The balance between
these subtraction constants is instrumental in reconciling the $k=-1$ sum rule
with the low-energy theorem (\ref{pLET}), as argued in Sec. \ref{subconst}.
Without the topological charge screening contributions this sum rule would
obviously be inconsistent and impossible to fit. The dependence on the LET
parameters (quark masses and condensate), incidentally, is much weaker than in
the scalar $k=-1$ sum rule since the phenomenological subtraction constant is
an order of magnitude smaller. The cancellation between the individually much
larger IOPE contributions is therefore a highly nontrivial consistency
requirement of the anomalous axial Ward identity (cf. Sec. \ref{topscr}).

In the higher moment sum rules, the screening contributions remain essential
and generate, besides the $\eta^{\prime}$ resonance, a clear $0^{-+}$ glueball
signal. The plots of the nonperturbative Wilson coefficients in Fig. 8 contain
the cancellations with the direct instanton contributions and demonstrate why
it would be detrimental to ignore the topological charge screening
contributions (cf. Sec. \ref{qualbehav}). The screening contributions affect
the behavior of the moments mostly at $\tau>0.5$ where their impact becomes
comparable to that of the perturbative and direct instanton contributions.
Positivity violations due to the direct instantons and the dissolution of the
glueball signal would manifest themselves in this $\tau$-region if the
screening contributions were ignored \cite{zha03}.

Despite the larger values of $s_{0}$, the size of the direct-instanton
contributions is not much smaller than in the scalar channel. Our discussion
of the finite-width distribution and renormalization effects in the previous
section applies to the most part here, too. However, due to the cancellations
among the nonperturbative contributions their overall impact on the
pseudoscalar sum rules is much more moderate. (The conventional criterion
(\ref{tmax1}) therefore determines $\tau_{\max}$ only in the $k=2$ sum rule.)
As a consequence, one expects the results to become closer to those of the
older $0^{-+}$ glueball sum rule analyses which neglected nonperturbative
Wilson coefficients altogether. Indeed, Ref. \cite{asn92} found $m_{P}%
=2.3\pm0.2$ GeV which is compatible with our result.

Our prediction for the glueball pole residue, however, is about twice as large
as the value $f_{P}=0.30\pm0.05$ GeV of Ref. \cite{asn92}. This enhancement is
due to the remaining nonperturbative contributions, the higher-order
perturbative corrections and the additional $\eta^{\prime}$ pole. Since
$f_{P}$ is related to the scale of the light-cone distribution amplitude for
the pseudoscalar glueball, our result can be used, e.g., to determine the
$\gamma\gamma\rightarrow G_{P}\pi^{0}$ cross section at large momentum
transfer \cite{pas03}. As in the scalar channel, the increased prediction for
the coupling $f_{P}$ and the consequently larger partial width of radiative
$J/\psi$ decays into pseudoscalar glueballs \cite{gou87} are of relevance for
experimental glueball searches.

Our prediction for the $0^{-+}$ glueball mass lies inside the range of
quenched lattice results, $m_{P}^{\left(  q\right)  }=2.1-2.5$ GeV
\cite{lee00}. (The range of values again reflects scale-setting ambiguities.)
Unquenched simulations are still at an exploratory stage, with correspondingly
large errors for the higher-lying glueball masses \cite{har02}. The
pseudoscalar mass, however, is close to that for the lowest tensor glueball
and in the same range as the quenched results. The fact that our prediction
for the $0^{-+}$ glueball mass is close to the quenched lattice results while
that for the $0^{++}$ glueball mass is significantly smaller may be related to
the smaller $0^{++}$ glueball size. Since sea-quark-induced vacuum
polarization, which is missing in the quenched approximation, affects mainly
the short-distance properties of the glueball wave functionals, one might
expect the quenched mass predictions to be subject to larger dynamical-quark
corrections in the scalar channel.

The $K$-matrix analysis of Ref. \cite{ani03} cannot identify narrow-resonance
candidates for a $0^{-+}$ glueball in the above mass range since both
$\eta\left(  1295\right)  $ and $\eta\left(  1440\right)  $ are found to lie
on linear quark-antiquark trajectories. The Gaussian sum rule analysis of
\cite{zha03} gives an about 20\% larger $0^{-+}$ mass than ours, outside of
the range of lattice results. Due to the absence of the crucial screening
contributions the underlying IOPE is inconsistent with the axial Ward
identity, however, and these results cannot be trusted.

In contrast to the results in the scalar channel, the $k$-dependence of the
pseudoscalar glueball mass shows no particular systematics, although the
coupling is again practically $k$-independent for $k\geq0$. It might be
tempting to speculate that this implies a less homogeneous distribution of the
$0^{-+}$ glueball strength, perhaps generated by two (or more) neighboring
resonances. In any case, it seems that no useful estimate for (or bound on)
the width of the pseudoscalar glueball can be delineated from the variation of
the pseudoscalar glueball properties with $k$. (The monotonic increase in the
predictions for the $\eta^{\prime}$ mass with $k$, as well as the larger
fluctuations in the coupling, are probably due to contaminations by the
increasingly dominant glueball signal.)

The derivative of the topological charge correlator at $Q^{2}=0$, plays an
important role in the analysis of the proton spin content and the related
structure function \cite{sho92}. The strong cancellations among the
nonperturbative contributions to the pseudoscalar IOPE might explain why QCD
sum rule estimates of $\chi^{\prime}$ on the basis of purely perturbative
Wilson coefficients seem to be sufficiently stable \cite{nar99}. Nevertheless,
this analysis should be repeated on the basis of the full IOPE.

\section{Summary and conclusions}

\label{sumconcl}

The central themes of this paper are the derivation of nonperturbative
short-distance contributions to the spin-0 glueball correlators and a
comprehensive sum-rule analysis of their predictions for glueball properties.
The dominant nonperturbative contributions to the operator product expansion
of these correlators turn out to be (semi-) hard, i.e. to reside in the Wilson
coefficients. Their soft counterparts, contained in the condensates, play a
comparatively minor role. Both sources of hard nonperturbative physics
considered in this paper, direct instantons and topological charge screening,
are associated with the topology of the vacuum gluon fields or, equivalently,
of the QCD gauge group. The main benefit of analyzing the hard nonperturbative
contributions by means of a short-distance expansion\ is that it allows an
analytical and largely model-independent treatment. The complementary bulk
information on the soft physics, which enters through a few condensates and
the instanton size distribution, can be straightforwardly imported from other sources.

Direct instanton contributions to other hadron correlators were calculated
previously, and several important effects with diverse physical impact were
found in the corresponding QCD sum rules (including those for the $0^{++}$
glueball). However, their evaluation was restrained by three major
approximations: (i) all instantons were taken to be of the same size, (ii) the
renormalization of the instanton-induced Wilson coefficients was ignored and
(iii) the contributions of only the instanton nearest to the interpolator
arguments were taken into account explicitly. The first of these
approximations becomes exact in instanton vacuum models at large $N_{c}$ and
the third can be regarded as the leading term in an expansion in the instanton
density. The second, although convenient, is not strictly associated with a
systematic expansion.

An important point on our agenda was to assess the validity of these
approximations and to improve upon them. We have removed restriction (i) by
developing and implementing minimal expressions for realistic, finite-width
instanton size distributions which incorporate the currently available
information on its scales, overall shape and limiting behavior. These
expressions are fully determined by the standard values of the instanton
density and the mean instanton size. To improve upon approximation (ii), we
have devised a simple renormalization procedure which explicitly restricts the
contributing instantons to the direct ones, i.e. to those which are smaller
than the inverse operator scale. These developments are channel independent
and will be useful in other hadron correlators as well.

The improvements in the direct-instanton sector have several notable effects
on the glueball sum rule analysis and its results. Finite-width distributions
considerably slow the asymptotic decay of the instanton contributions at large
momentum transfer, which reduces the glueball mass predictions. They
furthermore resolve artefacts in the IOPE which previously distorted the Borel
moments and contaminated the results for the scalar glueball. As an additional
benefit, realistic instanton size distributions tend to enlarge the fiducial
$\tau$-domain of the sum rule analysis. The gauge-invariant renormalization of
the instanton-induced Wilson coefficients significantly improves the
consistency with the underlying low-energy theorem in the scalar channel, and
sum-rule consistency in general. Renormalization also reduces the overall size
of the instanton contributions. The reanalysis of our previous IOPE sum rules
in the scalar glueball channel shows that the beneficial impact of the direct
instanton contributions is enhanced by the above improvements. The consistency
among different Borel moments and with the low-energy theorem is consolidated
and the previously deficient and usually discarded lowest-moment sum rule
becomes one of the most stable and reliable.

In the pseudoscalar channel the emerging pattern is more complex: the sum
rules with purely perturbative Wilson are found to be consistent with the
corresponding low-energy theorem (which had not been appreciated before since
the pertinent $\mathcal{R}_{P,-1}$ sum rule was not analyzed). However, this
consistency is lost entirely when direct instanton corrections are included
and additional physics, both (semi-) hard and nonperturbative, is needed to
restore it.

Our third improvement of the IOPE consists in identifying this missing physics
as due to topological charge screening and in implementing the screening
contributions. This is our conceptually farthest-reaching extension of the OPE
since it introduces for the first time nonperturbative physics beyond single
instantons into the IOPE coefficients. This physics turns out to be highly
channel-selective: it almost exclusively affects the $0^{-+}$ glueball
correlator (and the related $\eta^{\prime}$ correlator) which is proportional
to the topological charge correlator and thus specifically tuned to those
light-quark induced correlations which produce topological charge screening.
Under the suggestive assumption that the topological vacuum charge is mostly
due to instantons, the implementation of its screening can be regarded as an
improvement upon approximation (iii) mentioned above: the small instanton
packing fraction does not suppress the exceptionally strong and short-ranged
correlations between opposite topological charges as strongly as others.
Therefore, in this specific case the nearest-instanton approximation is
insufficient and the screening corrections have to be added in the $0^{-+}$
glueball (and $\eta^{\prime}$) channel.

We have found compelling evidence for the screening contributions to be an
indispensable complement to the direct instanton contributions. They restore
consistency with the axial Ward identity and thereby overcome, in particular,
the above-mentioned problem with the low-energy theorem. This manifests itself
e.g. in the screening contributions to the subtraction constant in the
lowest-moment sum rule which cancel most of the direct instanton
contributions. In the chiral limit, this cancellation becomes exact. It also
explains in a natural way how the equal size of the instanton-induced
subtraction constants in both spin-0 glueball channels can be reconciled with
the order-of-magnitude difference in the sizes of their phenomenological
counterparts. (In the absence of light quarks this difference would
practically disappear together with the screening contributions, so that the
sum rules would remain intact.)

The topological charge screening corrections lead to the emergence of a strong
$\eta^{\prime}$ resonance signal and necessitate a two-resonance analysis of
the $0^{-+}$ sum rules. Remarkably, the large gap between the $\eta^{\prime}$
and $0^{-+}$ glueball masses allows for a simultaneous prediction of all
associated resonance and threshold parameters. The impact of the direct
instanton contributions differs strongly in both spin-0 glueball channels. In
addition to their opposite sign (a consequence of the instanton's self-dual
field strength), the larger continuum threshold and the strong cancellations
with the screening corrections drastically modify the role of the instantons
in the $0^{-+}$ channel and counterbalance their repulsion. In fact, ignoring
the screening contributions would, besides violating the anomalous Ward
identity, lead to spectral positivity violations and the disappearance of the
$0^{-+}$ glueball signal. Pseudoscalar glueball sum rules with unscreened
direct instanton contributions are therefore invalid.

Our comprehensive numerical analysis of all eight Borel sum rules in both
spin-0 glueball channels reveals a rather diverse pattern of glueball
properties. The hard topological physics in the IOPE coefficients turns out to
strongly affect the sum rule results and to generate several new predictions.
In the scalar channel, the improved treatment of the direct-instanton sector
reduces our earlier (spike-distribution based) result for the $0^{++}$
glueball mass by about 20\%, to $m_{S}=1.25\pm0.2$ GeV. Although still
consistent within errors, our new central mass prediction is smaller than the
quenched lattice result. However, light-quark effects and especially
quarkonium admixtures are expected to reduce the quenched masses, and the
first unquenched simulations indeed show a tendency towards smaller scalar
glueball masses. Moreover, our mass prediction is consistent with the broad
glueball state found in a recent $K$-matrix analysis of the scalar meson
spectrum which includes the new states recently identified in the Crystal
Barrel data. The systematics in our results from different Borel-moment sum
rules indicates a rather large width of the scalar glueball, $\Gamma
_{S}\gtrsim0.3$ GeV. A similarly large width is found in $K$-matrix and mixing
analyses and could also be expected from unquenched lattice simulations with
realistic quark masses, due to the increased number of open decay channels.

Our prediction for the glueball decay constant, \ $f_{S}=1.05\pm0.1$ GeV, is
several times larger than the value obtained when ignoring the nonperturbative
Wilson coefficients. This result implies an exceptionally small glueball size,
in agreement with some lattice and instanton liquid model evidence. The strong
concentration of the $0^{++}$ glueball (Bethe-Salpeter) wave function is
therefore at least partially explained by the strong instanton-induced
attraction between gluons in the scalar channel. Moreover, its small size
makes the scalar glueball more susceptible to the momentum-dependence of the
color-dielectric constant arising from sea quarks. Since this part of the
vacuum polarization is missing in the quenched approximation, one would expect
larger errors in the quenched predictions for the $0^{++}$ glueball properties
than for their $0^{-+}$ counterparts. This might explain why our mass
prediction deviates more strongly from the quenched results in the scalar
channel. Furthermore, our enhanced prediction for $f_{S}$ implies
substantially larger partial widths of radiative $J/\psi$ and $\Upsilon$
decays into scalar glueballs. It is therefore of importance for experimental
glueball searches, in particular for the interpretation of the recent CLEO and
forthcoming CLEO-III data on $\Upsilon\rightarrow\gamma f_{0}$ and other decay
branches, and for related measurements of scalar glueball properties.

Although hints of some gluon-rich spectral strength in the $0^{++}$ channel
well below 1 GeV can be detected in the sum rule analysis, we find no clear
evidence for a low-lying and sufficiently narrow glueball resonance in the
region of the $f_{0}\left(  600\right)  $. (The $\mathcal{R}_{S,-1}$ sum rule
would ``predict'' a spurious low-lying resonance if the direct instanton
contributions were ignored. However, this result is inconsistent with both the
low-energy theorem and the higher-moment sum rules and therefore obsolete.)

In the pseudoscalar glueball channel, the topological charge screening
contributions do not only resurrect the sum rules but also have a strong
impact on their quantitative predictions. Due to the cancellations with the
direct instanton contributions, the overall size of the nonperturbative IOPE
coefficients remains below about 20\% of their perturbative counterparts and
the detrimental problems encountered when ignoring the screening contributions
are resolved. Although the origin of the individual nonperturbative
contributions is more diverse and complex than in the scalar channel, their
overall impact is therefore smaller. This is one of the reasons for the
considerably weaker binding of the $0^{-+}$ glueball. Nevertheless, the hard
nonperturbative contributions modify qualitative features of the $0^{-+}$
Borel moments to which the matching analysis is particularly sensitive, and
they are vital\ for achieving consistency among all moment sum rules and with
the axial anomaly. The strong cancellations among the nonperturbative
contributions to the pseudoscalar IOPE may also explain why QCD sum rule
estimates of the derivative of the topological susceptibility, which plays an
important role in the analysis of the proton spin content, seem to be
sufficiently consistent without nonperturbative Wilson coefficients. Their
reanalysis on the basis of the full IOPE is in progress.

The quantitative sum rule analysis results in the values $m_{P}=2.2\pm1.5$ GeV
for the pseudoscalar glueball mass and $f_{P}=0.6\pm0.2$ GeV for the decay
constant. Our mass prediction lies inside the range obtained from quenched and
unquenched lattice data. The coupling is again enhanced by the nonperturbative
Wilson coefficients, but less strongly than in the scalar channel. The
consequently larger partial width of radiative quarkonium decays into
pseudoscalar glueballs and the enhanced $\gamma\gamma\rightarrow G_{P}\pi^{0}$
cross section at high momentum transfers will be relevant for the experimental
identification of the lowest-lying $0^{-+}$ glueball and help in measuring its properties.

Our extended IOPE should be useful for the calculation of other spin-0
glueball properties as well. Quantitative estimates of the already mentioned
production rates in gluon-rich channels (including $J/\psi$ and $\Upsilon$
decays) and characteristic glueball decay properties and signatures, including
$\gamma\gamma$ couplings, OZI suppression and branching fractions incompatible
with $q\bar{q}$ decay, would be particularly interesting.

\begin{acknowledgments}
\bigskip
\end{acknowledgments}

This work was supported by FAPESP. The author would like to thank Gast\~{a}o
Krein and the IFT-UNESP for their hospitality.

\newpage

\appendix                                                        

\section{Borel moments from perturbative Wilson coefficients}

\label{bpmoms}

In this appendix we list the contributions from the perturbative OPE
coefficients to the Borel moments $\mathcal{L}_{k}\left(  \tau,s_{0}\right)  $
with $k\in\left\{  -1,0,1,2\right\}  $ for both spin-0 glueball correlators
(before RG improvement). In terms of the continuum factors $\rho_{k}$ and the
exponential integral $E_{1}$, defined as%

\begin{align}
\rho_{k}\left(  x\right)   &  =e^{-x}\sum_{n=0}^{k}\frac{x^{n}}{n!},\qquad\\
E_{1}\left(  x\right)   &  =\int_{1}^{\infty}dt\frac{e^{-xt}}{t},
\end{align}
one obtains ($\gamma=0.5772$ is Euler's constant and the coefficients
$A_{i}-D_{i}$ are given in Section \ref{pwcs})

\begin{enumerate}
\item
\begin{align}
\mathcal{L}_{G,-1}^{\left(  pc\right)  }\left(  \tau,s_{0}\right)   &
=-\frac{A_{0}}{\tau^{2}}\left[  1-\rho_{1}\left(  s_{0}\tau\right)  \right]
+\frac{2A_{1}}{\tau^{2}}\left[  \ln\left(  \tau\mu^{2}\right)  +\gamma
-1+E_{1}\left(  s_{0}\tau\right)  +e^{-\tau s_{0}}+\ln\left(  \frac{s_{0}}%
{\mu^{2}}\right)  \rho_{1}\left(  s_{0}\tau\right)  \right] \nonumber\\
&  -B_{0}+B_{1}\left[  \ln\left(  \tau\mu^{2}\right)  +\gamma+E_{1}\left(
s_{0}\tau\right)  \right]  -C_{0}\tau+C_{1}\tau\left[  \ln\left(  \tau\mu
^{2}\right)  +\gamma-1-\frac{e^{-\tau s_{0}}}{\tau s_{0}}+E_{1}\left(
s_{0}\tau\right)  \right]  -\frac{D_{0}}{2}\tau^{2}%
\end{align}

\item
\begin{align}
\mathcal{L}_{G,0}^{\left(  pc\right)  }\left(  \tau,s_{0}\right)   &
=-\frac{2A_{0}}{\tau^{3}}\left[  1-\rho_{2}\left(  s_{0}\tau\right)  \right]
\nonumber\\
&  +\frac{4A_{1}}{\tau^{3}}\left[  \ln\left(  \tau\mu^{2}\right)
+\gamma-\frac{3}{2}+E_{1}\left(  s_{0}\tau\right)  +\rho_{0}\left(  s_{0}%
\tau\right)  +\frac{1}{2}\rho_{1}\left(  s_{0}\tau\right)  +\ln\left(
\frac{s_{0}}{\mu^{2}}\right)  \rho_{2}\left(  s_{0}\tau\right)  \right]
\nonumber\\
&  -\frac{B_{1}}{\tau}\left[  1-\rho_{0}\left(  s_{0}\tau\right)  \right]
+C_{0}-C_{1}\left[  \ln\left(  \tau\mu^{2}\right)  +\gamma+E_{1}\left(
s_{0}\tau\right)  \right]  +D_{0}\tau
\end{align}

\item
\begin{align}
\mathcal{L}_{G,1}^{\left(  pc\right)  }\left(  \tau,s_{0}\right)   &
=-\frac{6A_{0}}{\tau^{4}}\left[  1-\rho_{3}\left(  s_{0}\tau\right)  \right]
\nonumber\\
&  +\frac{12A_{1}}{\tau^{4}}\left[  \ln\left(  \tau\mu^{2}\right)
+\gamma-\frac{11}{6}+E_{1}\left(  s_{0}\tau\right)  +\rho_{0}\left(  s_{0}%
\tau\right)  +\frac{1}{2}\rho_{1}\left(  s_{0}\tau\right)  +\frac{1}{3}%
\rho_{2}\left(  s_{0}\tau\right)  +\ln\left(  \frac{s_{0}}{\mu^{2}}\right)
\rho_{3}\left(  s_{0}\tau\right)  \right] \nonumber\\
&  -\frac{B_{1}}{\tau^{2}}\left[  1-\rho_{1}\left(  s_{0}\tau\right)  \right]
+\frac{C_{1}}{\tau}\left[  1-\rho_{0}\left(  s_{0}\tau\right)  \right]  -D_{0}%
\end{align}

\item
\begin{align}
\mathcal{L}_{G,2}^{\left(  pc\right)  }\left(  \tau,s_{0}\right)   &
=-\frac{24A_{0}}{\tau^{5}}\left[  1-\rho_{4}\left(  s_{0}\tau\right)  \right]
+\frac{48A_{1}}{\tau^{5}}\left[  \ln\left(  \tau\mu^{2}\right)  +\gamma
-\frac{25}{12}+E_{1}\left(  s_{0}\tau\right)  \right.  +\rho_{0}\left(
s_{0}\tau\right)  +\frac{1}{2}\rho_{1}\left(  s_{0}\tau\right) \nonumber\\
&  +\left.  \frac{1}{3}\rho_{2}\left(  s_{0}\tau\right)  +\frac{1}{4}\rho
_{3}\left(  s_{0}\tau\right)  +\ln\left(  \frac{s_{0}}{\mu^{2}}\right)
\rho_{4}\left(  s_{0}\tau\right)  \right]  -\frac{2B_{1}}{\tau^{3}}\left[
1-\rho_{2}\left(  s_{0}\tau\right)  \right]  +\frac{C_{1}}{\tau^{2}}\left[
1-\rho_{1}\left(  s_{0}\tau\right)  \right]  .
\end{align}
\end{enumerate}

\section{Instanton integrals}

\label{iint}

An explicit expression for $\Pi_{S}^{\left(  I+\bar{I}\right)  }\left(
x^{2}\right)  $ can be obtained by Fourier transforming its dispersive
representation (\ref{idisprel}):%
\begin{align}
\Pi_{S}^{\left(  I+\bar{I}\right)  }\left(  x^{2}\right)   &  =\int
\frac{d^{4}Q}{\left(  2\pi\right)  ^{4}}e^{-iQx}\Pi^{\left(  I+\bar{I}\right)
}\left(  Q^{2}\right)  =\frac{1}{\pi}\int_{0}^{\infty}ds\operatorname{Im}%
\Pi^{\left(  I+\bar{I}\right)  }\left(  -s\right)  \int\frac{d^{4}Q}{\left(
2\pi\right)  ^{4}}\frac{e^{-iQx}}{s+Q^{2}}\label{int1}\\
&  =\frac{1}{4\pi^{3}}\frac{1}{x}\int_{0}^{\infty}ds\operatorname{Im}%
\Pi^{\left(  I+\bar{I}\right)  }\left(  -s\right)  \sqrt{s}K_{1}\left(
\sqrt{s}x\right) \\
&  =-4\pi\int d\rho n\left(  \rho\right)  \rho^{4}\frac{1}{x}\int_{0}^{\infty
}dss^{5/2}J_{2}\left(  \sqrt{s}\rho\right)  Y_{2}\left(  \sqrt{s}\rho\right)
K_{1}\left(  \sqrt{s}x\right) \\
&  =\frac{2^{8}3}{7}\int d\rho\frac{n\left(  \rho\right)  }{\rho^{4}}%
\,_{2}F_{1}\left(  4,6,\frac{9}{2},-\frac{x^{2}}{4\rho^{2}}\right)  ,
\end{align}
as anticipated in Eq. (\ref{instx}). The hypergeometric function $\,_{2}%
F_{1}\left(  a,b,c,z\right)  $ \cite{abr,gra} is defined as the analytical
continuation of Gau\ss' hypergeometric series (except if $c$ is a nonpositive
integer $-n$ and neither $b$ nor $c$ equal an integer $-m$ with $m<n$)
\begin{equation}
\,_{2}F_{1}\left(  a,b,c,z\right)  =\frac{\Gamma\left(  c\right)  }%
{\Gamma\left(  a\right)  \Gamma\left(  b\right)  }\sum_{n=0}^{\infty
}\frac{\Gamma\left(  a+n\right)  \Gamma\left(  b+n\right)  }{\Gamma\left(
c+n\right)  }\frac{z^{n}}{n!}.
\end{equation}
For convenience, we recall the limits%
\begin{equation}
\,_{2}F_{1}\left(  4,6,\frac{9}{2},-\frac{x^{2}}{4\rho^{2}}\right)
\longrightarrow\left\{
\begin{tabular}
[c]{l}%
$1-\frac{4}{3}\frac{x^{2}}{\rho^{2}}+O\left(  \left(  \frac{x^{2}}{\rho^{2}%
}\right)  ^{2}\right)  $ \ \ \ \ for $x^{2}\ll\rho^{2}$\\
$14\left(  \frac{\rho^{2}}{x^{2}}\right)  ^{4}+O\left(  \left(  \frac{\rho
^{2}}{x^{2}}\right)  ^{5}\right)  $ \ \ \ \ for $\rho^{2}\ll x^{2}$%
\end{tabular}
\ \ \ \ \ \right.
\end{equation}
which imply, in particular,%
\begin{equation}
\Pi_{S}^{\left(  I+\bar{I}\right)  }\left(  x^{2}=0\right)  =\frac{2^{8}3}%
{7}\int d\rho\frac{n\left(  \rho\right)  }{\rho^{4}}\,.
\end{equation}
The integral (\ref{int1}) can alternatively be done by several other methods,
e.g., by introducing Feynman parameters, which leads to \cite{sch98}%
\begin{equation}
\Pi_{S}^{\left(  I+\bar{I}\right)  }\left(  x^{2}\right)  =-2^{11}\int d\rho
n\left(  \rho\right)  \rho^{8}\left(  \frac{\partial}{\partial x^{2}}\right)
^{3}\frac{\xi^{6}}{x^{6}}\left[  \frac{3}{\xi}\arctan h\left(  \xi\right)
+\frac{5-3\xi^{2}}{\left(  1-\xi^{2}\right)  ^{2}}\right]  \qquad
\text{where}\qquad\xi^{2}=\frac{x^{2}}{x^{2}+4\rho^{2}},
\end{equation}
but is generally less convenient for practical calculations and for the study
of analyticity properties.

A straightforward way to calculate the Fourier transform of $\Pi_{G}^{\left(
I+\bar{I}\right)  }\left(  x^{2}\right)  $ starts from the original integral%
\begin{equation}
\Pi_{S}^{\left(  I+\bar{I}\right)  }\left(  x^{2}\right)  =\frac{2^{9}3^{2}%
}{\pi^{2}}\int d\rho n\left(  \rho\right)  \int d^{4}x_{0}\frac{\rho^{8}%
}{[\left(  x-x_{0}\right)  ^{2}+\rho^{2}]^{4}[x_{0}^{2}+\rho^{2}]^{4}}%
\end{equation}
and makes use of the fact that (in Euclidean space-time)%
\begin{align}
\Pi_{S}^{\left(  I+\bar{I}\right)  }\left(  Q^{2}\right)   &  =\frac{2^{9}%
3^{2}}{\pi^{2}}\int d\rho n\left(  \rho\right)  \int d^{4}xe^{iQx}\int
d^{4}x_{0}\frac{\rho^{8}}{[\left(  x-x_{0}\right)  ^{2}+\rho^{2}]^{4}%
[x_{0}^{2}+\rho^{2}]^{4}}\\
&  =\frac{2^{9}3^{2}}{\pi^{2}}\int d\rho n\left(  \rho\right)  \left(  \int
d^{4}xe^{iQx}\frac{\rho^{4}}{\left(  x^{2}+\rho^{2}\right)  ^{4}}\right)
^{2}.
\end{align}
With%
\begin{equation}
\int d^{4}x\frac{e^{iQx}}{\left(  x^{2}+\rho^{2}\right)  ^{4}}=\frac{4\pi^{2}%
}{Q}\int_{0}^{\infty}dx\frac{x^{2}J_{1}\left(  Qx\right)  }{\left(  x^{2}%
+\rho^{2}\right)  ^{4}}=\frac{\pi^{2}}{12}\frac{Q^{2}}{\rho^{2}}K_{2}\left(
Q\rho\right)  \label{soln1}%
\end{equation}
one then immediately obtains%
\begin{equation}
\Pi_{G}^{\left(  I+\bar{I}\right)  }\left(  Q^{2}\right)  =2^{5}\pi^{2}\int
d\rho n\left(  \rho\right)  \left(  \rho Q\right)  ^{4}K_{2}^{2}\left(
Q\rho\right)  . \label{piqbis}%
\end{equation}

In order to perform the Borel transform of (\ref{piqbis}), it is convenient to
start from an integral representation\ \ (\cite{gra}, Eq. 8.486.15) for the
McDonald function,
\begin{equation}
Q^{2}K_{2}\left(  Q\rho\right)  =2\rho^{2}\int_{0}^{\infty}d\alpha\alpha
e^{-\frac{Q^{2}}{4\alpha}-\alpha\rho^{2}}.
\end{equation}
For the calculation of the lowest ($k=-1$) Borel moment we then write%
\begin{align}
\frac{\Pi_{S}^{\left(  I+\bar{I}\right)  }\left(  Q^{2}\right)  }{-Q^{2}}  &
=-2^{5}\pi^{2}\int d\rho n\left(  \rho\right)  \rho^{4}Q^{2}K_{2}^{2}\left(
Q\rho\right) \\
&  =-2^{7}\pi^{2}\int d\rho n\left(  \rho\right)  \rho^{8}\int_{0}^{\infty
}d\alpha\alpha\int_{0}^{\infty}d\beta\beta\int_{0}^{\infty}d\gamma
e^{-\frac{Q^{2}}{4}\left(  \frac{1}{\alpha}+\frac{1}{\beta}+4\gamma\right)
-\left(  \alpha+\beta\right)  \rho^{2}}%
\end{align}
and make use of
\begin{equation}
\hat{B}e^{-aQ^{2}}=\delta\left(  a-\tau\right)
\end{equation}
to obtain
\begin{align}
\mathcal{L}_{-1}^{\left(  I+\bar{I}\right)  }\left(  \tau\right)   &  =\hat
{B}\left[  \frac{\Pi_{S}^{\left(  I+\bar{I}\right)  }\left(  Q^{2}\right)
}{-Q^{2}}\right] \\
&  =-2^{7}\pi^{2}\int d\rho n\left(  \rho\right)  \rho^{8}\int_{0}^{\infty
}d\alpha\alpha\int_{0}^{\infty}d\beta\beta\int_{0}^{\infty}d\gamma e^{-\left(
\alpha+\beta\right)  \rho^{2}}\delta\left(  \frac{\alpha+\beta}{4\alpha\beta
}+\gamma-\tau\right)  .
\end{align}
The three remaining parameter integrals are elementary. Their evaluation leads
to%
\begin{equation}
\mathcal{L}_{-1}^{\left(  I+\bar{I}\right)  }\left(  \tau\right)  =-2^{6}%
\pi^{2}\int d\rho n\left(  \rho\right)  \xi^{2}e^{-\xi}\left[  \left(
1+\xi\right)  K_{0}\left(  \xi\right)  +\left(  2+\xi+\frac{2}{\xi}\right)
K_{1}\left(  \xi\right)  \right]  ,
\end{equation}
where we have defined the dimensionless variable%
\begin{equation}
\xi=\frac{\rho^{2}}{2\tau}.
\end{equation}
The higher moments with $k\geq-1$ follow, as previously, by differentiation
with respect to $-\tau,$
\begin{equation}
\mathcal{L}_{k+1}^{\left(  I+\bar{I}\right)  }\left(  \tau\right)
=-\frac{\partial}{\partial\tau}\mathcal{L}_{k}^{\left(  I+\bar{I}\right)
}\left(  \tau\right)  \text{,}%
\end{equation}
and generate the expressions (\ref{inb0}) - (\ref{inb2}).

\newpage

{\Large Figure captions:}

\begin{enumerate}
\item[Fig. 1: ] Direct-instanton induced imaginary part of the $0^{++}$
glueball correlator, $\operatorname{Im}\Pi_{S}^{\left(  I+\bar{I}\right)
}\left(  -s\right)  $, obtained on the basis of (a) the spike and (b)
finite-width instanton size distributions. In (b) the results for the
exponential-tail distribution without (dotted) and with (dashed) large-$\rho$
cutoff as well as for the Gaussian-tail distribution with large-$\rho$ cutoff
(full line) are plotted. Note the difference in scale between (a) and (b).

\item[Fig. 2: ] The $k=0$ continuum-subtracted IOPE\ Borel moment of the
$0^{-+}$ glueball correlator, calculated on the basis of the spike
distribution (a) without (left panel) and (b) with (right panel) topological
charge screening contributions.

\item[Fig. 3: ] The continuum-subtracted IOPE Borel moments of the scalar
glueball correlator (calculated on the basis of the Gaussian-tail
distribution, and renormalized at the operator scale), as a function of $\tau$
($x$-axis) and $s_{0}$ ($y$-axis). All units are appropriate powers of GeV.

\item[Fig. 4: ] The continuum-subtracted IOPE Borel moments of the
pseudoscalar glueball correlator (calculated on the basis of the Gaussian-tail
distribution, and renormalized at the operator scale), as a function of $\tau$
($x$-axis) and $s_{0}$ ($y$-axis). All units are appropriate powers of GeV.

\item[Fig. 5: ] The $k=0$ continuum-subtracted IOPE\ Borel moments of the
$0^{++}$ (left panel) and $0^{-+}$ (right panel) glueball correlators, as
obtained from the spike distribution.

\item[Fig. 6:\ ] The square root of the ratio between the $k=2$ and $k=1$
continuum-subtracted IOPE Borel moments of the scalar glueball correlator.

\item[Fig. 7: ] The individual contributions to the optimized Borel sum rules
in the scalar glueball channel: the continuum-subtracted IOPE\ moments (full
line), the $0^{++}$ glueball pole (and subtraction constant, for $k=-1$)
contributions (bullets), the direct instanton contributions (dashed) and the
contributions of the perturbative Wilson coefficients (dash-dotted).

\item[Fig. 8: ] The individual contributions to the optimized Borel sum rules
in the pseudoscalar glueball channel: the continuum-subtracted IOPE\ moments
(full line), the $0^{-+}$ glueball and $\eta^{\prime}$ (and subtraction
constant, for $k=-1$) contributions (bullets), the contributions of the
nonperturbative Wilson coefficients (due to both direct instantons and
topological charge screening) (dashed) and the contributions of the
perturbative Wilson coefficients (dash-dotted). The figure for $k=-1$
additionally shows the screening contributions by themselves (dotted).
\end{enumerate}

\bigskip
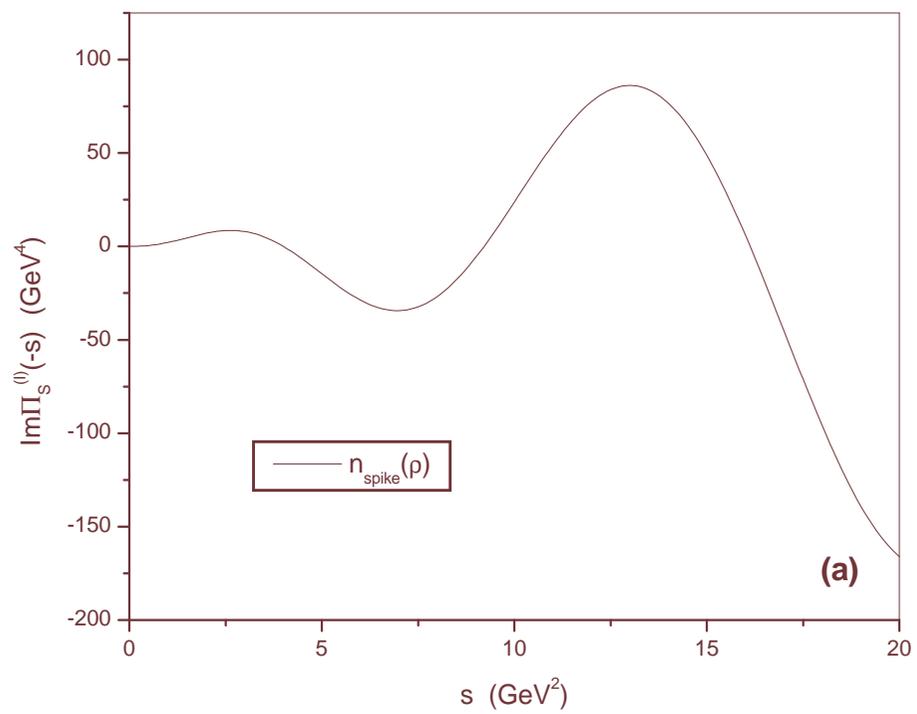
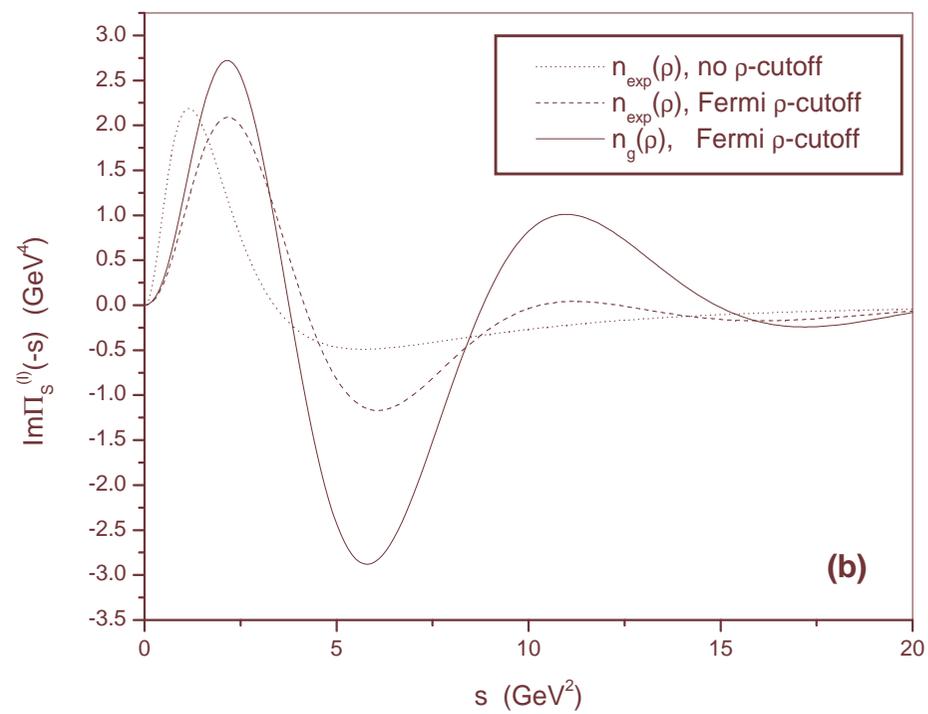
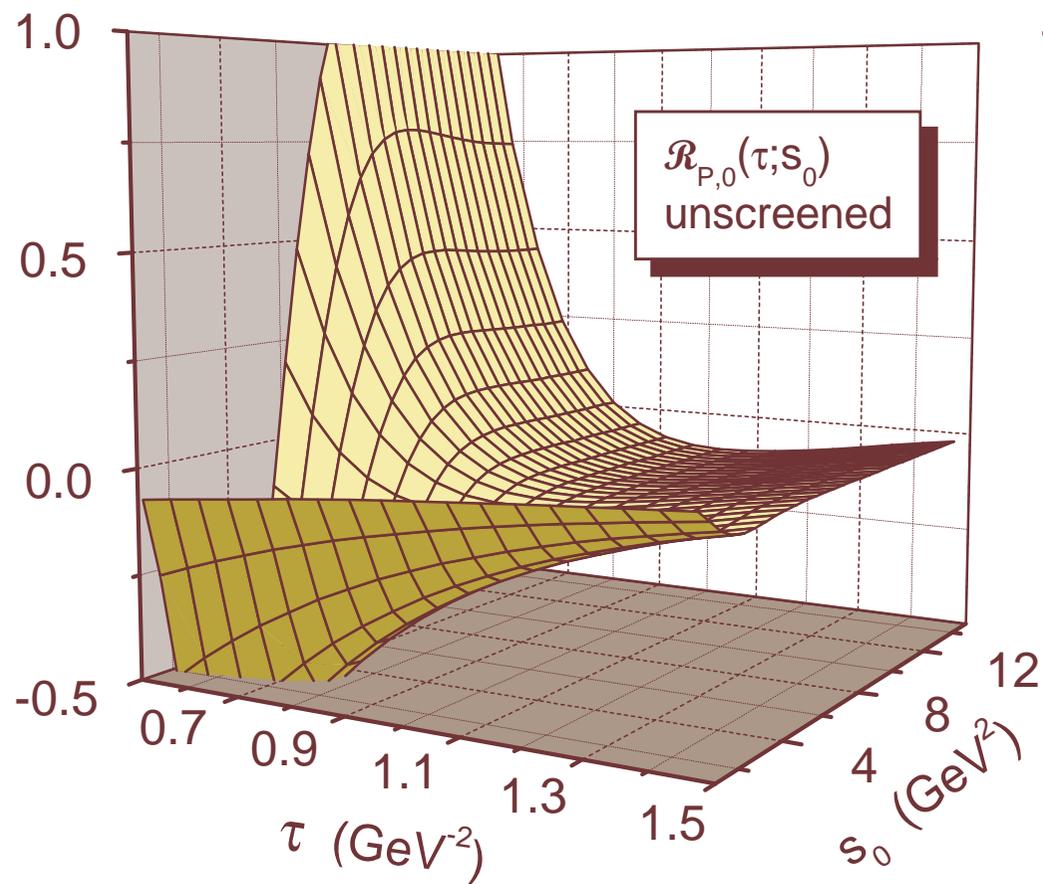
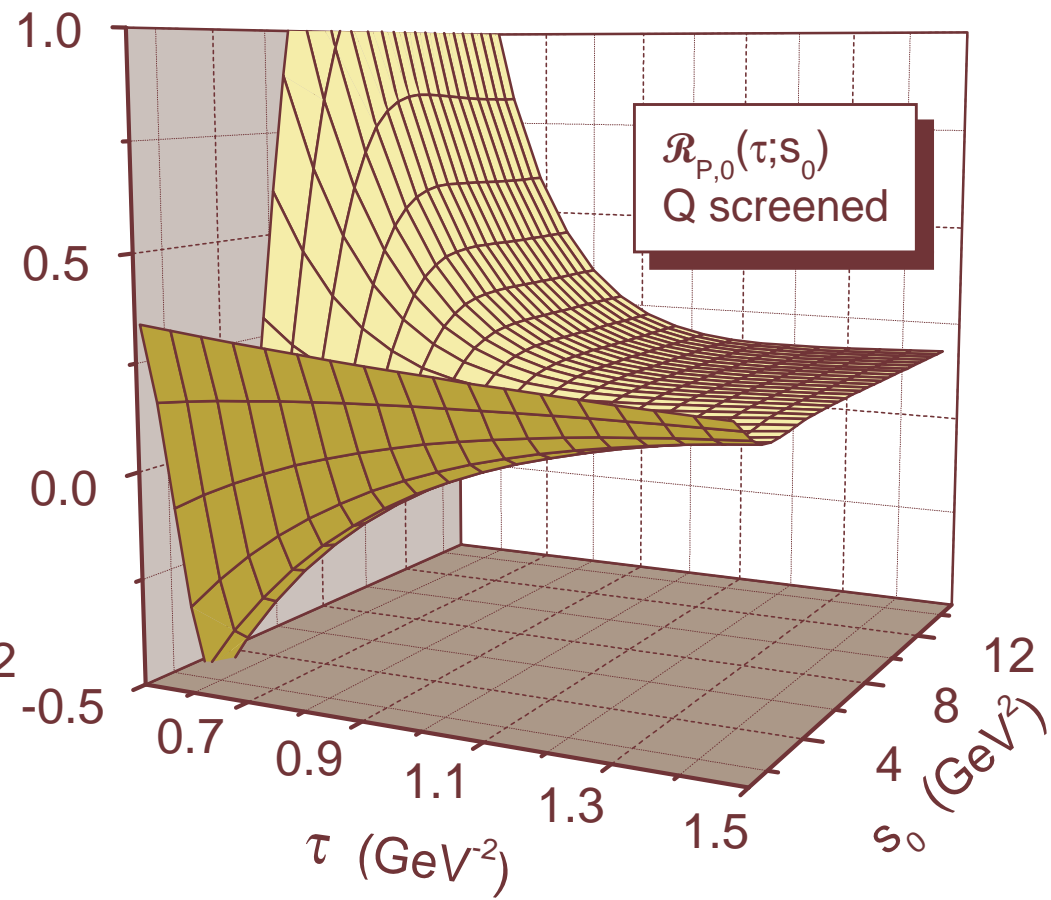
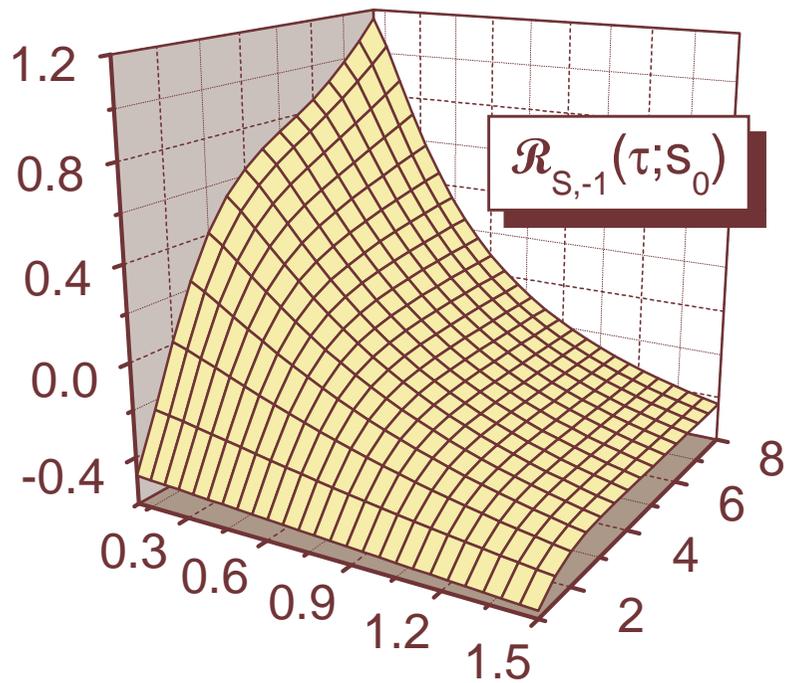
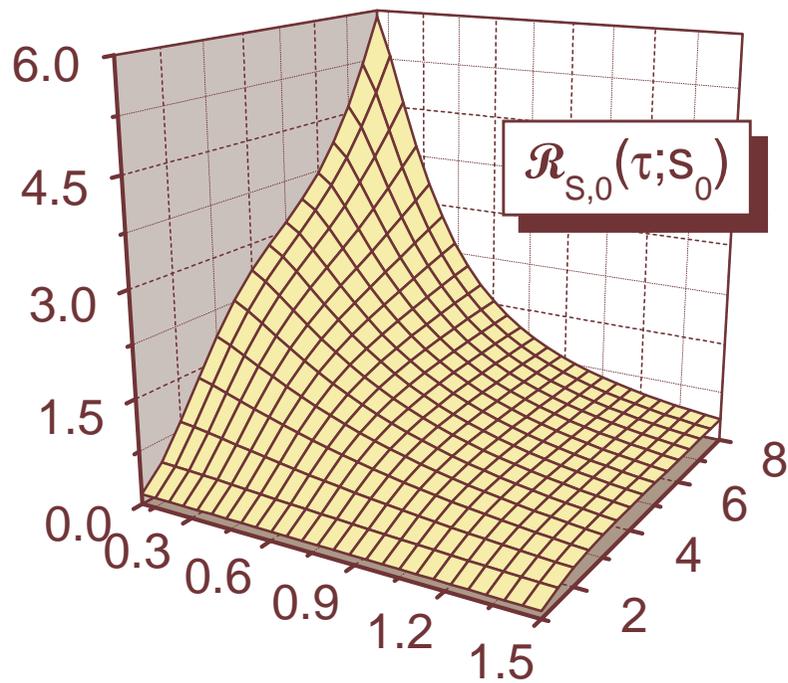
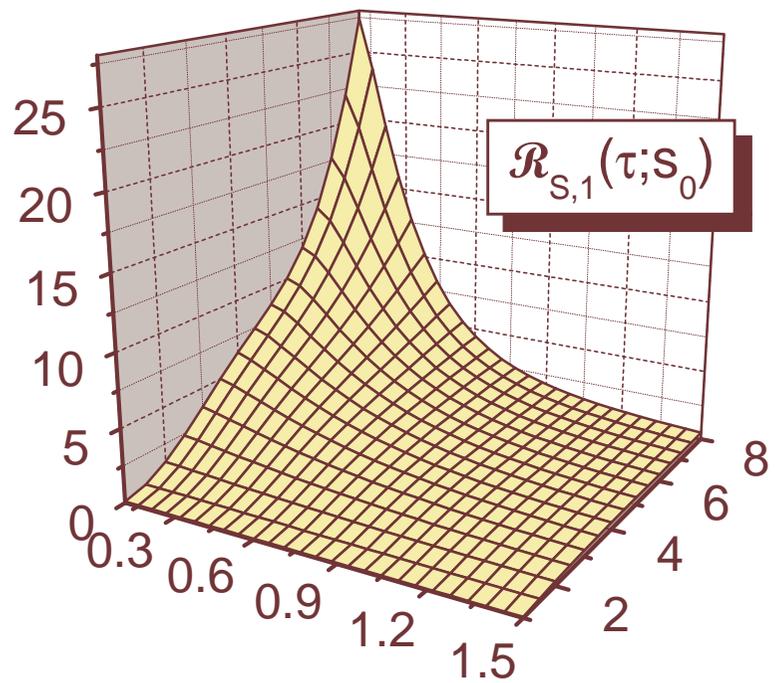
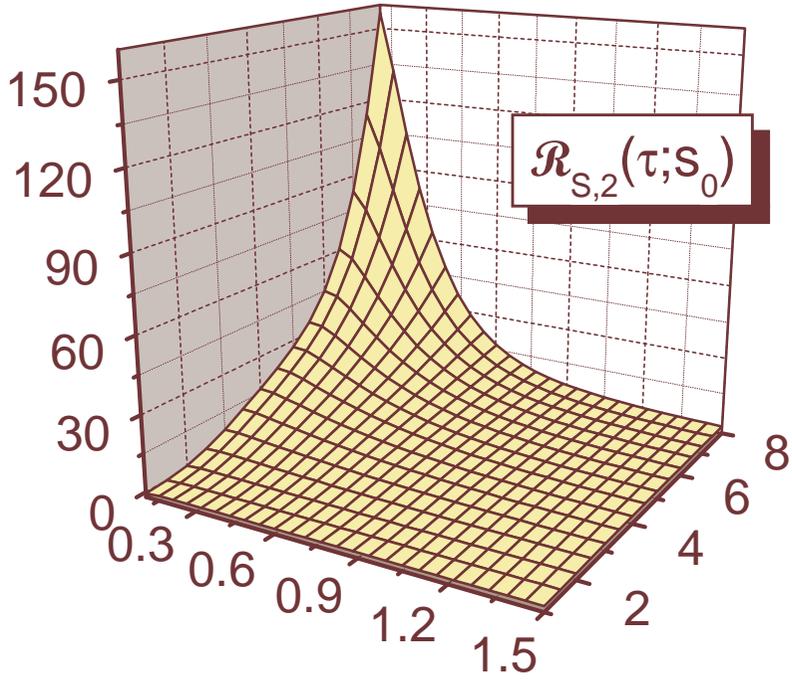
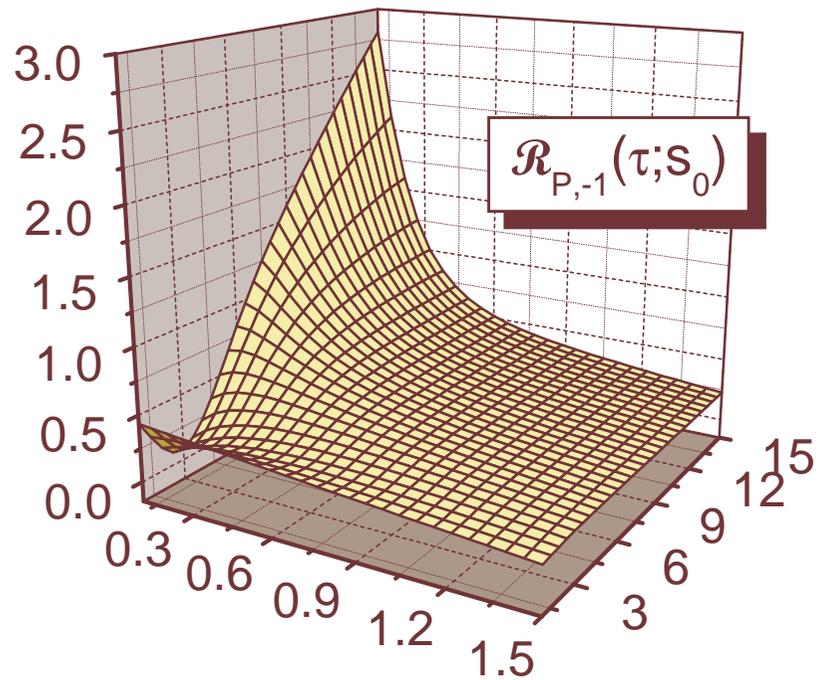
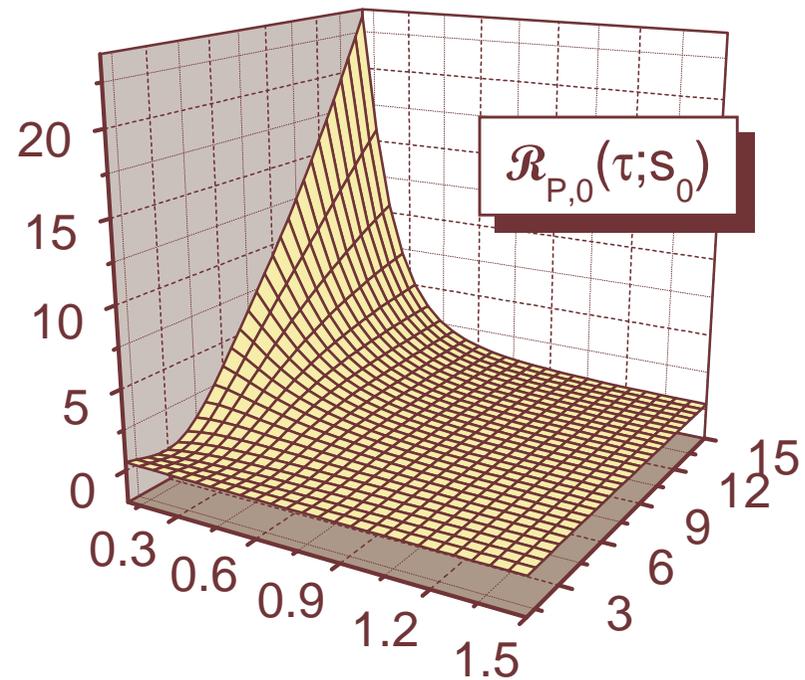
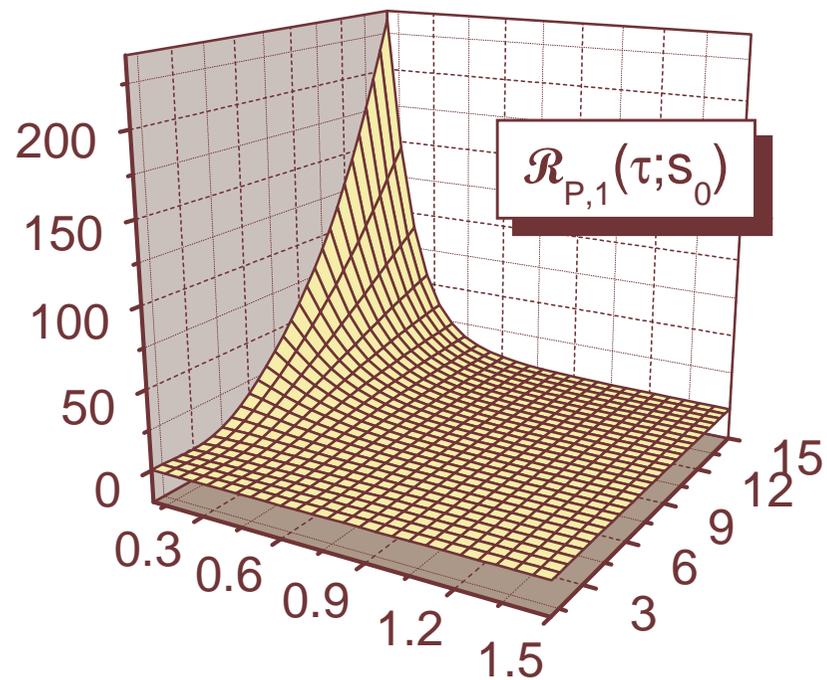
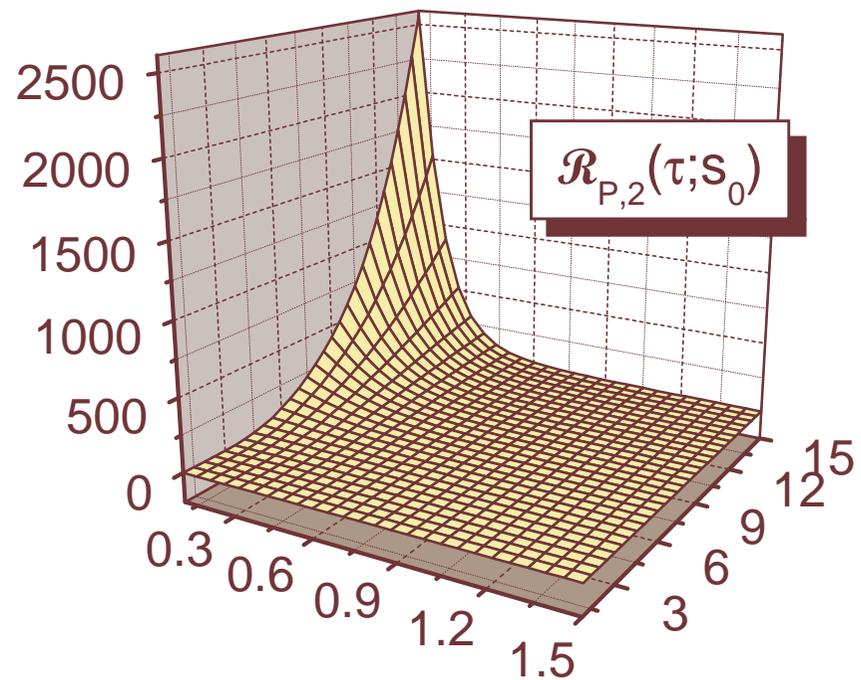
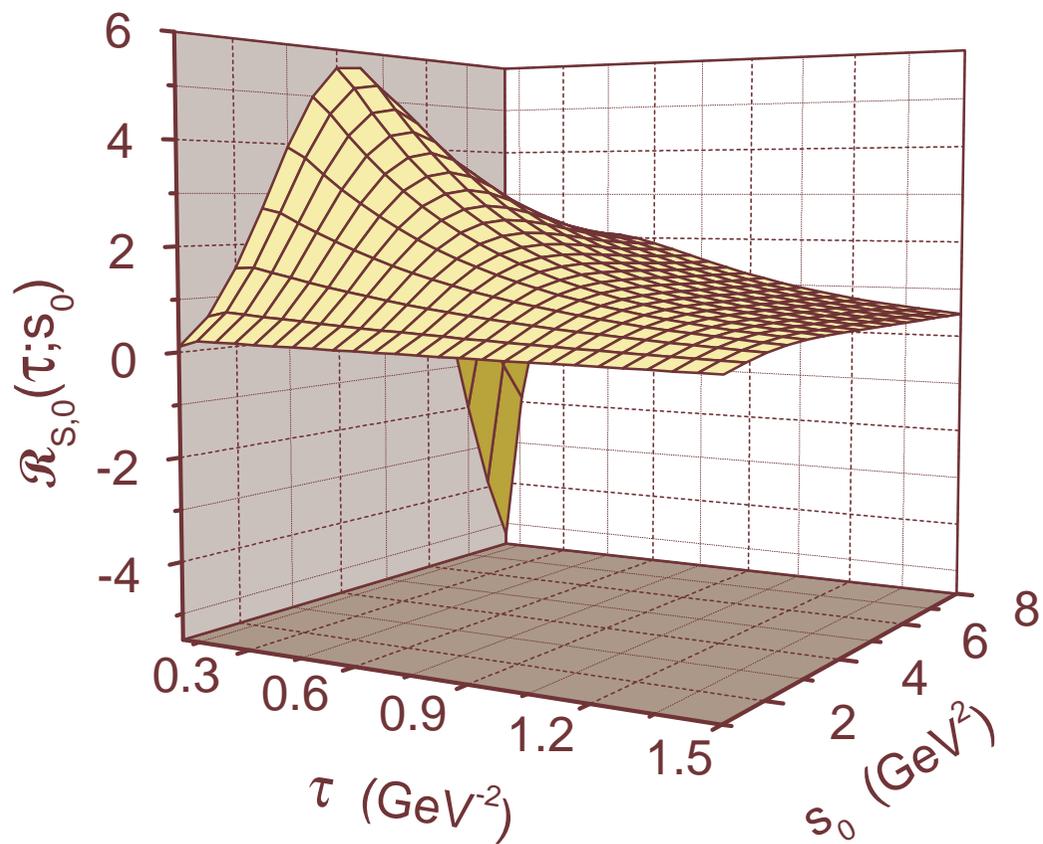
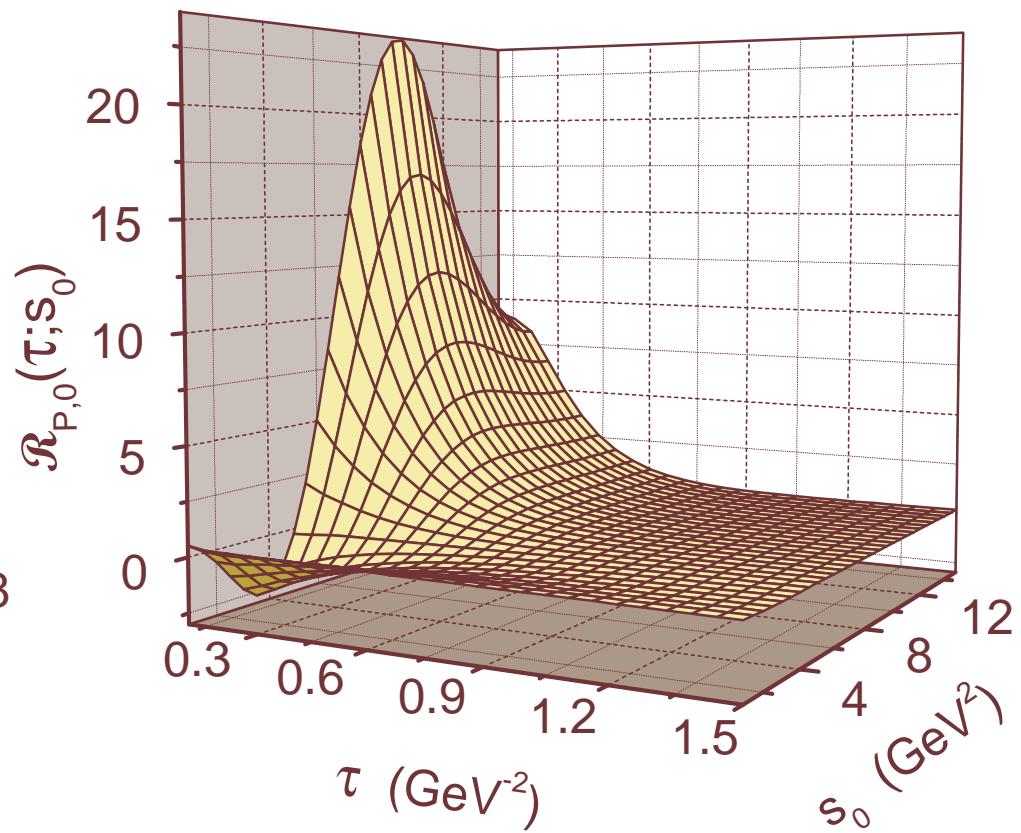
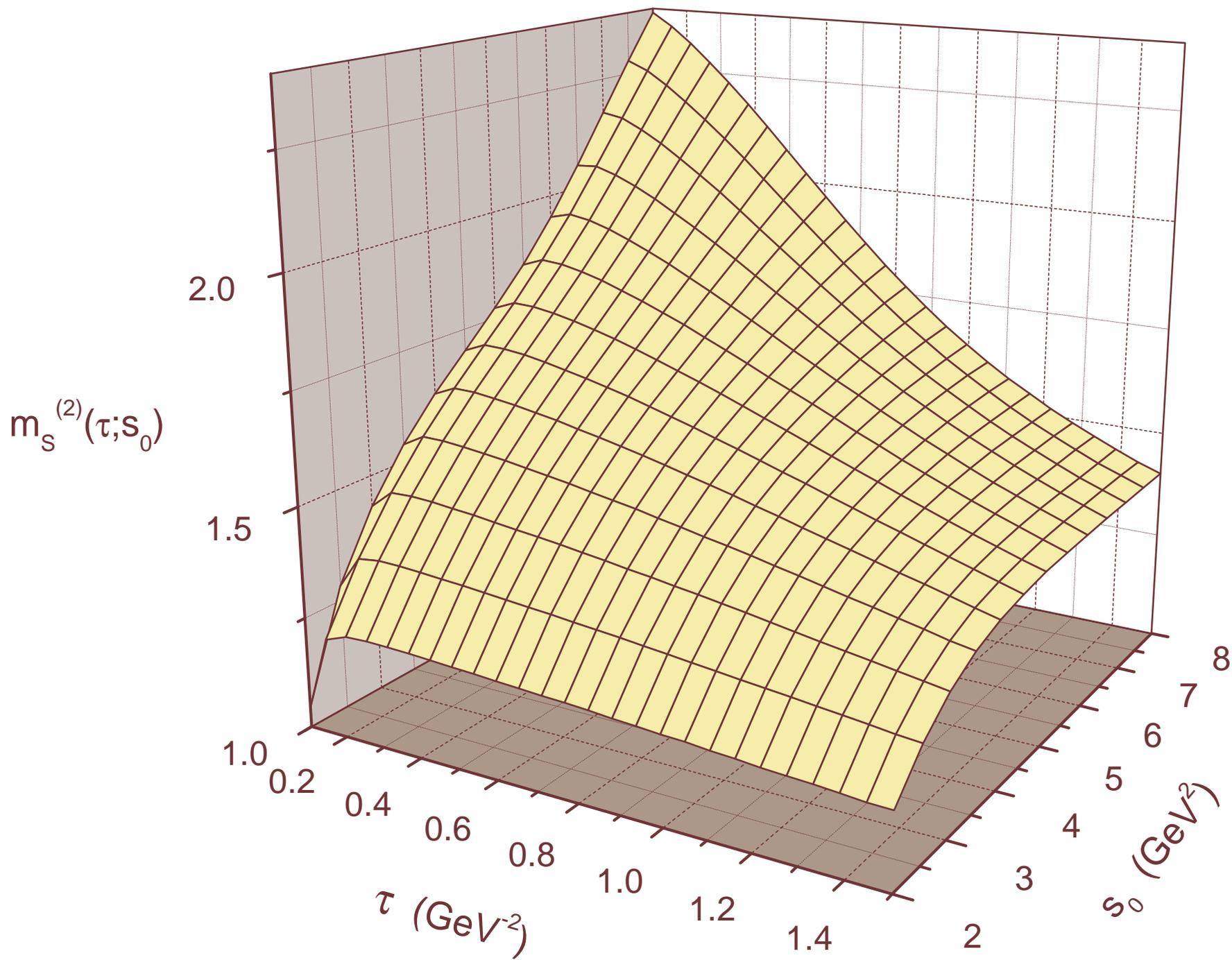
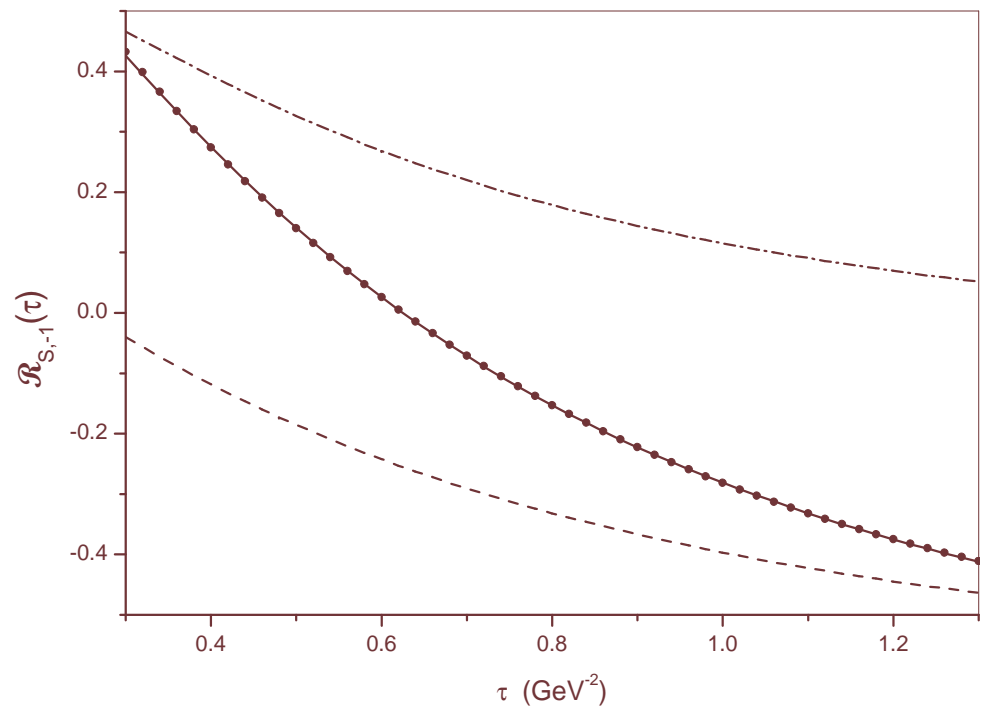
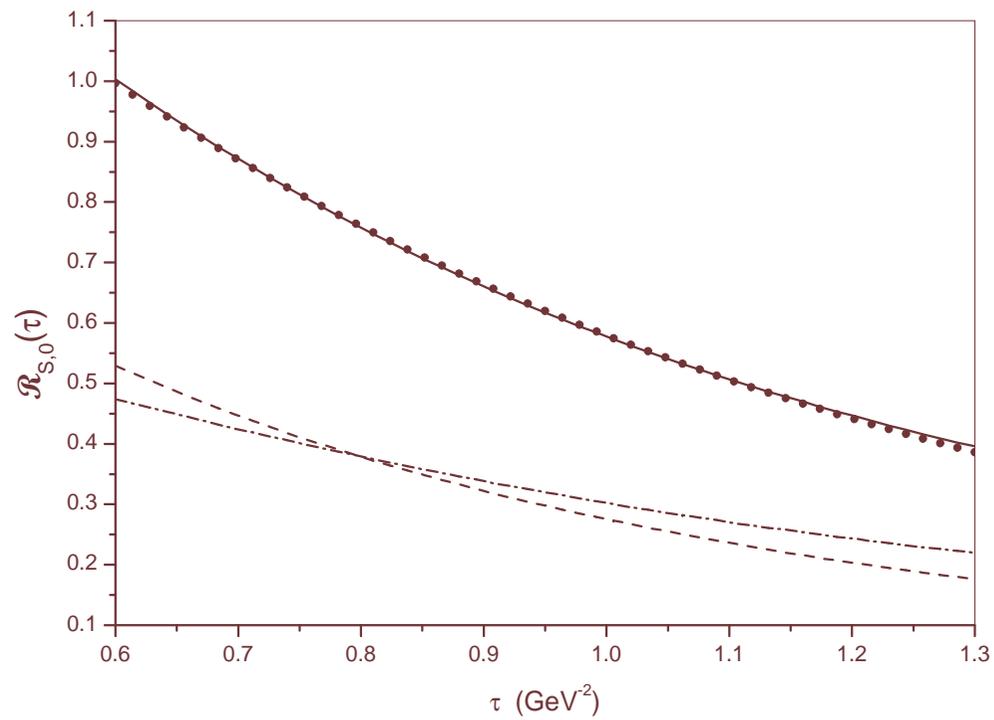
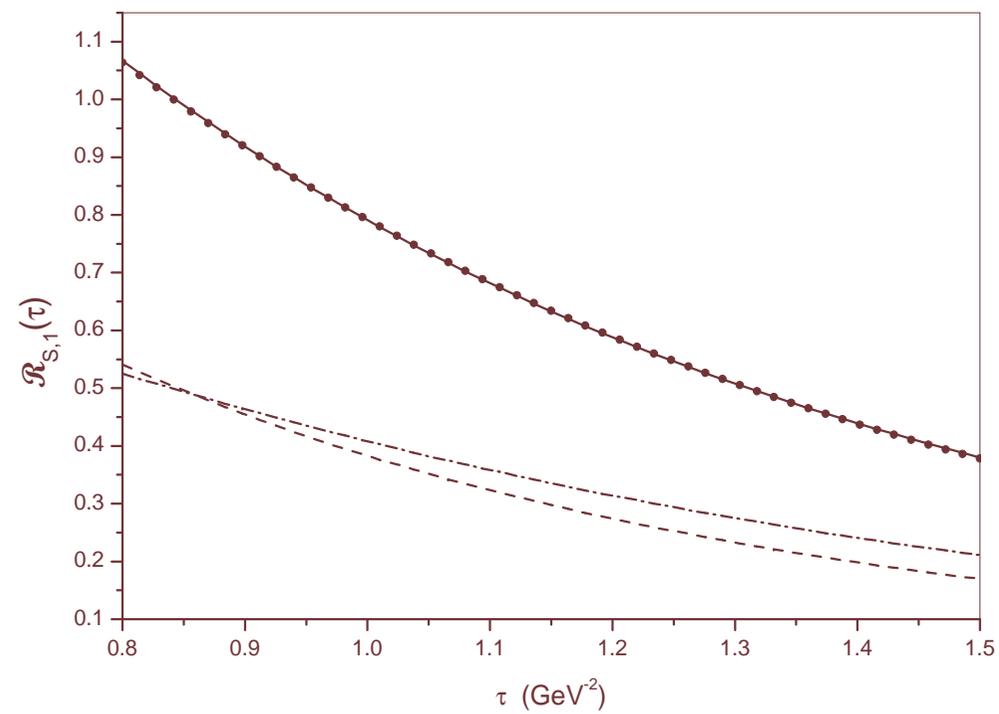
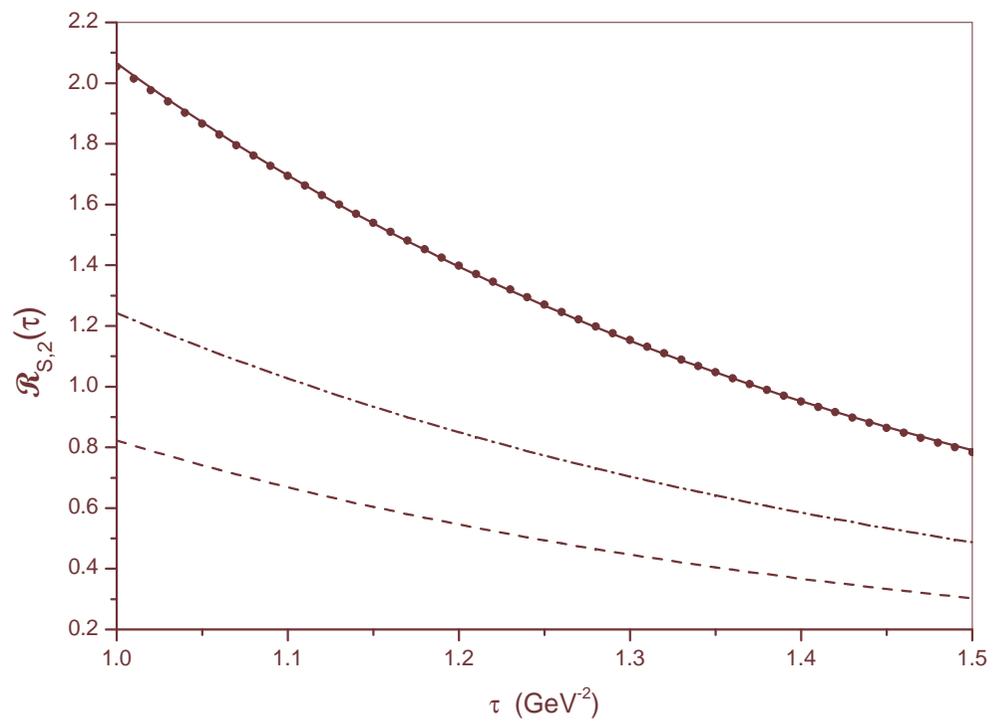
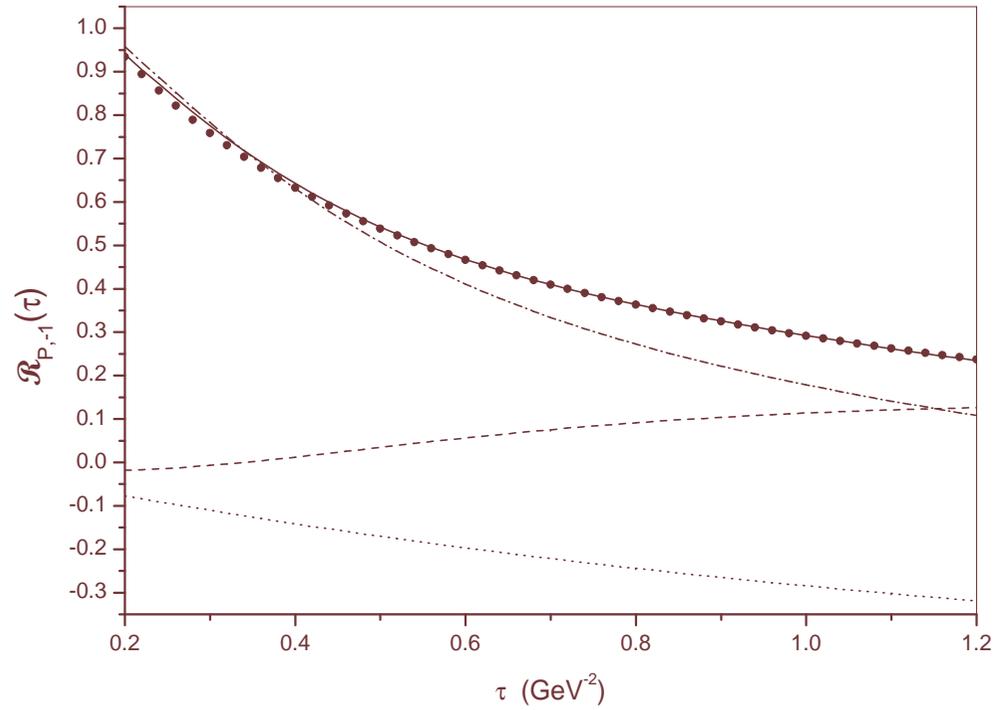
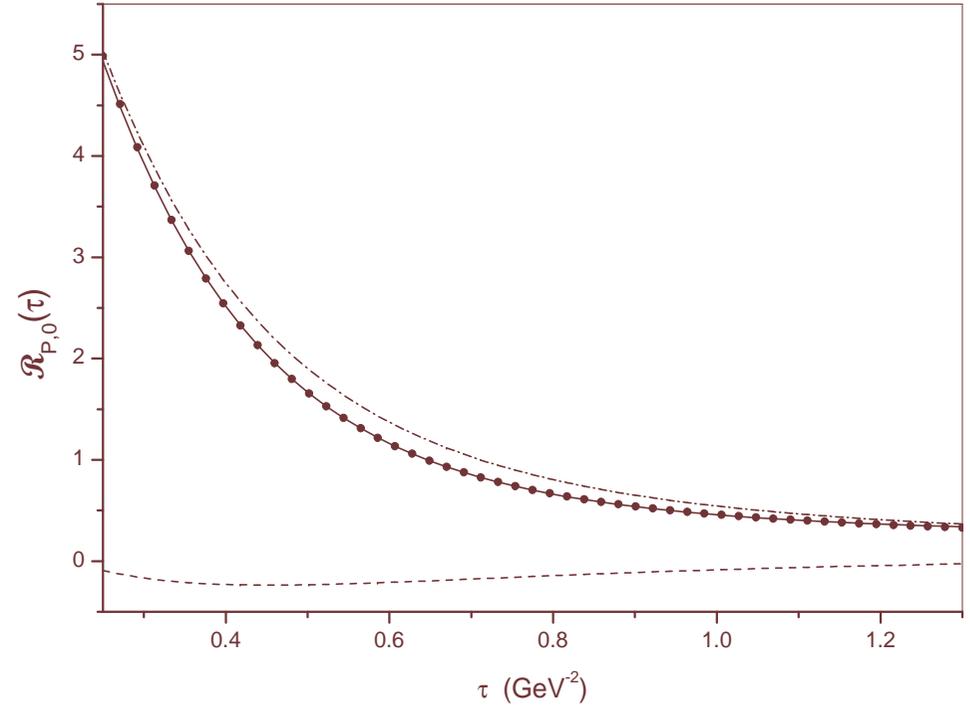
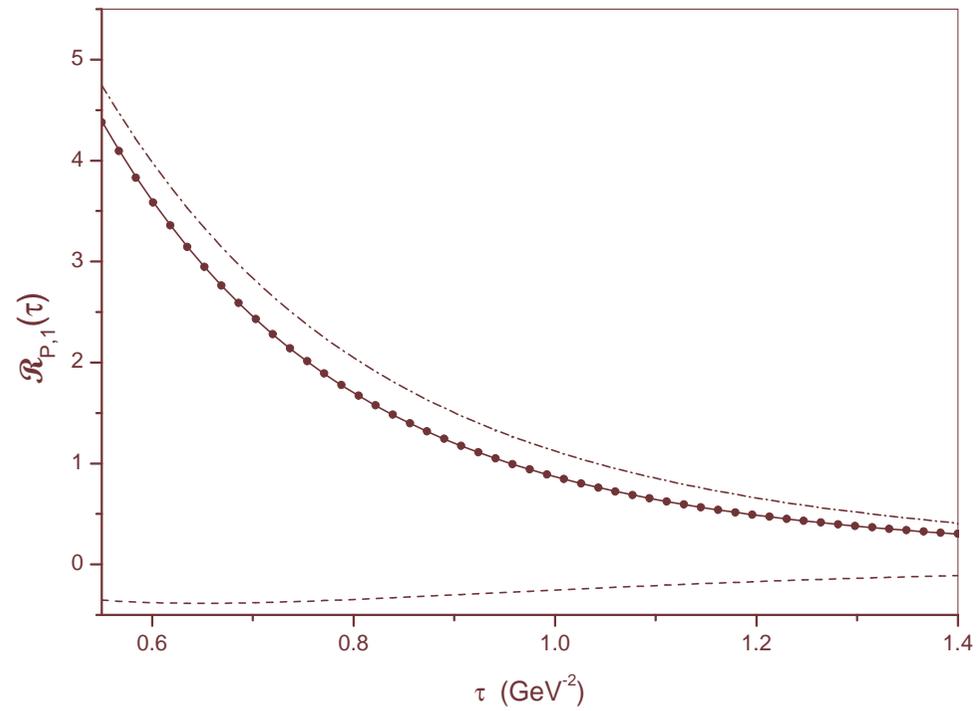
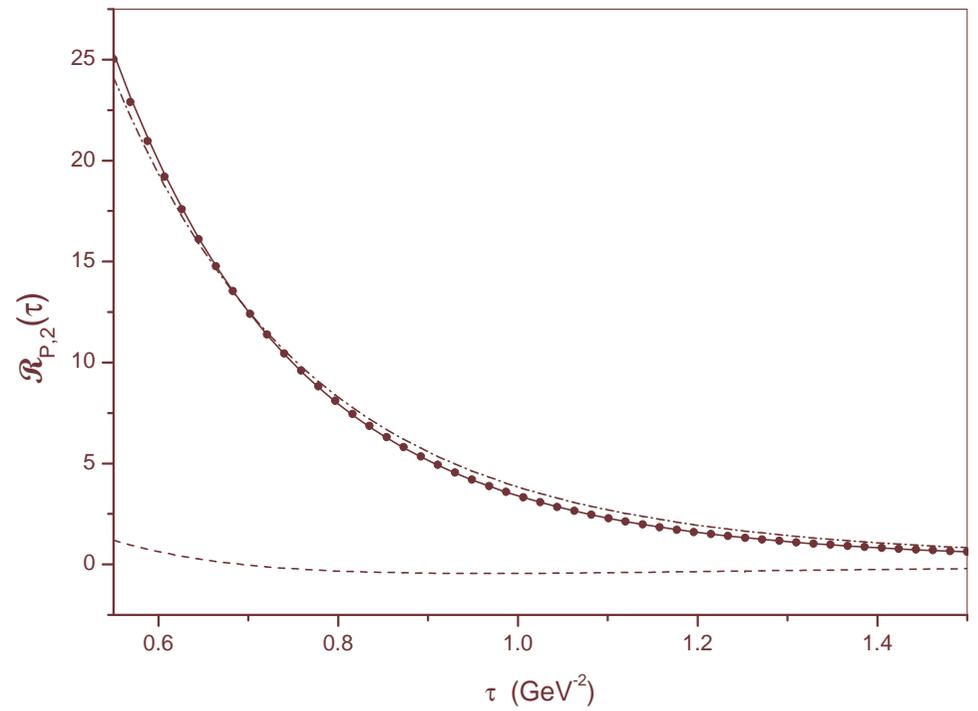
\end{document}